\documentstyle[11pt,fleqn,epsf,twoside,fancyhdr]{report}

\renewcommand{\baselinestretch}{1.50}

\newlength{\captsize}		\let\captsize=\normalsize
\newlength{\captwidth}		
\newlength{\beforetableskip} 	
\newcommand{\capt}[1]{
	\renewcommand{\baselinestretch}{0.9}		
        \begin{minipage}{\captwidth}
	\let\normalsize=\captsize
	\caption[#1]{\sf #1}
	\end{minipage}\\ \vspace{\beforetableskip}}

\makeatletter
	\long\def\@makecaption#1#2{\vskip 10\p@
	\setbox\@tempboxa\hbox{{\bf #1:} #2} %  %% new definition
	\ifdim \wd\@tempboxa >\hsize
           {\bf #1:} #2\par
        \else
           \hbox to\hsize{\box\@tempboxa\hfill}%         %% nocentering
        \fi}
\makeatother     

\newcommand{\SSfootnote}[1]{\renewcommand{\baselinestretch}{0.9}%
	    \footnote{#1}\renewcommand{\baselinestretch}{1.5}}

\begin{document}

\newcommand{\be}{\begin{equation}}
\newcommand{\ee}{\end{equation}}
\newcommand{\bea}{\begin{eqnarray}}
\newcommand{\eea}{\end{eqnarray}}
\newcommand{\sst}{\scriptscriptstyle}

\thispagestyle{empty}
\begin{center}
\vspace{-3.5cm}
{\Large  Universidad de Buenos Aires}\\
\vspace{0.3cm}
\hspace*{-0.5cm}{\Large Facultad de Ciencias Exactas y Naturales}\\
\vspace{0.3cm}
{\Large Departamento de F{\'\i}sica}\\
\vspace{3.0cm}
{\Huge Tesis Doctoral}\\
\vspace{1.5cm}
{\Huge\bf Correcciones Cu\'anticas y}\\
\vspace{0.5cm}
{\Huge\bf Acci\'on Efectiva en Teor\'{\i}a de Campos} \\
\vspace{3.cm}
{\Large Autor: Diego Alejandro Roberto Dalvit}\\
\vspace{0.3cm}
{\Large Director: Dr. Francisco Diego Mazzitelli}\\
\vspace{5.0cm}
{\large Trabajo de Tesis para optar por el t\'{\i}tulo de Doctor en Ciencias
F\'{\i}sicas}\\
\vspace{0.3cm}
{\large Julio de 1998}

\end{center}

\newpage

\thispagestyle{empty}
~
\newpage

\thispagestyle{empty}

\hfill \hfill {\it A los Willies} 

\newpage

\thispagestyle{empty}
~
\newpage

\thispagestyle{empty}

{\Huge\bf Resumen}

\vspace{1cm}

En esta Tesis estudiamos correcciones cu\'anticas a la din\'amica
cl\'asica de valores medios en teor\'{\i}a de campos. Para ello utilizamos
el formalismo de acci\'on efectiva de camino temporal cerrado, con el cual
obtenemos ecuaciones de movimiento reales y causales.

Introducimos una acci\'on efectiva de granulado grueso, que es de utilidad
en el estudio de transiciones de fase en teor\'{\i}a de campos. Derivamos
una ecuaci\'on exacta del grupo de renormalizaci\'on que describe c\'omo 
esta acci\'on var\'{\i}a con la escala de granulado grueso. Desarrollamos
distintos m\'etodos de aproximaci\'on para resolver dicha ecuaci\'on y
obtenemos mejoras no perturbativas para el potencial efectivo para una
teor\'{\i}a escalar autointeractuante. Discutimos adem\'as los aspectos
estoc\'asticos contenidos en esta acci\'on.

Por otro lado, usando la acci\'on efectiva, hallamos correcciones
cu\'anticas de bajas energ\'{\i}as y largas distancias al potencial de
interacci\'on gravitatorio, tratando a la relatividad como una teor\'{\i}a
efectiva de bajas energ\'{\i}as. Incluimos el efecto de campos
cu\'anticos escalares, espinoriales y de gravitones. La inclusi\'on
de fluctuaciones de la m\'etrica hace que las ecuaciones de Einstein
semicl\'asicas dependan de los par\'ametros de fijado de medida, 
y sean por ende no f\'{\i}sicas.
Resolvemos este problema identificando como observable f\'{\i}sico 
a la trayectoria de una part\'{\i}cula de prueba. Mostramos 
expl\'{\i}citamente que las ecuaciones geod\'esicas para dicha part\'{\i}cula
son independientes de los par\'ametros arbitrarios que aparacen en el
fijado de medida.

\vspace*{0.5cm}

\noindent{\sl Palabras claves:} 

Acci\'on efectiva - Teor\'{\i}a cu\'antica de campos - Gravedad 
semicl\'asica - Correcciones cu\'anticas - Grupo de renormalizaci\'on - 
Transiciones de fase - Cosmolog\'{\i}a 

\newpage

\thispagestyle{empty}
~
\newpage

\thispagestyle{empty}

{\Huge\bf Abstract}

\vspace{1cm}

In this Thesis we study quantum corrections to the classical dynamics 
for mean values in field theory. To that end we make use of the formalism
of the closed time path effective action to get real and causal equations
of motion.

We introduce a coarse grained effective action, which is useful in the study of
phase transitions in field theory. We derive an exact renormalization
group equation that describes how this action varies with the coarse
graining scale. We develop different approximation methods to solve that 
equation, and we obtain non perturbative improvements to the effective
potential for a self interacting scalar field theory. We also discuss the
stochastic aspects contained in this action.

On the other hand, using the effective action, we find low energy and
large distance quantum corrections for the gravitational potential, 
treating relativity as an effective low energy theory. We include
the effect of scalar fields, fermions and gravitons. The inclusion
of metric fluctuations causes Einstein semiclassical equations to depend
on the gauge fixing parameters, and they are therefore non physical.
We solve this problem identifying as a physical observable the
trayectory of a test particle. We explicitly show that the geodesic 
equation for such particle is independent of the arbitrary parameters
of the gauge fixing.

\vspace*{0.5cm}


\newpage

\thispagestyle{empty}
~
\newpage

\pagenumbering{roman}
\fancyhead[LE]{\thepage}
\fancyhead[RE]{\sl Indice}
\fancyhead[RO]{\bf \thepage}
\fancyfoot[CE,CO]{}

\tableofcontents
\clearpage

\renewcommand{\chaptermark}[1]{\markboth{\sl #1}{}}
\renewcommand{\sectionmark}[1]{\markright{\sl \thesection . \hspace{0.05cm} 
#1}}

\fancyhead[LE,RO]{\bf \thepage}
\fancyhead[RE]{\leftmark}
\fancyhead[LO]{\rightmark}

\chapter{Introducci\'on}

\pagenumbering{arabic}
\thispagestyle{empty}

El tema general de esta tesis es el c\'alculo de correcciones cu\'anticas
a la din\'amica cl\'asica en teor\'{\i}a de campos. Como punto de partida
para analizar la influencia de las fluctuaciones cu\'anticas, nos 
concentraremos en la acci\'on efectiva, la cual contiene toda la informaci\'on 
sobre los aspectos cu\'anticos del campo. Por un lado, a partir de la acci\'on
efectiva es posible hallar los elementos de matriz de dispersi\'on y estudiar
procesos de interacci\'on entre part\'{\i}culas en f\'{\i}sica de altas
energ\'{\i}as. Por otro lado, la acci\'on efectiva permite estudiar la
evoluci\'on de valores medios de campos cu\'anticos, lo cual es de utilidad
en distintas ramas de la f\'{\i}sica, como por ejemplo en cosmolog\'{\i}a
y mec\'anica estad\'{\i}stica. Para esta tesis es este segundo aspecto el
que ser\'a relevante.

Un rama de la f\'{\i}sica te\'orica en la cual la acci\'on efectiva es 
de gran utilidad es la gravedad 
semicl\'asica, en la cual el inter\'es radica en analizar los efectos 
de campos de materia cu\'anticos sobre el comportamiento de campos 
gravitatorios cl\'asicos \cite{birrell}. 
Las ecuaciones que rigen la din\'amica del
espacio-tiempo incluyendo tales efectos cu\'anticos se llaman ecuaciones 
de Einstein semicl\'asicas, que tienen como fuente el valor de expectaci\'on
del tensor de energ\'{\i}a-impulso de todos los campos existentes, y 
eventualmente tambi\'en el de las fluctuaciones de la m\'etrica. 
En ausencia de una teor\'{\i}a consistente de la gravedad 
cu\'antica, la justificaci\'on de la aproximaci\'on semicl\'asica se basa en 
la suposici\'on de la existencia de un r\'egimen de energ\'{\i}as para la 
gravedad cu\'antica en el cual el fondo gravitatorio se comporte 
cl\'asicamente. 
Las ecuaciones semicl\'asicas han sido utilizadas en la literatura para
estudiar distintos procesos f\'{\i}sicos en los cuales los efectos cu\'anticos
sobre el campo gravitatorio son importantes. Por un lado existe un amplio
conjunto de trabajos sobre la evoluci\'on de campos de materia cu\'anticos
en espacio-tiempo curvo, en los cuales se han estudiado con detalle las
soluciones a las ecuaciones semicl\'asicas \cite{anderson}. 
Los efectos cu\'anticos
pueden haber sido uno de los responsables de las caracter\'{\i}sticas 
observables de nuestro Universo actual, como por ejemplo la isotrop\'{\i}a
a gran escala. Si inicialmente el Universo era anis\'otropo, la expansi\'on
del mismo tuvo como consecuencia la creaci\'on de grandes cantidades de 
part\'{\i}culas, las cuales ``reaccionan'' sobre la m\'etrica del 
espacio-tiempo a trav\'es de las ecuaciones semicl\'asicas. Esto induce
t\'erminos disipativos en las ecuaciones para el campo gravitatorio, que
finalmente conducen a la disipaci\'on de las anisotrop\'{\i}as de partida
\cite{zeldovich, hartlehu, huanisot, jppanisot}. 
Otro campo de inter\'es es el estudio de los
procesos cu\'anticos que intervienen en el colapso gravitatorio y la 
formaci\'on de agujeros negros. Al colapsar una estrella para formar un
agujero negro se produce una perturbaci\'on gravitatoria que induce, por 
efectos cu\'anticos de los campos de materia presentes, la emisi\'on de
un flujo de radiaci\'on con espectro t\'ermico: el agujero negro no es
completamente negro, sino que emite radiaci\'on \cite{hawking}. 
La temperatura de la radiaci\'on es inversamente proporcional a la masa 
del agujero negro, de modo que mientras menor sea la masa de \'este, mayor 
ser\'a el n\'umero de especies de part\'{\i}culas que se crean: fotones, 
neutrinos, pares electr\'on-positr\'on, etc.

Dentro de los aspectos cosmol\'ogicos, el conocimiento de 
la forma exacta de la acci\'on efectiva en modelos inflacionarios juega un rol 
fundamental en la descripci\'on de las transiciones de fase que tuvieron lugar 
en el Universo temprano. Los primeros estudios de transiciones de fase en este
contexto se hicieron para configuraciones de campo constantes, en cuyo caso
la acci\'on efectiva se reduce al denominado potencial efectivo \cite{kolb}. 
Ello permite
deducir aspectos cualitativos de la transici\'on, como por ejemplo decidir
si es de primer o segundo orden. Sin embargo, cerca de la
temperatura cr\'{\i}tica a la cual se produce la transici\'on, el potencial
efectivo resulta insuficiente para describir los procesos fuera de equilibrio
que tienen lugar. En este caso resulta imprescindible evaluar la acci\'on 
efectiva para configuraciones dependientes del tiempo, y a partir de ella 
obtener la din\'amica del par\'ametro de orden. De esta forma es posible
estudiar procesos tales como la creaci\'on de part\'{\i}culas debida a la
amplificaci\'on de las fluctuaciones cu\'anticas, la influencia de dichas
part\'{\i}culas sobre la din\'amica del inflat\'on, y la din\'amica 
autoconsistente del espacio-tiempo \cite{boya1,steve}.

En f\'{\i}sica de altas energ\'{\i}as es tambi\'en de inter\'es estudiar
la evoluci\'on temporal de valores de expectaci\'on de campos en diversas
situaciones. Un ejemplo es la formaci\'on y ulterior evoluci\'on de los 
llamados condensados quirales desorientados que se formar\'{\i}an en
colisiones entre n\'ucleos pesados a altas energ\'{\i}as. 
El plasma de 
quarks y gluones que se genera en la colisi\'on se enfr\'{\i}a al expandirse
y atraviesa la transici\'on de fase quiral de la cromodin\'amica cu\'antica.
En esta forma se producen amplias regiones espaciales correlacionadas, donde
el condensado quiral tiene valor de expectaci\'on no nulo y est\'a orientado
en una direcci\'on incorrecta. El estudio de los aspectos de no equilibrio
de la transici\'on de fase quiral permitir\'{\i}a decidir si durante la 
evoluci\'on se generan inestabilidades que afectan a las fluctuaciones
cu\'anticas y producen eventualmente un crecimiento de las correlaciones y
la formaci\'on de dominios quirales \cite{jppdcc,boyadcc}. 
Relacionado con lo anterior, es tambi\'en
relevante el an\'alisis de los aspectos cu\'anticos y 
estad\'{\i}stico-cin\'eticos de sistemas multipart\'onicos fuera de equilibrio
en f\'{\i}sica de altas energ\'{\i}as. A partir de un formalismo
cin\'etico-cu\'antico basado en la acci\'on efectiva es posible desarrollar
una teor\'{\i}a de transporte para quarks y gluones con el fin de describir
procesos disipativos y dispersivos de sistemas multipart\'onicos en tiempo
real \cite{geiger}.

Todos los casos anteriores tienen en com\'un el hecho de que s\'olo ciertos
grados de libertad son los relevantes para estudiar la din\'amica, mientras
que la influencia de los otros grados de libertad sobre los primeros se 
considera en forma efectiva. Los grados de libertad relevantes forman el 
``sistema'' y los irrelevantes ``el entorno''. Por ejemplo, en gravedad
semicl\'asica, el
sistema de inter\'es es el fondo gravitatorio, mientras que los campos de
materia cu\'anticos sobre los cuales se integra forman el entorno, 
que afecta la din\'amica de la m\'etrica. En inflaci\'on, el sistema 
corresponde a los modos de longitudes de onda mayores 
que el horizonte, mientras
que el entorno est\'a formado por los de longitudes de onda menores
y por el resto de los campos que se acoplan al inflat\'on para producir
recalentamiento.
En el estudio de
las colisiones multipart\'onicas a altas energ\'{\i}as, importa analizar la
din\'amica de los modos livianos (excitaciones colectivas de largo alcance), 
mientras que los modos pesados (excitaciones de corto alcance) son tratados
en forma efectiva, y forman el entorno. Este proceso de extraer a partir de un
n\'umero grande, frecuentemente infinito, de grados de libertad unas pocas
variables que capturen la f\'{\i}sica esencial del sistema completo es 
t\'{\i}pico de la mec\'anica estad\'{\i}stica \cite{goldenfeld}. 
All\'{\i}, a partir del 
c\'alculo de la funci\'on de partici\'on, y previa identificaci\'on de las
variables relevantes del sistema de inter\'es, es posible hacer el promedio
sobre los grados de libertad irrelevantes que forman el entorno. El ejemplo
t\'{\i}pico es el movimiento Browniano cu\'antico, en el cual se estudia la
din\'amica de una part\'{\i}cula browniana pesada acoplada a un conjunto de
osciladores livianos \cite{caldeira,hpz}. 
Bas\'andose en la acci\'on efectiva resultante del 
proceso de promediado (tambi\'en conocida como acci\'on efectiva de granulado 
grueso, o funcional de influencia), se logra tener en cuenta c\'omo los
grados de libertad integrados ``reaccionan'' sobre el 
sistema \cite{feyvernon}. Dicho proceso
de separaci\'on sistema-entorno resulta en aspectos nuevos para la din\'amica
del sistema cu\'antico abierto. La caracter\'{\i}stica m\'as sobresaliente
es la aparici\'on de disipaci\'on, por transferencia de energ\'{\i}a del
sistema al entorno, y de ruido, por efecto de las fluctuaciones del entorno
sobre el sistema. La fuente de ruido puede provenir de fluctuaciones
t\'ermicas o de vac\'{\i}o, y est\'an relacionadas con la disipaci\'on a trav\'es
de la relaci\'on de fluctuaci\'on-disipaci\'on. Estos m\'etodos han sido
utilizados para analizar la transici\'on cu\'antico-cl\'asica mediante el
proceso de decoherencia, es decir la diagonalizaci\'on de la matriz densidad
reducida del sistema por efecto del acoplamiento con el entorno \cite{zurek}.

El esquema de aislaci\'on de un sistema de inter\'es y el tratamiento efectivo
de sus interacciones con el entorno no es exclusivo de la mec\'anica
estad\'{\i}stica. La t\'ecnica del grupo de renormalizaci\'on, usada
frecuentemente en teor\'{\i}a de campos y materia condensada, comparte 
b\'asicamente el mismo esp\'{\i}ritu \cite{wilson}. 
Originalmente esta t\'ecnica fue
introducida mediante la llamada transformaci\'on de bloques de esp\'{\i}n, 
con el objeto de estudiar sistemas de espines en la red \cite{kadanoff}. 
La idea subyacente
es partir de un sistema microsc\'opico de espines, identificar un conjunto
de grados de libertad relevantes (bloques de esp\'{\i}n), promediar
sobre los grados de libertad irrelevantes (espines en cada bloque) y deducir 
finalmente la interacci\'on efectiva entre estas nuevas variables. Se obtiene 
as\'{\i} un mapeo entre los par\'ametros del modelo a dos escalas de 
promediado diferentes. En este proceso el granulado grueso es discreto, 
pero puede generalizarse al caso continuo utilizando la teor\'{\i}a de campos 
basada en la acci\'on efectiva. Aqu\'{\i} se integran las fluctuaciones cuya
longitud de onda sea menor que una dada escala, y se obtiene as\'{\i} una nueva
acci\'on efectiva de granulado grueso. 
La correspondiente ecuaci\'on del grupo de renormalizaci\'on describe
la manera en la cual dicha acci\'on var\'{\i}a con la escala. 
La resoluci\'on de tal ecuaci\'on es una tarea formidable,
y en general requiere m\'etodos de aproximaci\'on, ya sean perturbativos o
no perturbativos. 

La presente tesis doctoral est\'a organizada en la siguiente forma: en el
Cap\'{\i}tulo 2 presentamos una revisi\'on de conceptos b\'asicos de 
teor\'{\i}a de campos y la definici\'on de la acci\'on efectiva. La 
formulaci\'on usual de la acci\'on efectiva adolece de dos problemas 
fundamentales a la hora de estudiar la evoluci\'on temporal de valores de 
expectaci\'on en teor\'{\i}a de campos. Por un lado, las ecuaciones que se
obtienen no son ni reales ni causales. 
Por este motivo describimos la denominada
acci\'on efectiva de camino temporal cerrado, apta para describir procesos
de no equilibrio. Por otro lado, en teor\'{\i}as de medida, 
la acci\'on efectiva usual es una cantidad
que depende (param\'etricamente) de la elecci\'on de las condiciones de medida
impuestas sobre las fluctuaciones cu\'anticas. Diferentes elecciones de esas
condiciones conducen a distintas acciones efectivas. M\'as en general, la
acci\'on efectiva usual no es una cantidad invariante ante 
reparametrizaciones de los campos. Describimos brevemente
la propuesta usualmente utilizada para tratar este problema
en la definici\'on de la acci\'on efectiva.

En el Cap\'{\i}tulo 3 nos ocupamos del primer problema de la formulaci\'on
usual de la acci\'on efectiva, y discutimos m\'etodos de aproximaci\'on no 
perturbativos para la acci\'on efectiva de camino temporal cerrado. 
Introducimos la acci\'on efectiva de granulado grueso,
que es de especial utilidad para estudiar transiciones de fase en teor\'{\i}a
de campos y desarrollamos
las t\'ecnicas del grupo de renormalizaci\'on exacto, que describen
c\'omo var\'{\i}a dicha acci\'on con la escala de granulado. 
En el Cap\'{\i}tulo 4 se describen m\'etodos de 
c\'alculo covariantes de la acci\'on efectiva. 
En la aproximaci\'on a un lazo en las fluctuaciones cu\'anticas,
el problema reside en evaluar determinantes funcionales de operadores
diferenciales de segundo orden. Debido a su complejidad, es necesario 
recurrir a t\'ecnicas de aproximaci\'on. Describimos la aproximaci\'on
(local) de Schwinger-DeWitt, que es una expansi\'on en derivadas del campo de
fondo, y otra t\'ecnica basada en una resumaci\'on de la anterior, que 
identifica los aspectos no locales de la acci\'on efectiva.

Los cap\'{\i}tulos restantes son aplicaciones de la acci\'on 
efectiva al c\'alculo de correcciones cu\'anticas en relatividad general.
Como hemos visto, en gravedad semicl\'asica se trabaja con escalas de
distancias mucho mayores que la distancia de Planck y energ\'{\i}as mucho 
menores que la energ\'{\i}a de Planck, lo cual permite
suponer que los efectos cu\'anticos del campo gravitatorio son despreciables,
y en consecuencia se trata a dicho campo en forma cl\'asica. Sin embargo,
esa suposici\'on no es estrictamente correcta. De acuerdo 
al principio de equivalencia, todas las formas de materia y energ\'{\i}a 
se acoplan a la gravedad con la misma intensidad. En particular, \'esto 
tambi\'en
incluye a la misma energ\'{\i}a gravitatoria: un gravit\'on se acopla 
a un campo gravitatorio externo al igual que cualquier otra excitaci\'on
cu\'antica. En consecuencia, siempre que un campo gravitatorio cl\'asico de
fondo produzca importantes efectos que involucren campos de materia cu\'anticos
(por ejemplo, fotones, ya sean reales o virtuales), tambi\'en debemos esperar
efectos igualmente relevantes relacionados con gravitones. Por lo tanto,
la gravedad cu\'antica entra en forma no trivial en todas las escalas de 
distancias y energ\'{\i}as. 
En resumen, la cuantizaci\'on del campo gravitatorio
es tan importante como la cuantizaci\'on de los campos de materia
\cite{dewitt}. A pesar
de la complicaci\'on que implica la cuantizaci\'on 
de la gravedad, todav\'{\i}a
es posible proceder con una descripci\'on semicl\'asica, escribiendo a la
m\'etrica como un fondo cl\'asico m\'as una fluctuaci\'on cu\'antica 
(gravitones) que se propaga en dicho fondo. Este campo de gravitones representa
perturbaciones linealizadas en el fondo cl\'asico, y puede incluirse junto 
con los otros campos cu\'anticos como parte de la materia en vez de la 
geometr\'{\i}a. Inmediatamente surge un problema con este procedimiento.
Debido al hecho que la constante de acoplamiento de la teor\'{\i}a posee
dimensiones, cada nuevo orden en teor\'{\i}a de perturbaciones conduce
a nuevas divergencias que no poseen la estructura de los ordenes anteriores.
Esto hace que la gravedad sea una teor\'{\i}a no renormalizable.
Ahora bien, la relatividad general, como
toda teor\'{\i}a f\'{\i}sica, es necesariamente provisoria, ya que su validez 
ha sido testeada experimentalmente para un rango limitado de distancias y
energ\'{\i}as. Pensando a la relatividad como una teor\'{\i}a efectiva de
bajas energ\'{\i}as, la no renormalizabilidad no representa un impedimento 
para calcular las correcciones cu\'anticas m\'as importantes a los resultados
cl\'asicos \cite{donoghue1}. 

A modo de ejemplificar los m\'etodos que usaremos para hallar correcciones
cu\'anticas en relatividad general, comenzamos en el 
Cap\'{\i}tulo 5 por el estudio de una caso sencillo de electrodin\'amica 
cu\'antica, correspondiente al apantallamiento de la carga el\'ectrica
por fluctuaciones del vac\'{\i}o. Seguidamente pasamos al tratamiento 
semicl\'asico de la gravedad, y calculamos las modificaciones cu\'anticas
al potencial newtoniano por campos de materia escalares y fermi\'onicos.
Finalmente, en el Cap\'{\i}tulo 6 consideramos el tratamiento de la gravedad
como una teor\'{\i}a efectiva, y nos ocupamos del segundo problema de la
acci\'on efectiva usual. 
Siguiendo los mismos m\'etodos del cap\'{\i}tulo 
anterior, hallamos la contribuci\'on de los gravitones a las ecuaciones
de Einstein y mostramos que su soluci\'on depende de los par\'ametros del
fijado de medida para las fluctuaciones cu\'anticas. Por lo tanto
la cantidad calculada no puede ser un observable. 
Proponemos identificar las cantidades f\'{\i}sicas, y consideramos
el movimiento de una part\'{\i}cula de prueba 
en presencia de gravitones. A partir del
c\'alculo de la acci\'on efectiva, mostramos expl\'{\i}citamente que
las ecuaciones geod\'esicas corregidas por efectos cu\'anticos y el
potencial newtoniano resultan 
independientes del fijado de medida. 
Finalmente en el Cap\'{\i}tulo 7 resumimos nuestras conclusiones.

\newpage

\thispagestyle{empty}
~
\newpage

\chapter{La acci\'on efectiva y sus variantes}

\thispagestyle{empty}

En este cap\'{\i}tulo introducimos algunos conceptos fundamentales de la
teor\'{\i}a de campos para estudiar la evoluci\'on de valores medios. 
Comenzamos con la definici\'on convencional de la acci\'on efectiva, apta
para describir problemas de dispersi\'on. El estudio de la evoluci\'on temporal
de valores medios basado en esta acci\'on efectiva presenta dos problemas.
Primero, las ecuaciones de movimiento efectivas no resultan ni reales ni 
causales, lo cual dificulta la interpretaci\'on de las mismas para describir
problemas a condiciones iniciales. 
Segundo, en el caso de teor\'{\i}as de medida, tanto la acci\'on como las
ecuaciones de movimiento dependen param\'etricamente de la forma en que se
fija la medida al integrar las fluctuaciones cu\'anticas, lo cual implica que
dichas ecuaciones de evoluci\'on no son f\'{\i}sicas. Como soluci\'on al primer
problema describimos el formalismo de la acci\'on efectiva de camino temporal
cerrado, apropiado para tratar sistemas fuera de equilibrio. En cuanto al 
segundo problema, describimos brevemente el formalismo de la acci\'on efectiva
de Vilkovisky-DeWitt.

%%%%%%%%%%%%%%%%%%%%%%%%%%%%%%%%%%%%%%%%%%%%%%%%%%%%%%%%%%%%%%%%%%%%%%%%%%%%

\section{La acci\'on efectiva usual}

Con el objeto de introducir la definici\'on usual de acci\'on efectiva nos
restringiremos al caso m\'as sencillo posible, es decir a una teor\'{\i}a 
(que no sea de medida) para
un \'unico campo bos\'onico en espacio-tiempo plano.
Consideremos pues un campo $\phi$ cuya acci\'on cl\'asica es $S(\phi)$.
La amplitud de persistencia del vac\'{\i}o 
en presencia de una fuente cl\'asica $J$ se define como
\be
Z[J]=\exp \left(\frac{i}{\hbar} W[J] \right) =
\langle 0, {\rm out} | 0, {\rm in} \rangle_J =
\langle 0, {\rm out} | {\rm T}  \exp \left( \frac{i}{\hbar} \int d^4x J(x) 
\phi_{\rm H}(x) \right) | 0, {\rm in} \rangle ,
\ee
donde $\phi_{\rm H}(x)$ 
es el operador de campo en representaci\'on de Heisenberg 
para la teor\'{\i}a sin fuente, y ${\rm T}$ 
es el operador de ordenamiento temporal.
El estado de vac\'{\i}o ``in'', $|0, {\rm in} \rangle$, est\'a definido en 
 $t \rightarrow -\infty$, mientras que el estado de vac\'{\i}o ``out'',  
 $|0, {\rm out} \rangle$, en el futuro remoto. En un contexto general fuera de
equilibrio, como por ejemplo un espacio-tiempo curvo dependiente del tiempo,
o para campos de fondo dependientes del tiempo, ambos estados no son 
necesariamente equivalentes. Podemos escribir tambi\'en una representaci\'on
de $Z[J]$ mediante integral de camino
\be
Z[J] = \int D \phi \exp\left[\frac{i}{\hbar} \left( S[\phi] + \int d^4x J(x) 
\phi(x) \right) \right] ,
\label{zetausual}
\ee
donde la integral funcional es la suma sobre las historias cl\'asicas del 
campo $\phi$ que en el pasado asint\'otico tienen frecuencias negativas
(es decir, los modos espaciales de Fourier de $\phi$ tienen una dependencia
temporal de la forma $\exp(i w t)$, con $w>0$), y frecuencias positivas
($\propto \exp(-i w t)$) en el futuro asint\'otico 
\SSfootnote{Estas condiciones
de contorno son equivalentes a agregar una peque\~na parte imaginaria 
 $-i \epsilon \phi^2$ en la acci\'on cl\'asica, con $\epsilon>0$.}.
Se asume que la interacci\'on con la fuente se apaga en estas regiones 
asint\'oticas.

Diferenciando la funcional $W[J]$ respecto a la fuente se genera el elemento
de matriz del operador de campo entre estados asint\'oticos de vac\'{\i}o
\be
\varphi(x) = \frac{\delta W[J]}{\delta J(x)} = 
\frac{ \langle 0, {\rm out}| \phi(x) | 0, {\rm in} \rangle_J}
{ \langle 0, {\rm out} | 0, {\rm in} \rangle_J} . 
\label{campmedusual}
\ee
Asumiendo que la relaci\'on anterior se puede invertir para expresar la
fuente $J$ como una funcional del campo (cl\'asico) $\varphi$, la acci\'on 
efectiva usual queda definida como la transformada de Legendre de la funcional
 $W[J]$ en la forma
\be
S_{\rm ef}[\varphi] = W[J[\varphi]] - \int d^4x J[\varphi](x) \varphi(x) .
\label{EAinout}
\ee
A partir de esta ecuaci\'on podemos derivar la din\'amica para el campo
efectivo $\varphi$ en la forma
\be
\frac{\delta S_{\rm ef}[\varphi]}{\delta \varphi} = - J ,
\label{eqninout}
\ee
que expresa las correcciones cu\'anticas a la ecuaci\'on cl\'asica como un 
problema variacional.
En particular, para $J=0$, obtenemos la ecuaci\'on din\'amica para el campo
cl\'asico $\varphi[J=0]$ en ausencia de fuentes.

La acci\'on efectiva puede escribirse como la siguiente integral de camino
\be
S_{\rm ef}[\varphi] = -i \hbar \ln 
\left\{
\int D \phi \exp \left[
\frac{i}{\hbar} \left(
S[\phi] - \frac{\delta S_{\rm ef}[\varphi]}{\delta \varphi} (\phi - \varphi)
\right)
\right]
\right\} ,
\label{AEusual}
\ee
que es una ecuaci\'on integro-diferencial para la acci\'on efectiva, y que 
posee una soluci\'on formal
\be
S_{\rm ef}[\varphi] = S[\varphi] - \frac{i \hbar}{2} \ln {\rm det} (A^{-1}) +
\Gamma_1[\varphi] ,
\ee
donde $A(x,x')$ es la segunda derivada funcional de la acci\'on cl\'asica
respecto al campo. El segundo t\'ermino es la correcci\'on a un lazo, mientras
que el tercero representa las correcciones a lazos superiores.
Esta acci\'on efectiva (o acci\'on efectiva ``in-out'') es la generatriz 
de funciones de v\'ertice de $n$ puntos
(diagramas de una part\'{\i}cula irreducible sin l\'{\i}neas externas),
incluyendo tanto sus componentes cl\'asicas como sus correcciones cu\'anticas.
A di\-fe\-ren\-cia de las fun\-ciones de v\'er\-tice cl\'a\-si\-cas, 
que en ge\-ne\-ral aparecen
en una cantidad finita y son locales, el n\'umero de funciones de v\'ertice 
cu\'anticas es siempre infinito y son no locales \cite{odibook}. Utilizando
estas funciones de v\'ertice y las correspondientes reglas de Feynman 
se pueden calcular todos los distintos elementos de matriz de dispersi\'on.
Es \'este el \'ambito en el cual la acci\'on efectiva usual tiene sus
mayores aplicaciones, permitiendo el estudio sistem\'atico de procesos
de dispersi\'on en la f\'{\i}sica de part\'{\i}culas. 

Es importante 
remarcar algunos aspectos: por un lado $\varphi[J=0]$ es 
un elemento de matriz entre estados asint\'oticos
en ausencia de fuente, y es en general una cantidad compleja. Las cantidades
de inter\'es para estudiar evoluciones din\'amicas son en cambio valores de
expectaci\'on de observables tomados respecto al mismo estado. Por otro lado,
la ecuaci\'on de movimiento (\ref{eqninout}) 
no posee una estructura causal. Por
la propia definici\'on de la acci\'on efectiva in-out, la soluci\'on a las 
ecuaciones efectivas es un problema de contorno (estados asint\'oticos
in y out) en vez de un problema de condiciones iniciales. 
Finalmente, el formalismo usual de acci\'on efectiva 
s\'olo permite analizar situaciones de vac\'{\i}o asint\'otico - situaciones m\'as 
generales, como por ejemplo estados t\'ermicos asint\'oticos, no est\'an 
contenidas. 

%%%%%%%%%%%%%%%%%%%%%%%%%%%%%%%%%%%%%%%%%%%%%%%%%%%%%%%%%%%%%%%%%%%%%%%%%%%%
 
\section{La acci\'on efectiva de camino temporal cerrado}

Como vimos anteriormente, es conveniente introducir un nuevo formalismo
que permita obtener ecuaciones de movimiento reales y causales para valores de
expectaci\'on, dado un conjunto de datos iniciales de Cauchy. Dicho formalismo
se conoce en la actualidad con el nombre general de Schwinger-Keldysh
\cite{ctp}. 
Versiones no relativistas del mismo han sido aplicadas a problemas de 
mec\'anica estad\'{\i}stica y materia condensada. Aqu\'{\i} describiremos
la formulaci\'on relativista de teor\'{\i}a de campos para el caso sencillo
de un campo bos\'onico en espacio-tiempo plano.

La idea es analizar c\'omo evoluciona el estado de vac\'{\i}o
\SSfootnote{La formulaci\'on permite incluir estados m\'as generales descriptos
por una matriz densidad $\rho$. A lo largo de esta Tesis nos restringiremos
a estados iniciales de vac\'{\i}o.}
in (asociado al estado asint\'otico en $t \rightarrow -\infty$) 
en presencia de dos 
fuentes externas distintas $J_+(x)$ y $J_-(x)$ y comparar los resultados en 
una base com\'un $ \{ |\psi \rangle \}$ en un tiempo futuro $T$. Definamos
la funcional $Z[J_+,J_-]$ como
\be
Z[J_+,J_-] = \exp (\frac{i}{\hbar} W[J_+,J_-]) =
 \sum_{| \psi \rangle} \langle 0, {\rm in} | \psi \rangle_{J_-} 
 \langle \psi | 0, {\rm in} \rangle_{J_+} , 
\ee
que tambi\'en puede escribirse en la forma
\be
\hspace{-0.5cm}
Z[J_+,J_-] = \int D\psi \langle 0, {\rm in} | {\hat{\rm T}}
e^{-\frac{i}{\hbar} \int_{-\infty}^T dt \int d^3x J_-(x) \phi_H(x)} | \psi 
\rangle 
\langle \psi | {\rm T} e^{\frac{i}{\hbar} \int_{-\infty}^T dt \int d^3x
J_+(x) \phi_H(x)} | 0, {\rm in} \rangle ,
\ee
donde $\hat{\rm T}$ 
denota el operador de ordenamiento antitemporal. Podemos pensar
a esta funcional como la integral sobre 
un camino temporal cerrado en el plano
complejo temporal. En efecto, dicho camino $\cal C$ se ilustra en la 
figura 2.1: 
el camino va desde el vac\'{\i}o in en el infinito pasado hasta el estado
 $| \psi \rangle$ definido sobre una superficie com\'un $\Sigma$ de tiempo
constante $T$ en el futuro, en presencia de una fuente $J_+$ definida en la
rama positiva del camino. Luego regresa por la rama negativa desde $T$ hasta
el vac\'{\i}o in en presencia de la fuente $J_-$. 

A partir de esta funcional podemos
calcular valores de expectaci\'on de observables f\'{\i}sicos a un tiempo 
finito. Por ejemplo, las funciones de Green de 2 puntos
\bea
G_{++}(x,x') = & i \left. \frac{\partial}{i \partial J_+(x)} 
                    \frac{\partial}{i \partial J_+(x')} Z 
                    \right|_{J_+=J_-=0}  
& = i \langle 0, {\rm in}| {\rm T} \phi_H(x) \phi_H(x') |
0, {\rm in} \rangle \nonumber \\
G_{--}(x,x') = & i  \left. \frac{\partial}{-i \partial J_-(x)} 
                    \frac{\partial}{-i \partial J_-(x')} Z 
                    \right|_{J_+=J_-=0}  
& = i \langle 0, {\rm in}| \hat{\rm T}  
\phi_H(x) \phi_H(x') | 0, {\rm in} \rangle
\nonumber \\
G_{+-}(x,x') = & i \left. \frac{\partial}{i \partial J_+(x)} 
                    \frac{\partial}{-i \partial J_-(x')} Z 
                    \right|_{J_+=J_-=0}  
& = i \langle 0, {\rm in}| \phi_H(x') \phi_H(x) | 0, {\rm in} 
\rangle \nonumber \\
G_{-+}(x,x') = & i  \left. \frac{\partial}{-i \partial J_-(x)} 
                     \frac{\partial}{i \partial J_+(x')} Z 
                     \right|_{J_+=J_-=0}  
& = i \langle 0, {\rm in}| \phi_H(x) \phi_H(x') | 0, {\rm in} \rangle ,
\eea
que corresponden, en la teor\'{\i}a libre, a los propagadores de Feynman,
Dyson, y de Wightman de frecuencias positivas y negativas, respectivamente.
En consecuencia, a diferencia de la formulaci\'on usual, en este nuevo 
formalismo las funciones de Green tienen un nuevo \'{\i}ndice, que indica
los posibles ordenamientos temporales a lo largo del camino temporal cerrado.
As\'{\i}, por ejemplo, hay cuatro funciones de Green de dos puntos, que
pueden representarse en una matriz $G_{ab}$ de $2 \times 2$.

\begin{figure}[t]
\centering \leavevmode
\epsfxsize=10cm
\epsfbox{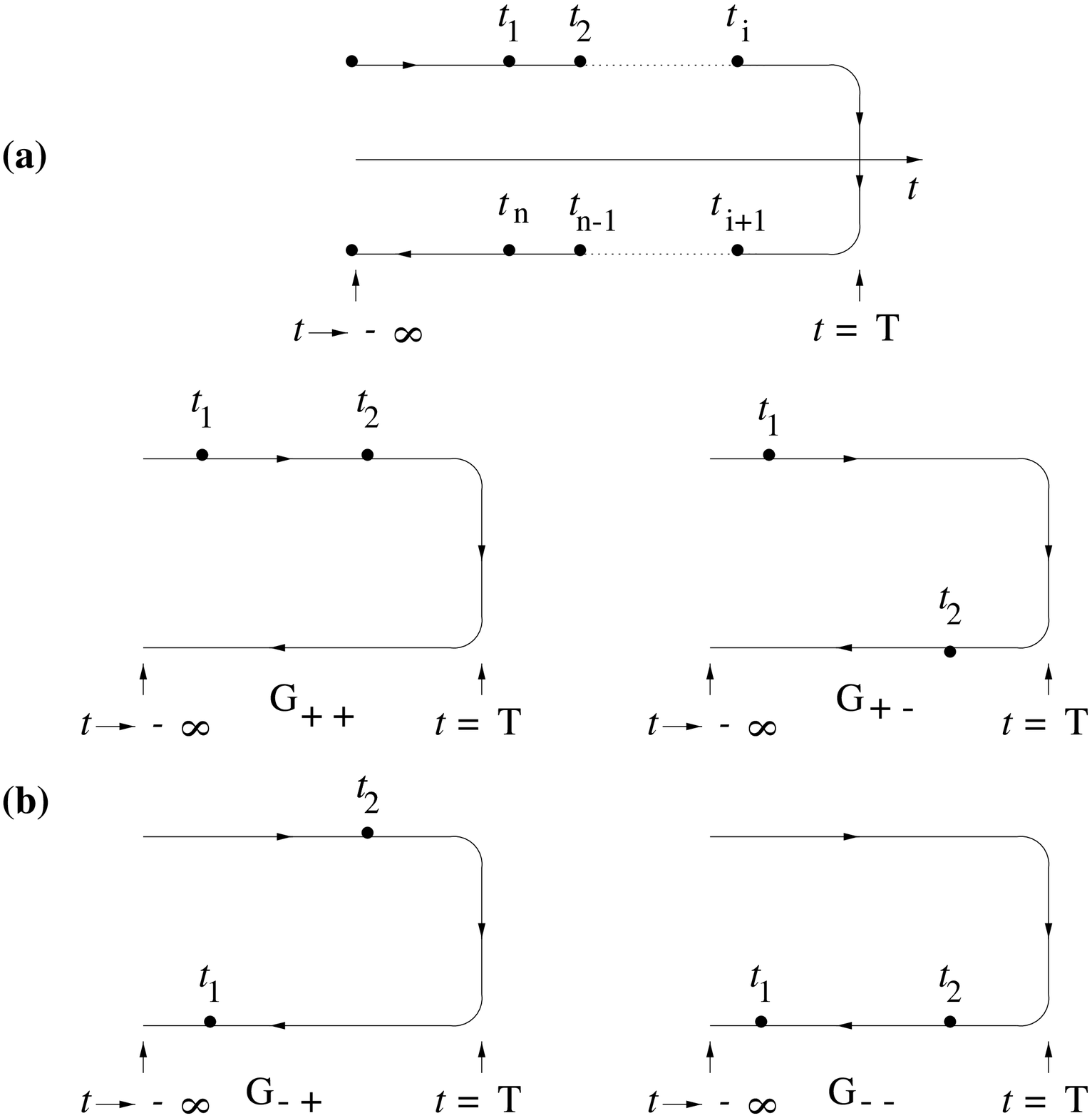}
\setlength{\captwidth}{12cm}
\capt{(a) El camino temporal cerrado en el plano complejo $t$ para la
evoluci\'on de valores de expectaci\'on. Cualquier punto en la rama positiva
 $(t \rightarrow -\infty, t=T)$ corresponde a un instante 
anterior a cualquier
punto en la rama negativa $(t=T , t \rightarrow -\infty)$. 
(b) Los cuatro posibles
ordenamientos temporales $(t_1,t_2)$ en los argumentos de las funciones de 
Green de dos puntos $G(x,y)=G(t_1,{\bf {x}};t_2, {\bf {x'}})$, correspondientes
a $G_{++}(x,x')$, $G_{+-}(x,x')$, $G_{-+}(x,x')$ y $G_{--}(x,x')$.}
\end{figure}

La funcional $Z[J_+,J_-]$ tiene tambi\'en una representaci\'on funcional
\be
Z[J_+,J_-] = \int D \phi_+ D \phi_- e^{ \frac{i}{\hbar} 
( S[\phi_+] + \int d^4x J_+(x) \phi_+(x) )} e^{- \frac{i}{\hbar} 
( S[\phi_-] + \int d^4x J_-(x) \phi_-(x) )} .
\ee
La integral funcional es sobre todas las configuraciones tales que i)
 $\phi_+=\phi_-$ en $t=T$, ii) $\phi_+$ ($\phi_-$) consiste 
en modos de frecuencia negativa (positiva) en $t \rightarrow -\infty$. 
N\'otese que las integrales funcionales sobre $\phi_+$ y $\phi_-$ en la
expresi\'on anterior no son independientes, pues est\'an conectadas por la 
condici\'on de contorno en la hipersuperficie en com\'un en el futuro remoto.
Diferenciando obtenemos los campos cl\'asicos
\be
\varphi_{\pm} (x) = \pm \frac{\delta W[J_+,J_-]}{\delta J_{\pm}(x)}
= \pm \frac{ \langle 0,{\rm in} | \phi_{\pm}(x) | 0, {\rm in} \rangle}
           { \langle 0,{\rm in} | 0, {\rm in} \rangle} ,
\ee
cuya acci\'on efectiva de camino temporal cerrado es
\be
S_{\rm ef}[\varphi_+,\varphi_-] = W[J_+,J_-] - \int d^4x J_+(x) \varphi_+(x) +
\int d^4x J_-(x) \varphi_-(x) ,
\ee
que conduce a las ecuaciones para los campos medios (en presencia de fuentes)
\be
\frac{\delta S_{\rm ef}[\varphi_+,\varphi_-]}{\delta \varphi_{\pm}}
= \mp J_{\pm} .
\label{ecCTP}
\ee
Si las fuentes son iguales, $J_+=J_-\equiv J$, entonces los campos medios 
tambi\'en lo son, $\varphi_+=\varphi_- \equiv \varphi$, que es el valor de 
expectaci\'on del campo en representaci\'on de Heisenberg con respecto al 
estado evolucionado a partir del $| 0, {\rm in} \rangle$ en presencia de la
fuente $J$. En particular, para el caso sin fuentes, obtenemos las ecuaciones
din\'amicas del valor de expectaci\'on $\varphi[J=0]$. Notar que estas 
ecuaciones din\'amicas no resultan inmediatamente de un simple principio
variacional, a diferencia de las ecuaciones in-out. Estrictamente hablando,
tenemos ecuaciones efectivas para valores medios $\varphi$, pero no tenemos
una acci\'on efectiva para los mismos. Dos puntos importantes
a recalcar son los siguientes: primero, $\varphi$ es real, por ser un valor de
expectaci\'on. Esto implica que las ecuaciones de movimiento que satisface
tambi\'en son reales. Segundo, dichas ecuaciones de movimiento tienen una
estructura causal, lo cual permite analizar problemas a condiciones iniciales 
\cite{jordan86}. 

La acci\'on efectiva de camino temporal cerrado tambi\'en admite una 
representaci\'on funcional
\be
S_{\rm ef}[\varphi_+,\varphi_-] = - i \hbar \ln 
\left\{
\int_{{\rm CTC}} D \phi_+ D \phi_-
\exp \left[
\frac{i}{\hbar}
\left(
S[\phi_+] - S[\phi_-] - \frac{\delta S_{\rm ef}}{\delta \varphi_a} 
( \phi_a - \varphi_a)
\right)
\right]
\right\} ,
\ee
donde $a=\pm$ y el s\'{\i}mbolo CTC 
denota que la integral funcional debe hacerse
con las particulares condiciones de contorno del camino temporal cerrado.
La expansi\'on en lazos de esta ecuaci\'on integro-diferencial es
\be
S_{\rm ef}[\varphi_+,\varphi_-] = S[\varphi_+] - S[\varphi_-] 
- \frac{i \hbar}{2} \ln {\rm det} (A_{ab}^{-1}) + 
\Gamma_1[\varphi_+,\varphi_-] ,
\ee
donde $A_{ab}$ es la matriz de segundas derivadas de la acci\'on cl\'asica.
El segundo t\'ermino es la correcci\'on a un lazo, y el tercero a lazos 
superiores.

Formalmente la acci\'on efectiva de camino temporal cerrado es an\'aloga a la
usual, excepto que los propagadores son ahora matrices $2 \times 2$ que
satisfacen las condiciones de contorno arriba mencionadas. 
Correspondientemente, las reglas de Feynman son las mismas, pero cada 
l\'{\i}nea
de un propagador en un diagrama puede corresponder a cualquiera de las
cuatro componentes de las funciones de Green de dos puntos. Adem\'as es la
generadora de diagramas de una part\'{\i}cula irreducible. 

Como vimos, la acci\'on efectiva da la din\'amica para valores medios de la 
teor\'{\i}a, es decir para las funciones de Green de un punto. Si se deseara 
adem\'as estudiar la din\'amica de las funciones de Green de dos o m\'as
puntos (varianzas, etc.), entonces ser\'{\i}a necesario generalizar la 
definici\'on de la funci\'on de partici\'on $Z$ introduciendo m\'ultiples
fuentes no locales ($K_2(x,y)$,$K_3(x,y,z)$, etc.), lo que conduce a 
nuevas representaciones de la acci\'on efectiva \cite{tombo,calhu88}.
A partir de ellas es posible derivar el conjunto (infinito) de ecuaciones
acopladas para la din\'amica de las funciones de Green, que es el an\'alogo
en teor\'{\i}a cu\'antica de campos de la jerarqu\'{\i}a BBGKY de la mec\'anica
estad\'{\i}stica. El truncamiento a m\'as bajo orden corresponde a la 
din\'amica de los valores medios.
        
Existen dos m\'etodos alternativos para obtener las mismas ecuaciones de 
movimiento reales y causales para los valores medios, para el caso en el que
el estado inicial sea el vac\'{\i}o. El primer m\'etodo consiste en
calcular la acci\'on efectiva usual en signatura eucl\'{\i}dea 
 $S_{\rm ef}^{\rm E}[\varphi]$,
calcular luego la ecuaci\'on de movimiento eucl\'{\i}dea 
 $\delta S_{\rm ef}^{\rm E}[\varphi]/\delta \varphi = 0$,
y en la expresi\'on final ir a signatura lorentziana reemplazando todas las
funciones de Green eucl\'{\i}deas por las funciones de Green retardadas,
 $\Box^{-1}_{\rm E} \rightarrow \Box^{-1}_{\rm ret}$. 
Este reemplazo no debe ser
hecho a nivel de la acci\'on, sino a nivel de las ecuaciones de movimiento.
En esta forma se obtienen las mismas ecuaciones para valores medios deducidas
mediante el formalismo in-in \cite{gospel}.
El segundo m\'etodo parte del c\'alculo de la acci\'on efectiva usual en
signatura lorentziana, es decir la acci\'on efectiva in-out. Se procede a
hallar las correspondientes ecuaciones de movimiento que, seg\'un vimos,
no son ni reales ni causales. Finalmente se reemplazan todos los propagadores
por dos veces su parte real y causal. As\'{\i} se llega a las mismas ecuaciones
din\'amicas anteriores \cite{jordan87}. A modo de ejemplo, en el cap\'{\i}tulo
5 calcularemos mediante estos tres m\'etodos las ecuaciones efectivas (reales
y causales) para el campo electromagn\'etico en electrodin\'amica cu\'antica.

%%%%%%%%%%%%%%%%%%%%%%%%%%%%%%%%%%%%%%%%%%%%%%%%%%%%%%%%%%%%%%%%%%%%%%%%%%%%%

\section{La acci\'on efectiva de Vilkovisky-DeWitt}

Todas las distintas acciones efectivas definidas hasta ahora presentan un
problema en su propia definici\'on, que se pone de manifiesto cuando la 
teor\'{\i}a en consideraci\'on es de medida. Dicha patolog\'{\i}a consiste en
la dependencia de la acci\'on y de las ecuaciones de movimiento en la
elecci\'on de las condiciones de medida. 

Una parte del problema es el hecho que la acci\'on no es invariante de 
medida. Usualmente, cuando se fija la 
medida para los campos cu\'anticos, autom\'aticamente se fija una medida para 
los campos medios. Existe sin embargo la posibilidad de fijar la medida de los
primeros sin fijar la de los segundos, es decir obtener una acci\'on efectiva
invariante de medida. Ello se logra utilizando el m\'etodo de cuantizaci\'on
mediante campos de fondo \cite{dewitt67}. La otra parte del problema
radica en el hecho que la acci\'on efectiva as\'{\i} obtenida 
todav\'{\i}a depende
{\it en forma param\'etrica} de la elecci\'on de las condiciones de 
medida para las fluctuaciones. Diferentes elecciones de estas condiciones
conducen a diferentes acciones efectivas covariantes.

Una manifestaci\'on m\'as general de la no unicidad de la acci\'on efectiva
es su dependencia en la elecci\'on de la parametrizaci\'on para los campos
cu\'anticos. Este problema es inherente tanto a teor\'{\i}as de medida como
a teor\'{\i}as que no son de medida.
%\SSfootnote{Existe un procedimiento de abelianizaci\'on mediante el cual,
%bajo ciertas suposiciones sobre las condiciones de medida, es posible
%reducir una teor\'{\i}a de medida a una de no medida. Un cambio en las 
%condiciones de medida se traduce en una reparametrizaci\'on del campo
%en la nueva teor\'{\i}a \cite{batalin}.} 
A modo de ilustraci\'on, consideremos
un campo escalar complejo $\Phi$ con autointeracci\'on 
 $(\Phi^{\dagger} \Phi)^2$ en espacio-tiempo plano. La acci\'on cl\'asica es
\be
S= \int d^nx \left[\partial^{\mu} \Phi^{\dagger} \partial_{\mu} \Phi -
m^2 \Phi^{\dagger} \Phi - \frac{\lambda}{6} (\Phi^{\dagger} \Phi)^2 \right] .
\ee
Dos posibles parametrizaciones del campo son en t\'erminos de su parte real
e imaginaria, $\Phi= \frac{1}{\sqrt{2}} (\phi_1 + i \phi_2)$, o en
t\'erminos de su m\'odulo y fase, $\Phi= \frac{1}{\sqrt{2}} \rho e^{i \theta}$.
En la primera parametrizaci\'on, la acci\'on cl\'asica es
\be
S_1[\phi_1,\phi_2] = \int d^nx \left[ \frac{1}{2} \partial^{\mu} \phi^a 
\partial_{\mu} \phi^a - \frac{1}{2} m^2 \phi^a \phi^a - \frac{\lambda}{4!}
(\phi^2 \phi^2)^2 \right] ,
\ee
donde los \'{\i}ndices repetidos denotan suma sobre 1 y 2. En la segunda
parametrizaci\'on, la acci\'on cl\'asica es
\be
S_2[\rho,\theta] = \int d^nx \left[ \frac{1}{2} \partial^{\mu} \rho 
\partial_{\mu} \rho + 
\frac{1}{2} \rho^2 \partial^{\mu} \theta \partial_{\mu} \theta - 
\frac{1}{2} m^2 \rho^2 - \frac{\lambda}{4!} \rho^4 \right] .
\ee
Si bien $S_1$ y $S_2$ corresponden simplemente a distintas parametrizaciones
de la misma teor\'{\i}a cl\'asica, las respectivas acciones efectivas son
distintas. En \cite{kunst} se calcula el potencial efectivo
a un lazo y se obtienen resultados distintos. Para este ejemplo en particular,
si nos concentramos en la teor\'{\i}a no interactuante ($\lambda=0$), en la 
primera parametrizaci\'on la acci\'on efectiva a un lazo es exacta, pues
la teor\'{\i}a es cuadr\'atica, mientras que en la segunda parametrizaci\'on
la acci\'on cl\'asica no es cuadr\'atica, y por lo tanto la aproximaci\'on
a un lazo no es exacta. En general, la diferencia
entre las acciones efectivas calculadas en distintas parametrizaciones
es proporcional al extremal, es decir, a las ecuaciones de movimiento
cl\'asicas.

Otro ejemplo de la dependencia ante reparametrizaciones y fijado de medida
se encuentra al calcular el potencial efectivo $V_{\beta}(\varphi)$
dependiente de temperatura en electrodin\'amica escalar, y la correspondiente
temperatura cr\'{\i}tica $\beta_c=(k T_c)^{-1}$, definida como el valor
de $\beta$ para el cual la masa efectiva 
$m^2(\beta) = 2 \partial^2 V_{\beta} / \partial \varphi^2 |_{\varphi=0}$
se anula \cite{dolan}. 
En la parametrizaci\'on ``cartesiana'' $(\phi_1,\phi_2)$ 
para el campo escalar,
el potencial efectivo a un lazo depende de los par\'ametros de fijado de 
medida, mientras que la temperatura cr\'{\i}tica resulta independiente de
dichos par\'ametros. Por otro lado, en la parametrizaci\'on ``polar'' 
$(\rho,\theta)$,
tanto el potencial como la temperatura cr\'{\i}tica resultan independientes
del fijado de medida \cite{kunst2}. 
Sin embargo, el valor de $\beta_c$ que se obtiene
en esta parametrizaci\'on difiere del obtenido en la parametrizaci\'on
anterior. Se debe decidir entonces cu\'al es la respuesta ``correcta''.
Para que el c\'alculo en lazos sea consistente se debe verificar que los
efectos de orden superior sean despreciables en ambas parametrizaciones. Ello
se ha verificado en la parametrizaci\'on $(\phi_1,\phi_2)$, pero no as\'{\i}
en la parametrizaci\'on $(\rho,\theta)$, debido a problemas con la
renormalizabilidad. En conclusi\'on, no se debe confiar en el c\'alculo
de la temperatura cr\'{\i}tica en la parametrizaci\'on polar, pues los efectos
de orden superior no se pueden despreciar consistentemente.

A pesar de la no unicidad de la acci\'on efectiva usual, los elementos
de matriz de dispersi\'on son invariantes ante reparametrizaciones de los
campos y fijados de medida. Ello se debe al hecho que la matriz de 
dispersi\'on se obtiene a partir de la acci\'on efectiva evaluada en la
capa de masa, de modo que s\'{\i} est\'a un\'{\i}vocamente definida.

Existe una definici\'on de una acci\'on efectiva que resuelve los problemas
de la dependencia ante reparametrizaciones y el fijado de medida de la acci\'on
efectiva usual, y que se 
conoce con el nombre de acci\'on efectiva de Vilkovisky-DeWitt \cite{vdwea}.
Los detalles de la definici\'on y de sus propiedades pueden encontrarse en 
estas referencias, aqu\'{\i} daremos una somera descripci\'on del
formalismo. 

Representemos por $\phi^i$ 
\SSfootnote{Usamos la notaci\'on de DeWitt, en la cual el \'{\i}ndice 
$i$ representa tanto un \'{\i}ndice 
continuo (espacio-temporal) como discreto (interno). Indices 
repetidos denotan una integral sobre los \'{i}ndices 
continuos y una suma sobre 
los discretos.} a un campo bos\'onico gen\'erico cuya acci\'on
cl\'asica es $S[\phi]$. La formulaci\'on de Vilkovisky-DeWitt (VDW) es
geom\'etrica: se considera que el campo $\phi^i$ es una coordenada en una 
variedad de campos ${\cal F}$, y se asocia a dicha variedad una conexi\'on 
af\'{\i}n $\Gamma^i_{mn}(\phi)$ y una 
m\'etrica $G_{mn}$. Una reparametrizaci\'on
del campo equivale entonces a un difeomorfismo en la variedad ${\cal F}$. A
diferencia de la funcional generatriz usual $Z[J]$, en la cual el campo se
acopla linealmente con la fuente $J$ (ver ec.(\ref{zetausual})), en el nuevo
formalismo un acoplamiento del tipo $\phi^i J_i$ no es covariante, por ser el
campo una coordenada. Se debe entonces modificar la definici\'on de
funcional generatriz.

Teniendo una conexi\'on podemos construir una cantidad
 $\sigma^i(\phi,\varphi)$ que depende de dos puntos en la variedad, el
campo cu\'antico $\phi^i$ y el campo de fondo $\varphi^i$. Est\'a definida
como el vector tangente a la geod\'esica que conecta $\phi$ y $\varphi$, 
siendo tangente en el punto $\varphi$ y dirigido de $\phi$ a $\varphi$.
Adem\'as, es un vector respecto del punto $\varphi$ y un escalar respecto
del punto $\phi$. Su expansi\'on en potencias de $(\phi-\varphi)$ es
\be
- \sigma^i(\varphi,\phi) = (\phi^i-\varphi^i) + \frac{1}{2}
\Gamma^i_{mn}(\varphi) (\phi^m-\varphi^m) (\phi^n-\varphi^n) + \ldots .
\ee
La propuesta de VDW es reemplazar el acoplamiento 
(no covariante) $\phi^i J_i$ por
 $[\varphi^i - \sigma^i(\varphi,\phi)] J_i$, y definir la siguiente funcional
generatriz
\be
Z^{\rm VDW}[J] = \int D \phi \exp \frac{i}{\hbar} 
\left\{
S[\phi] + [\varphi^i - \sigma^i(\varphi,\phi)] J_i
\right\} .
\ee
La definici\'on del campo de fondo $\varphi$ viene dada por
\be
\langle \sigma^i(\varphi,\phi) \rangle =0 .
\ee
Dado que $\sigma^i(\varphi,\phi)$ es un escalar respecto del punto de 
integraci\'on $\phi$, esta definici\'on de campo de fondo es independiente
de la parametrizaci\'on. Notar que cuando $\Gamma^i_{mn} \equiv 0$, 
esta definici\'on
coincide con 
la definici\'on usual de campo medio (ver ec.(\ref{campmedusual})).
Entonces la acci\'on efectiva de Vilkovisky-DeWitt es (compararla con la
acci\'on efectiva usual, ec.(\ref{AEusual}))
\be
S_{\rm ef}^{\rm VDW}[\varphi] = -i \hbar \ln 
\int D \phi \exp \frac{i}{\hbar} 
\left\{
S[\phi] + \sigma^i(\varphi,\phi) \frac{\delta S_{\rm ef}^{\rm VDW}[\varphi]}
{\delta \varphi^i}
\right\} .
\ee
Por construcci\'on, esta acci\'on efectiva no depende de la parametrizaci\'on
del campo cu\'antico ni del fijado de medida. Para definirla completamente
falta determinar la conexi\'on afin y la m\'etrica. Para ello Vilkovisky
impone tres condiciones sobre la conexi\'on: a) debe determinarse a partir
de la acci\'on cl\'asica mediante alguna regla universal; b) debe anularse
para un campo libre, en la parametrizaci\'on en la cual la acci\'on cl\'asica
es cuadr\'atica; y c) debe ser ``ultralocal'', es decir debe contener s\'olo
campos y funciones deltas sin diferenciar. 

En muchos casos existe una elecci\'on natural para la m\'etrica $G_{mn}$ de la
variedad ${\cal F}$, que se lee del t\'ermino cin\'etico de $S[\phi]$. La
conexi\'on, en tales situaciones, es un s\'{\i}mbolo de Christoffel,
\be
\Gamma^i_{mn} = \frac{1}{2} G^{ik} 
( \partial_m G_{km} + \partial_n G_{km} - \partial_k G_{mn}) .
\ee
Por ejemplo, en el caso de la electrodin\'amica escalar que hemos mencionado
m\'as arriba, para la parametrizaci\'on $(\phi_1,\phi_2)$, la componente
de la m\'etrica que involucra al campo escalar 
es $G_{mn} = \delta_{mn}$, ya que dicha parametrizaci\'on es el an\'alogo
a coordenadas cartesianas en la va\-rie\-dad $\cal F$ 
\SSfootnote{En coordenadas 
cartesianas, la acci\'on efectiva de Vilkovisky-DeWitt
para teor\'{\i}as de Yang-Mills es igual a la acci\'on efectiva usual calculada
en la medida de Landau-DeWitt \cite{rebhan}. 
Dado que la temperatura cr\'{\i}tica $\beta_c$
obtenida mediante la acci\'on efectiva usual en parametrizaci\'on cartesiana
es independiente del fijado de medida, entonces se sigue que la acci\'on
de Vilkovisky-DeWitt reproducir\'a dicho resultado, para {\it todas} las
parametrizaciones.}.
Sin embargo, hay muchos otros casos en los cuales la elecci\'on no es
un\'{\i}voca, 
quedando par\'ametros arbitrarios en la definici\'on de la m\'etrica
 $G_{mn}$ \cite{odivdw,kantvdw}. Entonces, si bien $S_{\rm ef}^{\rm VDW}$ 
resuelve el problema de la no invariancia ante reparametrizaciones y de
la dependencia en el fijado de medida para las fluctuaciones cu\'anticas, 
mantiene cierta arbitrariedad en la elecci\'on de la m\'etrica y la conexi\'on
af\'{\i}n: la acci\'on de Vilkovisky-DeWitt no est\'a un\'{\i}vocamente 
determinada.
Estos par\'ametros arbitrarios aparecen tanto en la acci\'on como en las
ecuaciones de movimiento para los campos de fondo. Por lo tanto, la
acci\'on de VDW  no es una soluci\'on satisfactoria puesto que no es
invariante ante el fijado de la m\'etrica de la variedad de campos. Para 
decirlo en forma cruda, en este sentido, el formalismo de VDW es m\'as un 
placebo que una panacea \cite{panacea}.

Ahora bien, ?` cu\'al es la verdadera necesidad de tener un formalismo
para una acci\'on efectiva que sea invariante ante el fijado de medida?
La dependencia en los par\'ametros de fijado de medida de la acci\'on efectiva
(in-out o in-in) y de las ecuaciones de movimiento para los campos medios no 
es en s\'{\i} necesariamente inaceptable, pues bien podr\'{\i}a ocurrir que 
las distintas elecciones describan una \'unica y misma teor\'{\i}a. 
El punto importante
es identificar las verdaderas cantidades f\'{\i}sicas observables, las cuales,
naturalmente, no deber\'{\i}an
depender de dichos par\'ametros arbitrarios. 
Un ejemplo de \'esto es la temperatura cr\'{\i}tica $\beta_c$ en 
electrodin\'amica escalar, que supuestamente es un observable. Otro claro
ejemplo se encuentra al calcular el potencial efectivo para un
campo de Higgs (que es una teor\'{\i}a de medida) \cite{aitchison}. 
Tanto el potencial efectivo como el estado base dependen del par\'ametro 
 $\lambda$ con el cual se fija la medida de las fluctuaciones cu\'anticas,
es decir $V=V(\phi,\lambda)$ y $\bar{\phi}= \bar{\phi}(\lambda)$. Sin embargo,
todas las cantidades f\'{\i}sicas son independientes de $\lambda$. Por ejemplo,
el valor del m\'{\i}nimo del potencial, $V(\bar{\phi},\lambda)$, que est\'a 
relacionado con la escala de ruptura de simetr\'{\i}a, y la masa del Higgs,
 $m^2(\bar{\phi},\lambda)$, resultan independientes
del par\'ametro arbitrario pues la dependencia expl\'{\i}cita en $\lambda$
se cancela con la dependencia impl\'{\i}cita a trav\'es de $\bar{\phi}$
gracias a las llamadas identidades de Nielsen \cite{nielsen} 
\bea
\frac{d V}{d \lambda} = 
\left. \frac{\partial V}{\partial \lambda}\right)_{\bar{\phi}} +
\left. \frac{\partial V}{\partial {\bar{\phi}}}\right)_{\lambda} 
\frac{d {\bar{\phi}}}{d \lambda} = 0  & ~~~~ &
\frac{d m^2}{d \lambda} = 
\left. \frac{\partial m^2}{\partial \lambda}\right)_{\bar{\phi}} +
\left. \frac{\partial m^2}{\partial {\bar{\phi}}}\right)_{\lambda} 
\frac{d {\bar{\phi}}}{d \lambda} = 0 .
\eea

En el cap\'{\i}tulo 6 
calcularemos la acci\'on efectiva (in-out/in-in) para la gravedad, que es una
teor\'{\i}a de medida. Si bien dicha acci\'on y las ecuaciones del movimiento
para el fondo gravitatorio dependen del fijado de medida, veremos que un 
observable particular (las geod\'esicas de una part\'{\i}cula de prueba)
es independiente de los par\'ametros de medida. Es de esperar que el mismo
c\'alculo realizado a partir de la acci\'on efectiva de Vilkovisky-DeWitt 
conduzca a una cancelaci\'on de los par\'ametros de la m\'etrica en el espacio
de campos y lleve a los mismos resultados obtenidos con la acci\'on efectiva
(in-out/in-in).

\newpage

\thispagestyle{empty}
~
\newpage

\chapter{La acci\'on efectiva de granulado grueso}

\thispagestyle{empty}

En este cap\'{\i}tulo definimos la acci\'on efectiva de granulado grueso,
que es de especial utilidad en el estudio de transiciones de fase en 
teor\'{\i}a cu\'antica de campos. Partimos de la acci\'on de camino temporal
cerrado (CTC) e integramos las fluctuaciones cu\'anticas de longitudes de onda
menores que un valor cr\'{\i}tico. Derivamos una ecuaci\'on exacta CTC del
grupo de renormalizaci\'on, que describe c\'omo la acci\'on efectiva depende
de la escala de granulado grueso \cite{nosctp}. 
Esta es una complicada ecuaci\'on 
integro-diferencial que requiere de m\'etodos de aproximaci\'on. Mediante una
expansi\'on en derivadas del campo (aproximaci\'on adiab\'atica) analizamos
la forma en que la acci\'on efectiva de granulado grueso fluye ante el
grupo de renormalizaci\'on. Tambi\'en discutimos los 
aspectos estoc\'asticos contenidos en la acci\'on de granulado grueso.

%%%%%%%%%%%%%%%%%%%%%%%%%%%%%%%%%%%%%%%%%%%%%%%%%%%%%%%%%%%%%%%%%%%%%%%%%%%%

\section{Su definici\'on}

El estudio de transiciones de fase en teor\'{\i}a cu\'antica de campos es de 
gran inter\'es en cosmolog\'{\i}a y f\'{\i}sica de part\'{\i}culas. Los
primeros trabajos sobre el tema se basaron en el uso del potencial efectivo,
que se obtiene a partir de la acci\'on efectiva al evaluarla sobre 
configuraciones constantes. Dicho potencial es \'util s\'olo en situaciones
cuasiest\'aticas, y con \'el no es posible analizar los aspectos fuera de
equilibrio de la transici\'on. Para ello es necesario considerar la acci\'on
efectiva completa, y en particular su versi\'on de camino temporal cerrado.

Usualmente las transiciones de fase ocurren v\'{\i}a 
la formaci\'on y crecimiento 
de dominios espaciales. Dentro de estos dominios, el par\'ametro de orden
de la transici\'on evoluciona din\'amicamente y el inter\'es yace en calcular
esa evoluci\'on temporal. La forma habitual de hacerlo es partir de la
acci\'on efectiva CTC y evaluarla para configuraciones de campo que dependan 
s\'olo del tiempo. Presumiblemente dichas configuraciones corresponden al
par\'ametro de orden dentro de un dominio t\'{\i}pico 
\cite{bdv,hartree,gleiser}. Nosotros procederemos
de una manera diferente, inspir\'andonos en la metodolog\'{\i}a que se sigue
en el contexto de materia condensada \cite{goldenfeld,gss}. 
Haremos un granulado grueso del campo
hasta una escala de longitud
 $\Lambda^{-1}$ comparable con el tama\~no inicial de un dominio. 
Por simplicidad consideraremos un campo escalar autointeractuante
 $\lambda \phi^4$ en espacio-tiempo plano.
Definimos una acci\'on efectiva de granulado grueso CTC,
 $S_{\Lambda}(\phi_{+},\phi_{-})$, que es b\'asicamente
la acci\'on CTC en la cual solamente ciertos modos (los de 
 $|{\bf q}| > \Lambda$) han sido integrados. Su definici\'on es
\SSfootnote{De ahora en m\'as usaremos unidades $\hbar=c=1$.}
\be
e^{i S_{\Lambda}(\phi_+,\phi_-)} \equiv \int_{\rm CTC}
\prod_{\Lambda_0>|{\bf q}| > 
\Lambda} {\cal D}[\phi_+({\bf q},t)] 
{\cal D}[\phi_-({\bf q},t)] \; e^{i S_{\rm cl}(\phi_+,\phi_-)} ,
\ee 
donde $S_{\rm cl}(\phi_+,\phi_-) = S_{\rm cl}(\phi_+) - S_{\rm cl}(\phi_-)$ 
es la 
diferencia de las acciones cl\'asicas en las dos ramas del camino temporal
cerrado. El granulado grueso lo logramos mediante una frecuencia de corte
 $\Lambda$. $\Lambda_0$ es una frecuencia de corte ultravioleta que
usamos para regularizar. Las integrales funcionales sobre los modos de
longitud de onda corta se hacen siguiendo la prescripci\'on de  
integraci\'on en el camino temporal cerrado: los campos $\phi_+$ 
y $\phi_-$ tiene frecuencias negativas y positivas respectivamente, en el 
pasado $-T$, y coinciden en el futuro $T$. 
Como resultado de la traza sobre los modos de longitud de onda corta (grados
de libertad del ``entorno''), la acci\'on efectiva de granulado grueso 
(que depende de los modos de longitud de onda larga, el ``sistema'') posee
una parte imaginaria, relacionada con el ruido, y una parte real, asociada
a la disipaci\'on. 

Esta acci\'on efectiva de granulado grueso fue introducida originalmente
en \cite{silarg} para estudiar cosmolog\'{\i}a inflacionaria, y evaluada
en forma perturbativa en \cite{fernando} para analizar la decoherencia
del sector de longitudes de onda larga en la teor\'{\i}a $\lambda \phi^4$.
En esp\'{\i}ritu es similar a la versi\'on eucl\'{\i}dea propuesta con 
anterioridad \cite{wegner,wette1,liao1,polsch,hasen,morris1,boni}. La principal
diferencia entre ambas acciones es que la acci\'on de granulado grueso
eucl\'{\i}dea promedia el campo sobre volumenes espacio-temporales, mientras
que la versi\'on de camino temporal cerrado lo hace sobre volumenes 
espaciales, y por lo tanto, es m\'as adecuada para analizar situaciones 
fuera de equilibrio. Ambas versiones tienen en com\'un el hecho de que 
interpolan entre la teor\'{\i}a desnuda en $\Lambda=\Lambda_0$ y la teor\'{\i}a
f\'{\i}sica en la escala de granulado grueso. Existen 
otras versiones de acciones efectivas de granulado grueso, como por 
ejemplo la de \cite{atta}, que interpola entre la teor\'{\i}a f\'{\i}sica
a temperatura nula $T=0$ y la teor\'{\i}a f\'{\i}sica a temperatura finita
$T \neq 0$.

%%%%%%%%%%%%%%%%%%%%%%%%%%%%%%%%%%%%%%%%%%%%%%%%%%%%%%%%%%%%%%%%%%%%%%%%%%%

\section{Aproximaci\'on a un lazo}

Al igual que con las diferentes versiones de acciones efectivas
del cap\'{\i}tulo anterior, en el presente caso es tambi\'en necesario 
utilizar m\'etodos de aproximaci\'on. Un m\'etodo posible es hacer 
perturbaciones en la constante de acoplamiento $\lambda$, suponi\'endola
peque\~na \cite{fernando}. 
Otra posibilidad, que seguiremos aqu\'{\i} a modo ilustrativo,
es hacer una aproximaci\'on a un lazo, en cuyo caso la acci\'on puede
escribirse como $S_{\Lambda}(\phi_+,\phi_-) = S_{\rm cl}(\phi_+) - 
S_{\rm cl}(\phi_-) + \Delta S_{\Lambda}(\phi_+,\phi_-)$, 
donde el \'ultimo t\'ermino es lineal
en $\hbar$. Para ello expresamos el campo como
 $\phi_{\pm} \rightarrow \phi_{\pm}(t) + \varphi_{\pm}$, donde las 
fluctuaciones $\varphi_{\pm}$ contienen modos espaciales con 
 $|{\bf q}| > \Lambda$. Notar que estamos asumiendo que el \'unico modo
con $|{\bf q}| < \Lambda$ es el modo espacialmente homog\'eneo
(${\bf q}=0$), que depende del tiempo. La correcci\'on a un lazo es entonces
\begin{eqnarray}
e^{i \Delta S_{\Lambda}(\phi_+(t),\phi_-(t))}& = &
\int_{\rm CTC} \prod_{\Lambda_0>|{\bf q}| > 
\Lambda} {\cal D}[\varphi_+] {\cal D}[\varphi_-] \;
e^{
\frac{i}{2} \int dt \int \frac{d^3q}{(2\pi)^3} \left[
\varphi_+ \; \frac{\delta^2 S_{\rm cl}}{\delta \phi_+ \delta \phi_+} \; 
\varphi_+ \; - \;
\varphi_- \; \frac{\delta^2 S_{\rm cl}}{\delta \phi_- \delta \phi_-} \;
\varphi_- \right] 
}  \nonumber \\
&&~~~~~~~~~~~~~~~~~~~~~~~~~~~~~~
\times e^{
\frac{i}{2} \int^{'} \frac{d^3q}{(2\pi)^3} \int dt
\frac{d}{dt} \left[
\varphi_+ \; \dot{\varphi}_+ \; - \; \varphi_- \; \dot{\varphi}_- 
\right] ,
}
\end{eqnarray}
donde las derivadas funcionales est\'an evaluadas en  $\varphi_{\pm}=0$. La
acci\'on para las fluctuaciones cu\'anticas es la de un campo escalar
libre con una masa  $M^2_{\pm}=V''(\phi_{\pm})$, donde $V$ es el potencial
en la acci\'on cl\'asica desnuda. Sus modos espaciales de Fourier son,
entonces, osciladores arm\'onicos con una frecuencia que depende del
tiempo,  $w^2_{q,{\pm}}(t) = q^2 + V''(\phi_{\pm}(t))$, con $q=|{\bf q}|$.
Al ser la integral funcional cuadr\'atica, su c\'alculo es sencillo, 
resultando 
\begin{equation}
\Delta S_{\Lambda}(\phi_+(t),\phi_-(t)) =
\frac{i}{2} \int_{\Lambda_{0}>\vert {\bf q}\vert >\Lambda}  
{d^3 q\over (2\pi)^3}
\ln [ g_-({\bf q},T) {\dot g}_+({\bf q},T) - 
      g_+({\bf q},T) {\dot g}_-({\bf q},T) ] ,
\label{1looP}
\end{equation}
donde los modos $g_{\pm}$ son soluciones de ${\ddot g_{\pm}}({\bf q},t) + 
w^2_{q,\pm}(t) g_{\pm}({\bf q},t)=0$ satisfaciendo las condiciones CTC
en el pasado y teniendo una normalizaci\'on arbitraria en el futuro.

La ecuaci\'on de movimiento real y causal para el campo $\phi(t)$,
se obtiene derivando funcionalmente la acci\'on respecto a $\phi_+$ y
poniendo $\phi_+=\phi_-=\phi$. Obtenemos
\begin{eqnarray}
{\ddot \phi} + V'(\phi) + \frac{1}{2} V'''(\phi) 
\int_{\Lambda_{0}>\vert\bf q\vert >\Lambda}  {d^3 q\over (2\pi)^3}
|g({\bf q},t)|^2 & = &0 \nonumber \\
{\ddot g}({\bf q},t) + (q^2 + V''(\phi)) g({\bf q},t) & = & 0 .
\end{eqnarray}

A\'un en la aproximaci\'on a un lazo, la acci\'on efectiva es un objeto    
muy complicado y debemos recurrir a aproximaciones adicionales para obtener
resultados anal\'{\i}ticos. La m\'as simple es la aproximaci\'on 
adiab\'atica \cite{birrell}, 
en la cual se desprecian las excitaciones del campo
 $\varphi_{\pm}$ debidas a la dependencia temporal del campo de fondo
 $\phi_{\pm}$. Empezamos entonces por escribir los modos en la forma
\begin{equation}
g_{\pm}({\bf q},t) = \frac{1}{\sqrt{2 W_q(\phi_{\pm}(t))}} e^{\pm i 
\int_{-T}^{t}  dt' W_q(\phi_{\pm}(t'))} ,
\end{equation}
donde las funciones $W_q(\phi_{\pm})$ satisfacen
\begin{equation}
W^2_q + {1\over 2}\left( {\ddot W_q\over W_q}-{3\over 2}\left(
{\dot W_q\over W_q}\right )^2\right)=w^2_q .
\label{W}
\end{equation}
La aproximaci\'on adiab\'atica consiste en resolver esta ecuaci\'on en una
expansi\'on en derivadas del campo de fondo. El resultado es
\begin{equation}
W^2_q= q^2 + V''+ {5\over 16}\left[
\left({V'''\over q^2+V''}\right )^2-{V''''\over 4(q^2+V'')}\right ]
\dot\phi^2-{V'''\over 4(q^2+V'')}\ddot \phi + \ldots ,
\label{Wapprox}
\end{equation}
donde los t\'erminos suspensivos denotan derivadas de orden superior.
Entonces la acci\'on de gra\-nu\-la\-do grueso CTC puede expresarse en la forma
\SSfootnote{Omitimos un 
t\'ermino de superficie evaluado en el tiempo final $T$.}
$\Delta S_{\Lambda}(\phi_+,\phi_-)=\Delta S_{\Lambda}(\phi_+)
- \Delta S_{\Lambda}(\phi_-)$, donde
\begin{equation}
\Delta S_{\Lambda}(\phi)={1\over 2}\int dt\int_{\Lambda}^
{\Lambda_{0}} {d^3 q\over (2\pi)^3}\left (
-\sqrt {q^2+V''}+{\dot\phi^2\over 32}{V'''^2\over (q^2+V'')^{5/2}}
\right ) .
\label {gamma1loop}
\end{equation}
Es importante recalcar que en esta aproximaci\'on la acci\'on de granulado
grueso CTC se separa en dos partes, una para cada rama del camino temporal
cerrado, y no hay t\'erminos que acoplen ambas ramas. Por lo tanto, se
pierden los importantes aspectos estoc\'asticos que est\'an contenidos en
la acci\'on de camino temporal cerrado. 

Incluyendo la parte cl\'asica, la acci\'on efectiva total resulta
\begin{equation}
S_{\Lambda}(\phi)=\int dt 
\left ( - V_{\Lambda}(\phi) + {1\over 2}(1+Z_{\Lambda}
(\phi ))\dot\phi^2 + \ldots \right) ,
\label{s1loop}
\end{equation}
donde
\begin{eqnarray}
V_{\Lambda}&=& V+{1\over 2}\int_{\Lambda}^{\Lambda_{0}}
{d^3 q\over (2\pi)^3}  \sqrt{q^2+V''}  \nonumber\\
&= & V + \frac{1}{4 \pi^2} \left[
\frac{\Lambda_0}{4} \sqrt{\Lambda_0^2+V''} (\frac{V''}{2}+\Lambda_0^2) -
\frac{V''^2}{8} \ln(\Lambda_0+\sqrt{\Lambda_0^2+V''})  \right. \nonumber \\
&& ~~~~~~~~~~~~~~\left. - \frac{\Lambda}{4} \sqrt{\Lambda^2+V''} 
(\frac{V''}{2}+\Lambda^2) +
\frac{V''^2}{8} \ln(\Lambda+\sqrt{\Lambda^2+V''}) \right] , 
\label{V1loop}
\end{eqnarray}
y
\begin{eqnarray}
Z_{\Lambda}&=&{1\over 32}\int_{\Lambda}^{\Lambda_{0}}
{d^3 q\over (2\pi)^3} {V'''^2\over (q^2+V'')^{5/2}} =
\frac{1}{192 \pi^2} \frac{V'''^2}{V''} \left[
\frac{\Lambda_0^3}{(\Lambda_0^2+V'')^{3/2}} -
\frac{\Lambda^3}{(\Lambda^2+V'')^{3/2}} 
\right] .
\label{Z1loop}
\end{eqnarray}
Mientras que la funci\'on $Z_{\Lambda}(\phi)$ es finita cuando
 $\Lambda_0 \rightarrow \infty$, el potencial $V_{\Lambda}(\phi)$ 
diverge en el UV. Para renormalizar, escribimos el potencial desnudo a un
lazo en la forma  
$V(\phi)=\frac{1}{2} (m_R^2+\delta m^2)\phi^2 + \frac{1}{4!} 
(\lambda_R+\delta \lambda) \phi^4$, donde la masa renormalizada y la constante
de acoplamiento renormalizada est\'an definidas como
\begin{eqnarray}
m_R^2 \equiv \left. \frac{\partial^2 V_{\Lambda}}{\partial \phi^2} 
\right|_{\Lambda=\phi=0}
& \; \; \;  \; \; \; &
\lambda_R \equiv \left. \frac{\partial^4 V_{\Lambda}}{\partial \phi^4} 
\right|_{\Lambda=\phi=0} ,
\end{eqnarray}
y los contrat\'erminos son
\begin{eqnarray}
\delta m^2 &=& -\frac{\lambda_R}{32 \pi^2} \left[
\frac{m_R^2}{2} + 2 \Lambda_0^2 + m_R^2 \ln(\frac{m_R^2}{4 \Lambda_0^2}) 
\right] \nonumber \\
\delta \lambda &=& -\frac{3 \lambda_R^2}{32 \pi^2} \left[
2 + \lambda_R^2 \ln(\frac{m_R^2}{4 \Lambda_0^2})
\right] .
\end{eqnarray}
Entonces el potencial renormalizado es
\begin{eqnarray}
V_{\Lambda}^{\rm ren} (\phi) &=&
\frac{1}{2} m_R^2 \phi^2 (1-\frac{\lambda_R}{64 \pi^2}) +
\frac{1}{4!} \lambda_R \phi^4 (1-\frac{3 \lambda_R}{16 \pi^2})  \nonumber \\
&& +\frac{1}{32 \pi^2} \left[
-\Lambda (2\Lambda^2 + m_R^2 + \frac{1}{2} \lambda_R \phi^2) 
\sqrt{\Lambda^2 + 
m_R^2 +\frac{1}{2} \lambda_R \phi^2} \;  \right. \nonumber \\
&& \left. ~~~~~~~~~~~ + (m_R^2+\frac{1}{2} \lambda_R \phi^2)^2 
\ln(\frac{\Lambda+\sqrt{\Lambda^2+m_R^2+\frac{1}{2} \lambda_R \phi^2}}{m_R})
\right] ,
\end{eqnarray}
y la renormalizaci\'on de la funci\'on de onda es
\begin{equation}
Z_{\Lambda}^{\rm ren} (\phi) =
\frac{1}{192 \pi^2} \frac{\lambda_R^2 \phi^2}{m_R^2+\frac{1}{2}
 \lambda_R \phi^2} \left[
1-\frac{\Lambda^3}{(\Lambda^2+m_R^2+\frac{1}{2} \lambda_R \phi^2)^{3/2}}
\right] .
\label{zren}
\end{equation}
A partir de las ecuaciones (\ref{V1loop}) y (\ref{Z1loop}), es inmediato
obtener el flujo con la escala de gra\-nu\-lado grueso del potencial efectivo  
 $V_{\Lambda}$ y de la renormalizaci\'on de la funci\'on de onda 
 $Z_{\Lambda}$, en la aproximaci\'on a un lazo,
\begin{eqnarray}
\Lambda {d V_{\Lambda}\over d \Lambda}&=&- {\Lambda^3 \over 4\pi^2}
\sqrt{\Lambda^2+V''}\nonumber \\
\Lambda {d Z_{\Lambda}\over d \Lambda} &=&-{\Lambda^3\over 64\pi^2}
{V'''^2\over (\Lambda^2+V'')^{5/2}} .
\label {1loopdiffeq}
\end{eqnarray}
La ecuaci\'on para el potencial efectivo fue obtenida con anterioridad
por otros autores usando una transformaci\'on de bloques (emparentada con los
bloques de esp\'{\i}n en el modelo de Ising) \cite{strick}.

El estudio del potencial $V_{\Lambda}$ muestra que es posible tener una 
estructura no trivial de dominios a\'un 
en la fase sim\'etrica de la teor\'{\i}a
($m_R^2>0$) \cite{strick}. En efecto, para cierto rango de los par\'ametros
de la teor\'{\i}a puede ocurrir que, a pesar que $m_R^2>0$, el cuadrado
de la masa desnuda sea negativo. En este caso, para escalas $\Lambda$ mayores
que un valor cr\'{\i}tico $\Lambda_{\rm cr}$, el potencial tiene la 
forma de un pozo doble con dos m\'{\i}nimos equidistantes 
del origen $\varphi=0$,
y para escalas menores, un \'unico m\'{\i}nimo en $\varphi=0$ (ver la figura
3.1). La 
interpretaci\'on
de este hecho es que el campo promediado fluct\'ua alrededor de cero para 
escalas 
 $\zeta> \Lambda_{\rm cr}^{-1}$, o alrededor de dos m\'{\i}nimos no nulos 
para escalas $\zeta<\Lambda_{\rm cr}^{-1}$. 
La fase sim\'etrica contiene entonces dominios de tama\~no $\zeta \approx 
\Lambda_{\rm cr}^{-1}$.
Este fen\'omeno ocurre en la fase sim\'etrica de la teor\'{\i}a, y no debe
ser confundido con la ruptura espont\'anea de simetr\'{\i}a (RES).

Por otra parte, cuando s\'{\i} hay ruptura espont\'anea de simetr\'{\i}a
 ($m_R^2<0$), tanto el potencial
a un lazo como la renormalizaci\'on de la funci\'on de onda desarrollan
una parte imaginaria para $\Lambda < \Lambda_{\rm RES}\equiv \sqrt 
{-m_R^2-\frac {1}{2}\lambda_R\phi^2}$. 
Estas partes imaginarias generan t\'erminos no reales en las ecuaciones de
movimiento, y son artefactos de la aproximaci\'on adiab\'atica. No est\'an
relacionadas con los t\'erminos de ruido de la acci\'on CTC que hemos
mencionado anteriormente. Adem\'as la renormalizaci\'on de la funci\'on de
onda diverge a medida que $\Lambda$ se aproxima a $\Lambda_{\rm RES}$ 
por arriba.
Todo \'esto 
muestra claramente que la aproximaci\'on adiab\'atica no es adecuada
para describir la evoluci\'on temporal del par\'ametro de orden, para escalas
pr\'oximas o menores que $\Lambda_{\rm RES}$.

Como se remarca en \cite{ww}, una parte imaginaria en el potencial efectivo
es una se\~nal de la aparici\'on de inestabilidades que conducen a la 
formaci\'on de dominios cuyo tama\~no es por lo menos $\sqrt{-m_R^2}$. Este 
hecho es analizado en \cite{hartree}, donde se concluye que para teor\'{\i}as
d\'ebilmente acopladas el tama\~no de los dominios puede ser mucho mayor
que la distancia de correlaci\'on a temperatura cero,  $\sqrt{-m_R^2}$.

%%%%%%%%%%%%%%%%%%%%%%%%%%%%%%%%%%%%%%%%%%%%%%%%%%%%%%%%%%%%%%%%%%%%%%%%%%%

\section{Ecuaci\'on del grupo de renormalizaci\'on exacto}

En esta secci\'on derivamos una ecuaci\'on exacta (no perturbativa) que
rige el flujo de la acci\'on de granulado grueso CTC ante el grupo de 
renormalizaci\'on. El m\'etodo que seguimos para derivarla es an\'alogo
al de Wegner y Houghton para la versi\'on eucl\'{\i}dea de 
la misma \cite{wegner}. Escribimos
la acci\'on de granulado grueso para una escala  $\Lambda - \delta \Lambda$, 
es decir
\begin{equation}
e^{i S_{\Lambda - \delta \Lambda}(\phi_+,\phi_-)} \equiv \int_{\rm CTC} 
\prod_{\Lambda_0>|{\bf q}| > \Lambda-\delta \Lambda}
 {\cal D}[\phi_+({\bf q},t)] 
{\cal D}[\phi_-({\bf q},t)] \; e^{i S_{\rm cl}[\phi_+,\phi_-]} .
\end{equation} 
Los modos a integrarse pueden dividirse en dos partes: unos dentro del
intervalo $\Lambda > |{\bf q}| > \Lambda - \delta \Lambda$ (que es una
especie de ``c\'ascara'' en el espacio de momentos) y otros modos cuyos
momentos satisfacen $\Lambda_0 > |{\bf q}| > \Lambda$. Expandiendo la
acci\'on en potencias de los modos dentro de la c\'ascara, tenemos
\begin{eqnarray}
e^{i S_{\Lambda - \delta \Lambda}(\phi_+,\phi_-)} &=&
e^{i S_{\Lambda}(\phi_+,\phi_-)} 
\int_{\rm CTC} 
\prod_{\Lambda >|{\bf q}| > \Lambda-\delta \Lambda} {\cal D}[\phi_+] 
{\cal D}[\phi_-] \; e^{i(S_1+S_2+ S_3)} \nonumber \\
&& ~~~~~~~~~~~~~~~~~~~~~~~ \times 
e^{\frac{i}{2} \int^{'} \frac{d^3q}{(2\pi)^3} \int dt 
\frac{d}{dt} ( \phi_a(-{\bf q},t) \dot{\phi}_b({\bf q},t) g_{ab} )} , 
\label{exp}
\end{eqnarray} 
donde
\begin{eqnarray}
S_1 &=& \int dt \int^{'} \frac{d^3q}{(2\pi)^3} \; \phi_a({\bf q},t) \;
\frac{ \partial S_{\Lambda} }{ \partial \phi_a (-{\bf q},t) } 
\nonumber \\
S_2 &=& \frac{1}{2} \int dt \; dt' \int^{'} \frac{d^3q}{(2\pi)^3} 
\; \phi_a({\bf q},t) \; \frac{\partial^2 S_{\Lambda}}
{\partial \phi_a(-{\bf q},t) \phi_b({\bf q},t')} \; 
\phi_b({\bf q},t') .
\end{eqnarray}
El significado de la prima en las integrales sobre los momentos es que \'estas
est\'an restringidas a la c\'ascara. Adem\'as, en las derivadas funcionales de
 $S_{\Lambda}$ (que contienen modos cuyos vectores de onda satisfacen
 $|{\bf q}| < \Lambda$) los modos dentro de la c\'ascara se toman iguales a
cero. La notaci\'on que usamos
es la usual para el formalismo de camino temporal cerrado,
\begin{eqnarray}
\phi_a({\bf q},t) = \left( \begin{array}{c}
                \phi_+({\bf q},t) \\ \phi_-({\bf q},t)
                \end{array}
         \right) & \;   \;& 
g_{ab} = \left( \begin{array}{cc}
                1 & 0 \\ 0 & -1
                 \end{array} \right) .
\end{eqnarray}

El t\'ermino $S_3$ es c\'ubico en los modos dentro de la c\'ascara,
y puede probarse que no contribuye en el l\'{\i}mite 
 $\delta \Lambda \rightarrow 0$ (b\'asicamente porque se est\'a haciendo
un c\'alculo a un lazo para los modos dentro de la c\'ascara). Las integrales
funcionales sobre dichos modos cumplen con las particulares condiciones de
contorno del camino temporal cerrado.

Para evaluar las integrales funcionales dividimos al campo en la forma
 $\phi_a={\bar \phi}_a + \varphi_a$ e imponemos las condiciones de contorno
sobre los campos cl\'asicos ${\bar \phi}_{\pm}$, i.e. se anulan en el pasado
 $-T$ (frecuencias negativas y positivas respectivamente) y coinciden en una
superficie de Cauchy al tiempo $T$ en el futuro. Las fluctuaciones $\varphi_a$
se anulan tanto en el pasado como en el futuro.
Los campos cl\'asicos son soluci\'on de
\begin{equation}
(- \frac{d^2}{dt^2} - q^2) g_{ab} \bar{\phi}_b({\bf q},t) + 
\int dt' 
\frac{\partial^2 S_{\rm int}}
{\partial \varphi_a(-{\bf q},t) \partial 
\varphi_b({\bf q},t')} \bar{\phi}_b({\bf q},t') = 0 ,
\label{mod1}
\end{equation}
donde hemos separado a la acci\'on de granulado grueso en la parte
cin\'etica y la de interacci\'on, 
$S_{\Lambda}(\phi_{\pm}) = S_{\rm cin}(\phi_{\pm})+S_{\rm int}(\phi_{\pm})$ con
\begin{equation}
S_{\rm cin} =  \int d^4x  \left[ \frac{1}{2}  (\partial_{\mu} \phi_+)^2 + 
\frac{i \epsilon}{2} \phi_+^2 \right] -
\int d^4x  \left[ \frac{1}{2} (\partial_{\mu} \phi_-)^2 -
\frac{i \epsilon}{2} \phi_-^2 \right] .
\end{equation}
Al igual que antes, en las derivadas funcionales los modos dentro de la
c\'ascara se toman iguales a cero.

Sean $h_a$ soluciones de la ec.(\ref{mod1}), que se anulan en el pasado y que
satisfacen una normalizaci\'on arbitraria en el futuro, y sea $\phi(\bf{q})$
el valor com\'un que toman los campos en cada rama del camino temporal cerrado
en el futuro. Podemos entonces escribir
\begin{equation}
\bar{\phi}_a({\bf q},t) = \phi({\bf q}) 
\frac{h_a({\bf q},t)}{h_a({\bf q},T)} .
\end{equation}
Integramos primero sobre el valor com\'un $\phi(\bf{q})$ y luego procedemos
con la integraci\'on funcional sobre las fluctuaciones $\varphi_a$ (ambas son
integrales gaussianas con t\'erminos de fuente). Finalmente resulta 
\begin{eqnarray}
\Lambda \frac{\partial S_{\Lambda}}{\partial \Lambda} &=& 
- \frac{i \Lambda}
{2 \delta \Lambda} \int^{'} \frac{d^3q}{(2\pi)^3} 
\ln \left( \frac{\dot{h}_{+}({\bf q},T)}{h_{+}({\bf q},T)} -
\frac{\dot{h}_{-}({\bf q},T)}
{h_{-}({\bf q},T)} \right)  \nonumber \\
&& + \frac{\Lambda}{2 \delta \Lambda}  \int^{'} \frac{d^3q}{(2\pi)^3} 
\left( \frac{\dot{h}_{+}({\bf q},T)}{h_{+}({\bf q},T)} -
\frac{\dot{h}_{-}({\bf q},T)}
{h_{-}({\bf q},T)} \right)^{-1} \, 
\left( \int dt \frac{h_a({\bf q},t)}{h_a({\bf q},T)} 
\frac{\partial S_{\Lambda}}{\partial \varphi_a(-{\bf q},t)} \right)^{2} 
 \nonumber \\
&& - \frac{i \Lambda}{2 \delta \Lambda} \ln {\rm det}'(A_{ab})  \nonumber \\
&& + \frac{\Lambda}{2 \delta \Lambda} \int dt \; dt' \int^{'} \frac{d^3q}
{(2\pi)^3} \frac{\partial S_{\Lambda}}{\partial 
\varphi_a({\bf q},t)} A_{ab}^{-1}(-{\bf q},t;{\bf q},t') 
\frac{\partial S_{\Lambda}}{\partial \varphi_b({\bf q},t')} . 
\label{exact}
\end{eqnarray}

La matriz de $2\times 2$ $A_{ab}$ posee los siguientes elementos
\begin{eqnarray}
A_{++}(-{\bf q},t;{\bf q\, '},t') &=& 
        (- \frac{d^2}{dt^2} - q^2 + i \epsilon) \delta(t-t') 
        \delta^3({\bf q}+{\bf q\,'}) + 
        \frac{\partial^2 S_{\rm int}}{\partial \varphi_+(-{\bf q},t) 
        \partial \varphi_+({\bf q\,'},t')} \nonumber \\
A_{--}(-{\bf q},t;{\bf q\,'},t') &=& 
        (\frac{d^2}{dt^2} + q^2 + i \epsilon) \delta(t-t') 
        \delta^3({\bf q}+{\bf q\,'}) + 
        \frac{\partial^2 S_{\rm int}}{\partial \varphi_-(-{\bf q},t) 
        \partial \varphi_-({\bf q\,'},t')} \nonumber \\
A_{+-}(-{\bf q},t;{\bf q\,'},t')&=& A_{-+}({\bf q\,'},t';{-\bf q},t)=
 \frac{\partial^2 S_{\rm int}}{\partial \varphi_+(-{\bf q},t) 
        \partial \varphi_-({\bf q\,'},t')} .
\end{eqnarray}
El determinante primado debe ser calculado como el producto de los autovalores
de $A_{ab}$ en un espacio de funciones cuyos vectores de onda se encuentren
dentro de la c\'ascara ($\Lambda -\delta \Lambda < |{\bf q}| < \Lambda$) 
y satisfagan condiciones de contorno nulas, tanto en el pasado como en el
futuro. Condiciones similares se deben usar para calcular la inversa
$A_{ab}^{-1}$.  

La ecuaci\'on del grupo de renormalizaci\'on exacto (\ref{exact}) es 
``exacta'' en el sentido de que hasta el momento
no se ha requerido ninguna aproximaci\'on perturbativa. Como ya hemos dicho, 
es similar a la versi\'on eucl\'{\i}dea \cite{wegner}, pero resulta m\'as
complicada por las peculiares condiciones de contorno del CTC. Contiene
toda la informaci\'on de la influencia de los modos de longitud de onda corta
sobre los de longitud de onda larga, y puede representar un punto de partida
para c\'alculos no perturbativos de decoherencia, disipaci\'on, formaci\'on 
de dominios y evoluci\'on fuera de equilibrio.
 
%%%%%%%%%%%%%%%%%%%%%%%%%%%%%%%%%%%%%%%%%%%%%%%%%%%%%%%%%%%%%%%%%%%%%%%%%%%

\subsection{Expansi\'on en derivadas}

La ecuaci\'on del grupo de renormalizaci\'on exacto es una ecuaci\'on 
integro-diferencial para la acci\'on efectiva de granulado grueso CTC.
La extrema complejidad de la misma hace necesario que se busquen m\'etodos
de aproximaci\'on para hallar soluciones. En la formulaci\'on eucl\'{\i}dea a 
la que hicimos referencia con anterioridad, los m\'etodos m\'as usuales
son una expansi\'on en derivadas del campo 
\cite{morris3,enrique,tetradis} o un desarrollo en potencias del campo
\cite{morris2}. En esta secci\'on aplicaremos la t\'ecnica del desarrollo
en derivadas para resolver la ecuaci\'on del grupo de renormalizaci\'on.
Este m\'etodo da un desarrollo local para la acci\'on, y la ec.(\ref{exact})
admite una soluci\'on de la forma  
\begin{equation}
S_{\Lambda}(\phi_+,\phi_-) = S_{\Lambda}(\phi_+) - 
S_{\Lambda}(\phi_-) .
\label{ans}
\end{equation} 
Claramente esta no es la forma m\'as general que puede adoptar la
acci\'on de granulado grueso de camino temporal cerrado, ya que no tiene
en cuenta los t\'erminos de acoplamiento entre los campos en las dos ramas del
camino temporal cerrado, y por ende los t\'erminos de disipaci\'on y ruido. 
Como ya vimos en la aproximaci\'on a un lazo, la 
causa de \'esto es el mismo desarrollo en derivadas, que al ser una expansi\'on
local pierde los importantes aspectos estoc\'asticos. Nosotros nos
conformaremos con usar este m\'etodo pues permite en forma sencilla obtener
soluciones a la ecuaci\'on RG. Para ir m\'as all\'a de la aproximaci\'on
en derivadas, se requiere una aproximaci\'on no local de la acci\'on, que
incluya t\'erminos reales e imaginarios asociados a la disipaci\'on y el
ruido, respectivamente. Esos t\'erminos estar\'an conectados a nivel de la
ecuaci\'on RG, y es de esperar que a partir de ella se pueda obtener una
relaci\'on de fluctuaci\'on-disipaci\'on no perturbativa, que describa c\'omo
los n\'ucleos de ruido y disipaci\'on var\'{\i}an con la escala de granulado
grueso (ver la discusi\'on al final del cap\'{\i}tulo).

El Ansatz de la ec.(\ref{ans}), consistente con la aproximaci\'on en 
derivadas, permite expresar el determinante de la matriz $A_{ab}$ como
el producto de dos determinantes, uno para $A_{++}$ y otro para $A_{--}$.
Bas\'andonos en \cite{cole}, podemos expresar ambos determinantes, 
 ${\rm det}' A_{++}$ y ${\rm det}' A_{--}$, 
como el producto sobre los momentos de la 
c\'ascara de los modos $h({\bf q},t)$ evaluados en el tiempo $t=T$. El 
\'ultimo t\'ermino de la ecuaci\'on RG puede ser reescrito como
\begin{equation}
\ln {\rm det}'(A_{ab}) = \ln [{\rm det}'(A_{++}) {\rm det}'(A_{--})] = 
\int^{'} \frac{d^3q}{(2\pi)^3} \ln( h_{+}({\bf q},T) h_{-}({\bf q},T) ) .
\end{equation}
Entonces la ecuaci\'on RG adopta finalmente la forma
\begin{eqnarray}
\Lambda \frac{\partial S_{\Lambda}}{\partial \Lambda} &=& 
- \frac{i \Lambda} {2 \delta \Lambda} \int^{'} \frac{d^3q}{(2\pi)^3} 
\ln({h_{-}({\bf q},T) \dot{h}_{+}({\bf q},T) - h_{+}}({\bf q},T) 
\dot{h}_{-}({\bf q},T)) 
\nonumber  \\
&&  + \frac{\Lambda}{2 \delta \Lambda}  \int^{'} \frac{d^3q}{(2\pi)^3} 
\left( \frac{\dot{h}_{+}({\bf q},T)}{h_{+}({\bf q},T)} -
\frac{\dot{h}_{-}({\bf q},T)}
{h_{-}({\bf q},T)} \right)^{-1} \, 
\left( \int dt \frac{h_a({\bf{q}},t)}{h_a({\bf{q}},T)} 
\frac{\partial S_{\Lambda}}{\partial \varphi_a(-{\bf q},t)} \right)^{2} 
 \nonumber \\
&&  + \frac{\Lambda}{2 \delta \Lambda} \int dt \; dt' \int^{'} \frac{d^3q}
{(2\pi)^3} \frac{\partial S_{\Lambda}}{\partial 
\varphi_a({\bf q},t)} A_{ab}^{-1}(-{\bf q},t;{\bf q},t') 
\frac{\partial S_{\Lambda}}{\partial \varphi_b({\bf q},t')} .
\label{Slambda}
\end{eqnarray}
Por otro lado, con el Ansatz precedente para la acci\'on, las ecuaciones 
que satisfacen los modos $h_+$ y $h_-$ (ec.(\ref{mod1})) 
se simplifican considerablemente, dado que ambas ecuaciones se desacoplan.
Lo que todav\'{\i}a resta por probar es que el miembro derecho de la
ecuaci\'on RG tambi\'en adopta la forma de una diferencia entre dos 
contribuciones, una para cada rama del camino temporal cerrado.

Consideramos la siguiente expansi\'on en derivadas del t\'ermino de
interacci\'on
\begin{equation}
S_{\rm int}(\phi_{\pm}) = \int d^4x [ - V_{\Lambda}(\phi_{\pm}) + 
\frac{1}{2} Z_{\Lambda}(\phi_{\pm})  \dot{\phi}^2_{\pm} -
\frac{1}{2} Y_{\Lambda}(\phi_{\pm})( {\bf \nabla} \phi_{\pm})^2 + \ldots ] .
\end{equation}
Dado que el granulado grueso rompe la invariancia Lorentz, hemos
tomado diferentes coeficientes para las derivadas temporales y espaciales.
El objetivo entonces es utilizar esta pro\-pues\-ta para la acci\'on
y obtener ecuaciones que digan c\'omo los coeficientes $V_{\Lambda}$, 
 $Z_{\Lambda}$ e $Y_{\Lambda}$ var\'{\i}an con la escala de granulado grueso.
Expandimos los campos alrededor de un fondo dependiente del tiempo
 $\phi_{\pm} = \phi_{\pm}(t) + \varphi_{\pm}({\bf x},t)$ y transformamos 
Fourier en las coordenadas espaciales. Resolvemos luego la ec.(\ref{mod1})
para los modos a orden cero en las fluctuaciones, es decir, igualamos
t\'erminos en las ecuaciones para $h_{\pm}$ que sean independientes de
 $\varphi_{\pm}$. Dado que la primer derivada funcional de la acci\'on
efectiva de granulado grueso ($S'$) es lineal en las fluctuaciones
 $\varphi_{\pm}$, ponemos $S'=0$ y mantenemos las contribuciones en 
 $S_{{\rm int}}''$ que sean independientes de $\varphi_{\pm}$. Luego de un
poco de \'algebra y derivadas funcionales, obtenemos
\begin{eqnarray}
\frac{\partial^2 S_{\rm int}}{\partial \varphi({\bf q},t) \partial 
\varphi(-{\bf q}',t')}
&=& [ -V'' - \frac{1}{2} \dot{\phi}^2 Z'' - Y q^2 -  Z' \dot{\phi}
\frac{d}{dt} - Z \frac{d^2}{dt^2} - \ddot{\phi} Z' + \ldots ] 
\nonumber \\
&& \times \delta(t-t') \delta^3({\bf q}-{\bf q}') ,
\end{eqnarray}
donde las primas denotas derivaci\'on respecto al campo, y los puntos 
suspensivos denotan t\'erminos lineales en las fluctuaciones. En esta 
expresi\'on y en forma sucesiva, omitiremos los \'{\i}ndices  ${\pm}$, tanto
en los campos de fondo $\phi_{\pm}(t)$, en el potencial
 $V_{\Lambda}(\phi_{\pm}(t))$, como en los factores de la funci\'on de onda
 $Z_{\Lambda}(\phi_{\pm}(t))$ e $Y_{\Lambda}(\phi_{\pm}(t))$. En esta 
aproximaci\'on en derivadas, las ecuaciones para los modos $h_a$ se localizan
y adoptan la forma de ecuaciones tipo oscilador arm\'onico (con una frecuencia
que depende del fondo variable $\phi(t)$) con un t\'ermino con primeras 
derivadas temporales (tipo disipativo). Si se definen nuevos modos 
 $f({\bf q},t) \equiv (1+Z_{\Lambda})^{1/2} h({\bf q},t)$, los 
t\'erminos lineales
en derivadas temporales se cancelan y los nuevos modos son osciladores
arm\'onicos de frecuencia
\begin{equation}
w_{q}^2(t)= q^2 \, \frac{1+Y_{\Lambda}}{1+Z_{\Lambda}} + 
\frac{V_{\Lambda}''}{1+Z_{\Lambda}} + 
\frac{1}{4} \frac{Z_{\Lambda}'^2}{(1+Z_{\Lambda})^{2}} \, \dot{\phi}^2 +
\frac{1}{2} \frac{Z_{\Lambda}'}{1+Z_{\Lambda}} \,  \ddot{\phi} .
\end{equation}
Usando la expansi\'on adiab\'atica para los modos,
\begin{equation}
h_{\pm}({\bf q},t) = (1+Z_{\Lambda})^{-1/2} 
\frac{1}{\sqrt{2 W_{\pm}({\bf q},t)}} e^{\pm i 
\int_{-T}^{t} W_{\pm}({\bf q},t') dt'} ,
\end{equation}
podemos evaluar el t\'ermino logar\'{\i}tmico del lado derecho de la
ecuaci\'on RG (ver ec.(\ref{Slambda})). Los otros t\'erminos de dicha
ecuaci\'on son cuadr\'aticos en las fluctuaciones y no contribuyen al orden
en que estamos trabajando.  Tenemos entonces

\clearpage
\begin{eqnarray}
&h_{-}({\bf q},T) \dot{h}_{+}({\bf q},T) - h_{+}({\bf q},T) 
\dot{h}_{-}({\bf q},T) =& \nonumber \\
&e^{i \int_{-T}^{T} [W_{+}({\bf q},t)-W_{-}({\bf q},t)] dt} \times 
\left\{
\frac{\left[
        -\frac{Z_+'}{2(1+Z_+)} \dot{\phi}_+ - \frac{\dot{W}_+}{2 W_+} 
+ i W_+
      \right]-
      \left[
        -\frac{Z_-'}{2(1+Z_-)} \dot{\phi}_- - \frac{\dot{W}_-}{2 W_-}
- i W_-
      \right] }
     {2 \sqrt{W_+ W_- (1+Z_+) (1+Z_-)} }
     \right\}_{t=T} .
\end{eqnarray}

Notar que, al igual que en el caso a un lazo, los campos $+$ y $-$ se 
desacoplan, a menos de un factor evaluado en $t=T$. Este factor es una 
contribuci\'on de superficie que, al tomar logaritmos, es irrelevante
para las ecuaciones de movimiento. Por el contrario, el primer factor 
s\'{\i} depende de toda la historia de los campos y su forma es consistente
con el Ansatz para la acci\'on, ec.(\ref{ans}). Entonces podemos escribir la
ecuaci\'on del grupo de renormalizaci\'on en la siguiente forma
\begin{eqnarray}
&&\Lambda \int dt 
\left\{
\left[
- \frac{d V_{\Lambda}(\phi_+)}{d \Lambda} + \frac{1}{2} 
\frac{d Z_{\Lambda}(\phi_+)}{d \Lambda} \dot{\phi}_+^2 
\right]
-
\left[
- \frac{d V_{\Lambda}(\phi_-)}{d \Lambda} + \frac{1}{2} 
\frac{d Z_{\Lambda}(\phi_-)}{d \Lambda} \dot{\phi}_-^2 
\right]
\right\}  \nonumber \\
&& = \frac{ \Lambda}{2 \delta
\Lambda} \int^{'} \frac{d^3q}{(2\pi)^3}  \int 
[W_{+}({\bf q},t)-W_{-}
({\bf q},t)]dt .
\end{eqnarray}
En la aproximaci\'on adiab\'atica 
\begin{equation}
W^2=A_{\Lambda} + B_{\Lambda}
\dot{\phi}^2(t)+ C_{\Lambda} \ddot{\phi}(t) ,
\end{equation} 
donde los coeficientes son
\begin{eqnarray}
A_{\Lambda} &=& \Lambda^2 \frac{1+Y_{\Lambda}}{1+Z_{\Lambda}} +
     \frac{V_{\Lambda}''}{1+Z_{\Lambda}} \nonumber \\
B_{\Lambda} &=& \frac{Z_{\Lambda}'^2}{4(1+Z_{\Lambda})^2} + 
     \frac{5 A_{\Lambda}'^2}{16 A_{\Lambda}^2} - \frac{A_{\Lambda}''}
{4 A_{\Lambda}} \nonumber \\
C_{\Lambda} &=& \frac{Z_{\Lambda}'}{2(1+Z_{\Lambda})} 
- \frac{A_{\Lambda}'}{4 A_{\Lambda}} .
\end{eqnarray}
Integrando por partes finalmente obtenemos
\begin{equation}
\int dt \left\{
- \Lambda \frac{dV_{\Lambda}}{d\Lambda} + 
\frac{1}{2} \Lambda \frac{dZ_{\Lambda}}{d\Lambda} {\dot \phi}^2 \right\}
=\frac{\Lambda^3}{4 \pi^2} \int dt 
\left\{
\sqrt{A_{\Lambda}} +
\frac{1}{2} \dot{\phi}^2 
\left[
\frac{B_{\Lambda}}{\sqrt{A_{\Lambda}}} - 
\left( \frac{C_{\Lambda}}{\sqrt{A_{\Lambda}}} \right)'
\right]
\right\} .
\end{equation}
A partir de esta expresi\'on podemos leer la dependencia 
de $V_{\Lambda}$ y $Z_{\Lambda}$ con la escala de granulado grueso,
esto es
\begin{eqnarray}
\Lambda \frac{dV_{\Lambda}}{d\Lambda} &=& -\frac{\Lambda^3}{4 \pi^2} 
\sqrt{\Lambda^2 \frac{1+Y_{\Lambda}}
{1+Z_{\Lambda}} + \frac{V_{\Lambda}''}{1+Z_{\Lambda}}} \nonumber \\
\Lambda \frac{dZ_{\Lambda}}{d\Lambda} &=& \frac{\Lambda^3}{4 \pi^2} 
\left[
\frac{B_{\Lambda}}{\sqrt{A_{\Lambda}}} - 
\left( \frac{C_{\Lambda}}{\sqrt{A_{\Lambda}}} \right)'
\right] .
\label{flowing}
\end{eqnarray}
Estas ecuaciones describen el flujo de la acci\'on de granulado grueso
con la escala de renor\-ma\-li\-za\-ci\'on, en la aproximaci\'on en derivadas.
De la comparaci\'on entre estas ecuaciones y las co\-rres\-pon\-dientes al caso
de la aproximaci\'on a un lazo, concluimos que los t\'erminos de orden 
superior en el desarrollo en derivadas modifican la ecuaci\'on diferencial
para el potencial efectivo. 

\begin{figure}[ht]
\centering \leavevmode
\epsfxsize=12cm
\epsfbox{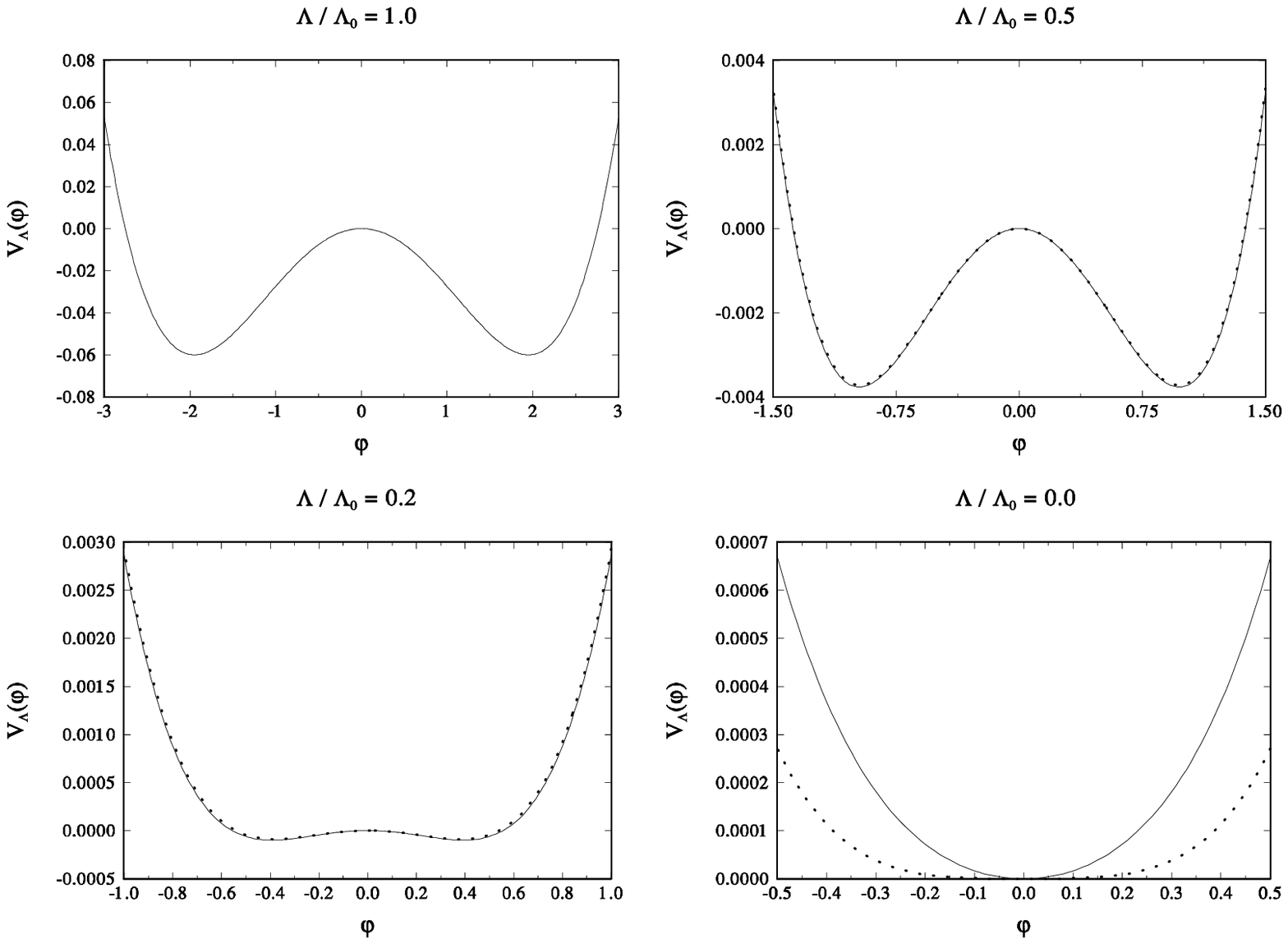}
\setlength{\captwidth}{12cm}
\capt{El potencial efectivo de granulado grueso  $V_{\Lambda}$ para
$\Lambda_0=10$, $m_R^2=10^{-4}$, y $\lambda_R=0.1$. Las l\'{\i}neas llena
y punteada corresponden al c\'alculo mediante el grupo de 
renormalizaci\'on y el c\'alculo a un lazo, respectivamente.}
\end{figure}

Hemos obtenido dos ecuaciones para las tres cantidades  
inc\'ognitas $V_{\Lambda}$, $Z_{\Lambda}$ e $Y_{\Lambda}$. Para hallar
una tercera ecuaci\'on para la variaci\'on de $Y_{\Lambda}$ con la escala,
se requiere ir hasta orden cuadr\'atico en las fluctuaciones a nivel de la
ecuaci\'on RG. En vez de hacer eso, y por simplicidad, asumiremos que 
 $Z_{\Lambda}$ e $Y_{\Lambda}$ son cantidades peque\~nas, y las pondremos
iguales a cero en el lado derecho de la ec.(\ref{flowing}). Esta suposici\'on
se ve confirmada por los c\'alculos num\'ericos que hemos llevado a cabo
para la fase sim\'etrica de la teor\'{\i}a. En esta 
aproximaci\'on la forma del potencial efectivo mejorado por el grupo de
renormalizaci\'on es la misma que la propuesta en  \cite{strick}. Adem\'as,
cuando en el miembro derecho de la ecuaci\'on RG reemplazamos las funciones
por sus valores cl\'asicos, $V_{\Lambda}=V \, , \, Z_{\Lambda}=Y_{\Lambda}=0$, 
recobramos las ecuaciones de evoluci\'on a un lazo, ec.(\ref{1loopdiffeq}).

Como ilustraci\'on consideremos una teor\'{\i}a $\lambda \phi^4$. Debemos 
resolver las ecuaciones diferenciales (\ref{flowing})
con las condiciones iniciales
cl\'asicas, $V_{\Lambda_{0}}=V$, $Z_{\Lambda_0}=0$
y $Y_{\Lambda_{0}}=0$, evolucionando las ecuaciones desde la escala
ultravioleta $\Lambda_0$ hasta la escala de granulado grueso $\Lambda$ de
inter\'es. Los resultados se muestran en las figuras 3.1 y 3.2, 
donde se compara
el resultado a un lazo con la mejora ante el grupo de renormalizaci\'on.
Notar que los resultados son consistentes con la suposici\'on 
 $Z_{\Lambda} \ll 1$.

\begin{figure}[h]
\centering \leavevmode
\epsfxsize=12cm
\epsfbox{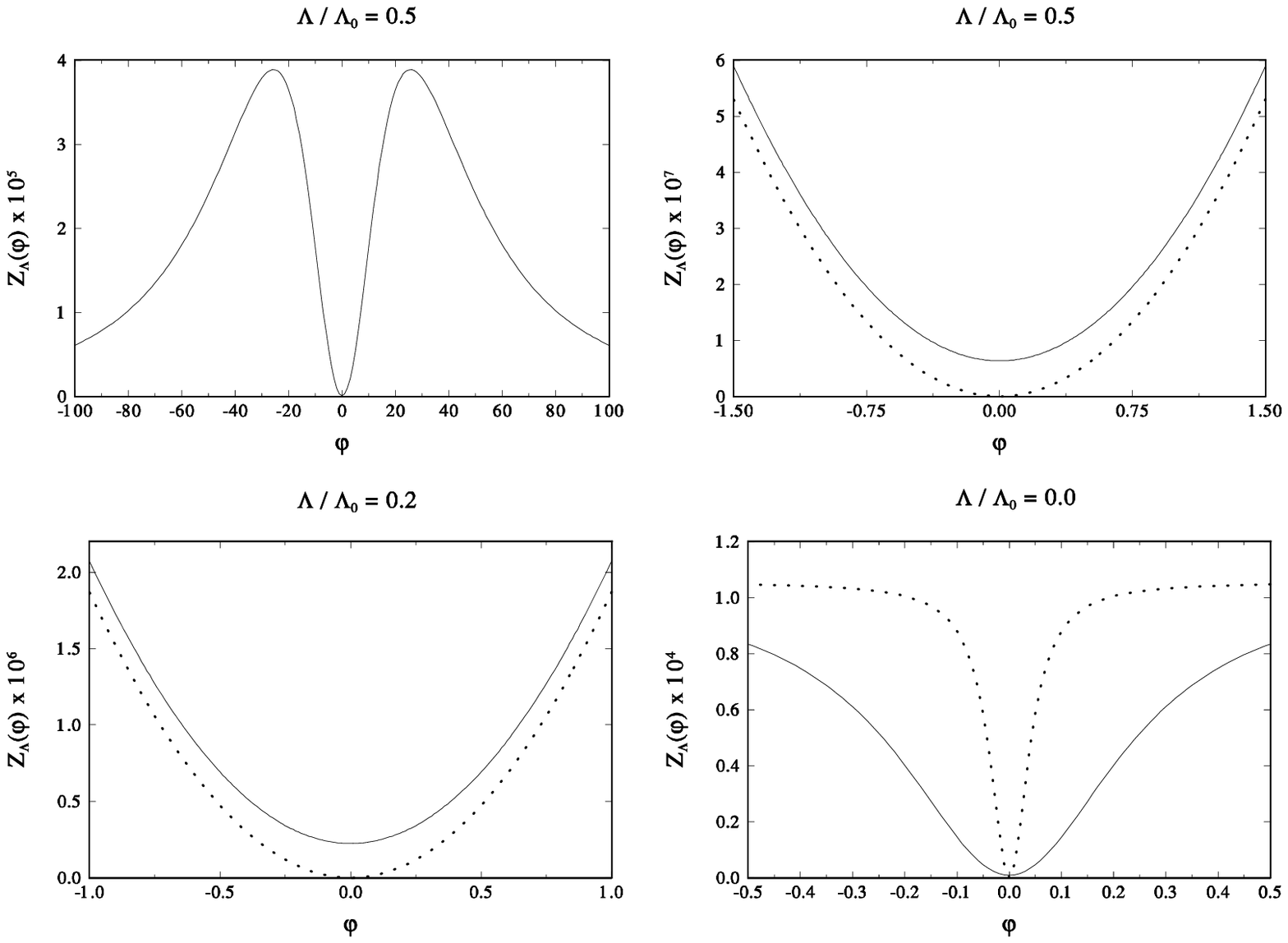}
\setlength{\captwidth}{12cm}
\capt{La renormalizaci\'on de la funci\'on de onda de granulado grueso
 $Z_{\Lambda}$ para
$\Lambda_0=10$, $m_R^2=10^{-4}$, y $\lambda_R=0.1$.
Las l\'{\i}neas llena
y punteada corresponden al c\'alculo mediante el grupo de 
renormalizaci\'on y el c\'alculo a un lazo, respectivamente.}
\end{figure}

Una vez conocidas las funciones $V_{\Lambda}$, $Z_{\Lambda}$ e
 $Y_{\Lambda}$, se puede escribir la ecuaci\'on din\'amica del par\'ametro 
de orden
\be
\Box \phi + V_{\Lambda}'(\phi) + Z_{\Lambda}(\phi) \ddot{\phi}
+ \frac{1}{2} Z_{\Lambda}'(\phi) \dot{\phi}^2 
- Y_{\Lambda}(\phi) \nabla^2 \phi - \frac{1}{2} Y_{\Lambda}'(\phi) 
(\nabla \phi)^2 = 0 ,
\ee
que es una mejora por el grupo de renormalizaci\'on a la ecuaci\'on
que se obtiene mediante la expansi\'on usual en derivadas.
Esta ecuaci\'on ser\'a v\'alida siempre que $\phi$ sea lentamente variable y 
 $Z_{\Lambda}, Y_{\Lambda} \ll 1$.
Al igual que en el caso a un lazo, la expansi\'on en 
derivadas resulta inadecuada si hay ruptura espont\'anea de simetr\'{\i}a,
para escalas pr\'oximas o menores que $\Lambda_{\rm RES}$. 

%%%%%%%%%%%%%%%%%%%%%%%%%%%%%%%%%%%%%%%%%%%%%%%%%%%%%%%%%%%%%%%%%%%%%%%%%%%%%

\section{Discusi\'on}

La acci\'on efectiva de granulado grueso de camino temporal cerrado 
 $S_{\Lambda}$ contiene toda la informaci\'on sobre la influencia de los
modos de longitud de onda corta sobre los de larga longitud de onda. En
principio, a partir de ella es posible encontrar no s\'olo la ecuaci\'on para
el valor medio del campo, sino tambi\'en una ecuaci\'on estoc\'astica
del tipo Langevin para el campo cl\'asico. Esta ecuaci\'on de Langevin puede
ser utilizada para analizar la formaci\'on y crecimiento de dominios, y, en
general, aspectos de no equilibrio de transiciones de fase.

En este cap\'{\i}tulo hemos definido la acci\'on $S_{\Lambda}$ y 
obtenido una ecuaci\'on de evoluci\'on exacta para la dependencia de dicha
acci\'on con la escala de granulado grueso (ec.(\ref{exact})), que
resolvimos mediante una aproximaci\'on en derivadas. Esta 
t\'ecnica nos sirvi\'o para hallar una soluci\'on a la ecuaci\'on RG
y describir mejoras por el grupo de renormalizaci\'on al potencial efectivo.

No podemos terminar este cap\'{\i}tulo sin discutir en forma cr\'{\i}tica
esta aproximaci\'on en derivadas. Como ya hemos mencionado,
su principal defecto es que 
es responsable de la falta de t\'erminos de ruido y disipaci\'on en
la acci\'on de granulado grueso, y por lo tanto no permite describir los
importantes aspectos estoc\'asticos en teor\'{\i}a de campos.
M\'as all\'a de esta aproximaci\'on adiab\'atica, es de esperar que a medida
que la escala de granulado grueso decrezca a partir de $\Lambda_0$, los 
t\'erminos de ruido y de disipaci\'on empiecen a crecer: $S_{\Lambda}$
desarrollar\'a una parte imaginaria, relacionada con el ruido, y una parte
real, asociada con la disipaci\'on, y en ambos casos habr\'a t\'erminos
de interacci\'on entre los campos $\phi_+$ y $\phi_-$ de cada rama temporal.
Todo \'esto puede ser f\'acilmente chequeado tanto a nivel de la ecuaci\'on
a un lazo (ec.(\ref{1looP})) como a nivel de la ecuaci\'on exacta 
RG (ec.(\ref{exact})). Las partes reales e imaginarias de la acci\'on
de granulado grueso no est\'an desacopladas: en la escala $\Lambda=\Lambda_0$
la acci\'on posee s\'olo t\'erminos reales, que inducir\'an una parte 
imaginaria a escalas menores. 

Dada la ecuaci\'on exacta RG es posible resolverla en forma perturbativa.
As\'{\i}, por ejemplo, el c\'alculo expl\'{\i}cito de la acci\'on de granulado
grueso para una teor\'{\i}a $\lambda \phi^4$, a segundo orden 
en $\lambda$ \cite{fernando}, da
 $S_{\Lambda}(\phi_+,\phi_-) = S_0(\phi_+) - S_0(\phi_-) + 
 \Delta S_{\Lambda}(\phi_+,\phi_-)$,
donde $S_0$ es la parte libre de la acci\'on cl\'asica, y
\bea
\hspace{-0.5cm}
{\rm Re} \Delta S_{\Lambda}(\phi_+,\phi_-) &=&  
-\lambda \int d^4x 
\left\{\frac{1}{12} P_-(x) + \frac{i}{2} G^{\Lambda}_{++}(0)
Q_-(x) \right\} 
+ \lambda^2 \int d^4x \int d^4y \theta(x^0-y^0) \nonumber \\
&& ~~ \times \left\{
-\frac{1}{18} R_+(x) {\rm Re}G_{++}^{\Lambda}(x-y) R_-(y) +
\frac{1}{4} Q_+(x) {\rm Im} G_{++}^{\Lambda 2}(x-y) Q_-(y) \right. \nonumber \\
&& ~~~~~~ \left. + \frac{1}{3} S_+(x) {\rm Re} G_{++}^{\Lambda 3}(x-y) S_-(y)
\right\}
\label{IFre}
\eea
\bea
\hspace{-0.5cm}
{\rm Im}  \Delta S_{\Lambda}(\phi_+,\phi_-) &=&
\lambda^2 \int d^4x \int d^4y \left\{ -\frac{1}{18} R_-(x) {\rm Im} 
G_{++}^{\Lambda}(x-y) R_-(y) \right. \nonumber \\
&& \left. - \frac{1}{4} Q_-(x) {\rm Re} G_{++}^{\Lambda 2}(x-y) Q_-(y) +
\frac{1}{3} S_-(x) {\rm Im} G_{++}^{\Lambda 3}(x-y) S_-(y)
\right\} ,
\label{IFim}
\eea
donde $G_{++}^{\Lambda}$ es el propagador de Feynman para los modos
con $|{\bf{q}}| > \Lambda$ y
\bea
P_{\pm} = \frac{1}{2}(\phi_+^4 \pm \phi_-^4)  & ~~~~ &
R_{\pm} = \frac{1}{2}(\phi_+^3 \pm \phi_-^3) \nonumber \\
Q_{\pm} = \frac{1}{2}(\phi_+^2 \pm \phi_-^2)  & ~~~~ &
S_{\pm} = \frac{1}{2}(\phi_+ \pm \phi_-) .
\eea
Para introducir los n\'ucleos de ruido, reescribimos la exponencial de la 
parte imaginaria como la integral funcional sobre fuentes de ruido (tres
en este ejemplo particular, $\nu(x)$, $\zeta(x)$ y $\eta(x)$) con 
distribuci\'on
de probabilidad gaussiana $P[\nu]$, $P[\zeta]$ y $P[\eta]$. 
Esto es posible ya que, en esta aproximaci\'on
perturbativa, 
la parte imaginaria de la acci\'on efectiva es suma de t\'erminos
gaussianos que involucran funciones de dos puntos. 
Mediante este truco, la acci\'on de granulado grueso queda
\be
S_{\Lambda}(\phi_+,\phi_-) = i \ln \int D\nu P[\nu] \int D\zeta P[\zeta] 
\int D\eta P[\eta] \exp i S_{\Lambda}^{\rm ef}(\phi_+,\phi_-,\nu,\zeta,\eta) ,
\ee
donde hemos definido una nueva acci\'on efectiva, real y dependiente de
las fuentes de ruido, cuya expresi\'on es
\bea
S_{\Lambda}^{\rm ef}(\phi_+,\phi_-,\nu,\zeta,\eta) &=& 
S_0(\phi_+) -S_0(\phi_-) + {\rm Re} \Delta S(\phi_+,\phi_-) \nonumber \\
&& - 
\int d^4x [R_-(x) \nu(x) + Q_-(x) \zeta(x) + S_-(x) \eta(x) ] .
\eea
A partir de esta acci\'on es posible hallar la ecuaci\'on de Langevin para
el campo, resultando
\be
\left. 
\frac{\delta (S_0 + {\rm Re} \Delta S_{\Lambda})}{\delta \phi_+} 
\right|_{\phi+=\phi-=\phi} = F_{\rm ruido} ,
\ee
donde la fuerza estoc\'astica es 
$F_{\rm ruido} = 3 \nu \phi^2 /2 + \zeta \phi + \eta /2$, es decir ruido 
multiplicativo y aditivo. Las fuentes de ruido est\'an caracterizadas por
sus funciones de dos puntos 
 $\ll\!\nu(x)\nu(y)\! \gg = (\lambda^2/9) {\rm Im}G^{\Lambda}_{++}(x,y) $,
 $\ll \! \zeta(x)\zeta(y) \! \gg = (\lambda^2/2) 
{\rm Re}G^{\Lambda 2}_{++}(x,y)$
y
 $\ll \! \eta(x)\eta(y) \! \gg = - (2 \lambda^2/3) 
{\rm Im}G^{\Lambda 3}_{++}(x,y)$, 
donde $\ll \! {\cal O} \! \gg \equiv \int D\nu D\zeta D\eta 
P[\nu] P[\zeta] P[\eta] \;{\cal O}$.
Mientras que la parte imaginaria de la acci\'on 
de granulado grueso induce efectos de ruido, la parte real renormaliza
la acci\'on cl\'asica e induce efectos disipativos.

Para lograr retener los aspectos estoc\'asticos en una aproximaci\'on 
{\it no perturbativa} de la ecuaci\'on exacta RG, es necesario proponer un 
Ansatz para la acci\'on de granulado grueso que sea m\'as general que el
utilizado para la aproximaci\'on en derivadas.
Un posible m\'etodo es proponer un desarrollo no local
de la acci\'on de granulado grueso en potencias de la suma 
 $\Sigma = \phi_+ + \phi_-$ y diferencia $\Delta = \phi_+ - \phi_-$ de
campos, que respete las propiedades generales que la acci\'on de granulado
grueso de camino temporal cerrado cumple: 
a) $S_{\Lambda}[\Sigma, \Delta=0] =0$ ; 
b) $S_{\Lambda}[\Sigma, \Delta] =- S_{\Lambda}^*[\Sigma, -\Delta]$ ;
c) ${\rm Im} S_{\Lambda}[\Sigma, \Delta] \ge 0$;
y d) estructura real y causal para las ecuaciones de movimiento
 $ \delta S_{\Lambda} / \delta \Delta |_{\Delta =0}=0$. El Ansatz m\'as
general es \cite{matacz}
\bea
\Delta S_{\Lambda}(\Delta,\Sigma) &=& 
\int d^4x_1 \Delta(x_1) C_1(x_1;\Sigma_{x_1}) \nonumber \\
&& + \frac{i}{2} \int d^4x_1 \int d^4x_2 \Delta(x_1) \Delta(x_2)
C_2(x_1,x_2;\Sigma_{x_1},\Sigma_{x_2}) \nonumber \\
&& - \frac{1}{3!} \int d^4x_1 \int d^4x_2 \int d^4x_3 
\Delta(x_1) \Delta(x_2) \Delta(x_2)
C_3(x_1,x_2,x_3;\Sigma_{x_1},\Sigma_{x_2},\Sigma_{x_3}) \nonumber \\
&& + \ldots ,
\eea
donde $C_n(x_1,\ldots,x_n;\Sigma_{x_1},\ldots,\Sigma_{x_n})$ son cantidades
reales, que son funci\'on de los puntos $x_i$ $(1 \leq i \leq n)$, y 
funcionales de los campos $\Sigma_{x_i}$ $(1 \leq i \leq n)$, entre los
l\'{\i}mites temporales $-\infty$ y $x_i^0$. Por ejemplo,
\bea
C_1(x_1) &=& \int^{x_1^0} d^4y_1 \Sigma(y_1) \mu_1^{(1)}(y_1) \nonumber \\
&& + \int^{x_1^0} d^4y_1 \int^{x_1^0} d^4y_2 \int^{x_1^0} d^4y_3
\Sigma(y_1) \Sigma(y_2) \Sigma(y_3) \mu_3^{(1)}(y_1,y_2,y_3) + \ldots \\
C_2(x_1,x_2) &=& \nu_0^{(2)}(x_1,x_2) + \int^{x_1^0} d^4y_1 \int^{x_2^0} d^4y_2
\Sigma(y_1) \Sigma(y_2) \nu_{2}^{(2)}(y_1,y_2) + \ldots \\
\vdots && \nonumber
\eea
donde $\mu_j^{(i)}$ y $\nu_j^{(i)}$ son n\'ucleos de disipaci\'on y de ruido,
respectivamente, de $j$ puntos
\SSfootnote{Es relativamente sencillo hallar la expresi\'on de estos n\'ucleos
en el caso perturbativo, a partir de la ecs.(\ref{IFre}) y (\ref{IFim}).
En el caso perturbativo, todos los n\'ucleos de m\'as de dos puntos pueden
expresarse como n\'ucleos de s\'olo dos puntos.}, que dependen de la escala
de granulado grueso $\Lambda$.
Dada esta forma general de la acci\'on de granulado grueso, y haciendo 
alg\'un tipo de truncamiento (por ejemplo, conservando s\'olo
unos pocos t\'erminos en la expansi\'on en potencias del campo), 
podr\'{\i}amos resolver en forma aproximada (pero no perturbativa) la
ecuaci\'on RG exacta. Hallar\'{\i}amos as\'{\i} ecuaciones RG 
para los n\'ucleos de ruido y disipaci\'on que aparecen en ese desarrollo 
no local. Es de esperar que ello permita encontrar en forma no 
perturbativa una relaci\'on de fluctuaci\'on-disipaci\'on que dependa 
de la escala de granulado grueso.
En la presente Tesis no continuaremos con esta l\'{\i}nea de trabajo.

\newpage

\thispagestyle{empty}
~
\newpage

\chapter{La acci\'on efectiva usual: m\'etodos de c\'alculo covariantes}

\thispagestyle{empty}

En este cap\'{\i}tulo desarrollamos dos t\'ecnicas de aproximaci\'on 
covariantes para calcular la acci\'on efectiva usual, en signatura 
eucl\'{\i}dea. Si bien trabajamos con la acci\'on efectiva a un lazo, los
m\'etodos que describimos pueden extenderse a mayor cantidad de lazos. Ambas
t\'ecnicas se basan en el denominado n\'ucleo de calor (``heat kernel''),
que permite calcular determinantes funcionales de operadores diferenciales
de segundo orden. Las t\'ecnicas de aproximaci\'on que describimos son:
i) la llamada expansi\'on de Schwinger-DeWitt, que da una expansi\'on local de
la acci\'on efectiva; ii) una t\'ecnica de resumaci\'on, basada en la anterior,
que conduce a una expansi\'on no local. Discutimos los rangos de validez de
ambas aproximaciones, sus ventajas y sus desventajas.

\section{El heat kernel}

En la aproximaci\'on a un lazo, la acci\'on efectiva usual
(con signatura eucl\'{\i}dea) est\'a dada por
\be
S_{\rm ef}[\varphi] = S[\varphi] + \frac{1}{2} \ln {\rm det} \frac{F}{\mu^2} =
S[\varphi] + \frac{1}{2} {\rm Tr} \ln \frac{F}{\mu^2} ,
\ee
donde $F$ es un operador diferencial de segundo orden que se obtiene tomando
las segundas derivadas de la acci\'on cl\'asica (eucl\'{\i}dea) $S$, 
${\rm Tr}$ es una
supertraza (sobre \'{\i}ndices de Lorentz e internos del campo de 
fondo $\varphi$), y $\mu$ es un par\'ametro con unidades de masa. Existe un
amplio conjunto de problemas para los cuales este operador toma la forma
\be
F = {\hat F} (\nabla) = - \Box {\hat 1} + m^2 {\hat 1} - 
( {\hat Q} - \frac{1}{6} R {\hat 1} ) ,
\label{Fminimo}
\ee
donde $\Box = g^{\mu\nu} \nabla_{\mu} \nabla_{\nu}$, $g_{\mu\nu}$ es la 
m\'etrica con signatura eucl\'{\i}dea $(++++)$, $\nabla_{\mu}$ es la derivada 
covariante, ${\hat Q}$ es una matriz que no involucra derivadas y que 
depende del campo de fondo y $m^2$ es un par\'ametro de masa. La convenci\'on 
que utilizaremos para definir los tensores de curvatura es
$R^{\mu}_{\, \cdot \, \nu\alpha\beta} = 
\partial_{\alpha} \Gamma^{\mu}_{\nu\beta} - \ldots$, 
$R_{\alpha\beta} = R^{\mu}_{\, \cdot \, \alpha\mu\beta}$ y
$R= g^{\alpha\beta} R_{\alpha\beta}$. 
Cuando el operador $F$ 
toma esta forma se dice que es m\'{\i}nimo. Operadores no m\'{\i}nimos poseen
estructuras m\'as complicadas en las derivadas, como veremos en el caso de la
cuantizaci\'on a un lazo de la gravedad.

Asumiendo que ${\hat F} (\nabla)$ es definido positivo, podemos escribir
\be
\frac{1}{ - \Box {\hat 1} + m^2 {\hat 1} - 
( {\hat Q} - \frac{1}{6} R {\hat 1} )} = \int_0^{\infty} ds 
\exp \left\{ -s \left[ - \Box {\hat 1} + m^2 {\hat 1} -
( {\hat Q} - \frac{1}{6} R {\hat 1} ) \right] \right\} ,
\ee
con lo cual
\be
{\rm  Tr} \ln \left[ - \Box {\hat 1} + m^2 {\hat 1} -
( {\hat Q} - \frac{1}{6} R {\hat 1}) \right] = - \int_0^{\infty} \frac{ds}{s}
{\rm Tr} K(s) + {\rm constante} ,
\label{hk}
\ee
donde $K(s) \equiv e^{-s m^2} \exp \left\{ s [ \Box {\hat 1} + 
( {\hat Q} - \frac{1}{6} R{\hat 1}) ] \right\}$ es el n\'ucleo del
calor (``heat kernel''). La constante no depende de los par\'ametros del 
operador $\Box$ y puede
reabsorberse en el par\'ametro de masa $\mu$. Obviamente, el heat kernel 
se puede calcular en forma exacta s\'olo en casos de campos de fondo
muy particulares. Sin embargo, interesa tener una expresi\'on general de la
acci\'on efectiva a partir de la cual, por derivadas funcionales, se obtengan 
las ecuaciones de movimiento corregidas de los campos de fondo. Por este motivo
es necesario desarrollar m\'etodos de aproximaci\'on en un caso general. 
Adem\'as, resulta conveniente que tales m\'etodos sean manifiestamente 
covariantes, en el sentido que preserven la covariancia general de la 
teor\'{\i}a orden a orden.

\section{La expansi\'on de Schwinger-DeWitt}

La expansi\'on de Schwinger-DeWitt (SDW) \cite{sdw}
es un desarrollo asint\'otico de la traza del heat kernel
para tiempos cortos, $s \rightarrow 0$, en la forma
\be
{\rm Tr} K(s) = (4 \pi s)^{-w} e^{-s m^2} \sum_{n=0}^{\infty}
s^n \int d^{2w}x \sqrt{g} {\rm Tr} a_n(x) ,
\label{hksdw}
\ee
donde $d=2w$ es la dimensi\'on del espacio-tiempo, $g= {\rm det} g_{\mu\nu}$ y
 $a_n(x)$ es el l\'{\i}mite de coincidencia de los denominados coeficientes
de Schwinger-DeWitt. Estas son cantidades locales constru\'{\i}das a 
partir de las
curvaturas de fondo $\cal R$. Denotaremos con $\cal R$ en forma indistinta
al tensor de Riemann $R_{\mu\nu\alpha\beta}$ (o a cualquiera de sus 
contracciones con la m\'etrica), al conmutador de sus derivadas covariantes
 ${\cal R}_{\mu\nu}$ ($[\nabla_{\mu}, \nabla_{\nu}] \phi = 
 {\cal R}_{\mu\nu} \phi$) o a la matriz ${\hat Q}$. 
El c\'alculo de los coeficientes
de Schwinger-DeWitt en un fondo arbitrario es desde el punto de vista t\'ecnico
una tarea compleja, y s\'olo se han podido calcular en forma expl\'{\i}cita 
los primeros coeficientes. La estructura general del en\'esimo coeficiente es
\be
a_n(x) = \nabla^{2n-2} {\cal R} + {\cal R} \nabla^{2n-4} {\cal R} + \ldots +
\nabla \nabla {\cal R}^{n-1} + {\cal R}^n .
\label{coefsdw}
\ee
N\'otese que tiene dimensi\'on de $({\rm longitud})^{-2n}$ y que aparecen a lo
sumo $n$ curvaturas de fondo ${\cal R}$. La expresi\'on (integrada) de los 
primeros tres coeficientes es \cite{sdw}
\bea
\int d^{2w}x \sqrt{g} {\rm Tr} a_0(x) &=& \int d^{2w}x \sqrt{g} {\rm Tr} 
{\hat 1} \\
\int d^{2w}x \sqrt{g} {\rm Tr} a_1(x) &=& \int d^{2w}x \sqrt{g} {\rm Tr} 
{\hat Q} \\
\int d^{2w}x \sqrt{g} {\rm Tr} a_2(x) &=& \int d^{2w}x \sqrt{g} {\rm Tr} 
\left\{ \frac{1}{2} {\hat Q}^2 + \frac{1}{12} {\hat{\cal R}}_{\mu\nu} 
{\hat{\cal R}}^{\mu\nu} \right. \nonumber \\
&& ~~~~~~~~~~~~~~~~~ +
\left. \left[
\frac{1}{180} R_{\alpha\beta\mu\nu} R^{\alpha\beta\mu\nu} -
\frac{1}{180} R_{\mu\nu} R^{\mu\nu} \right] {\hat 1}
\right\} .
\eea

La expansi\'on de SDW es v\'alida en el caso de campos cu\'anticos
masivos en campos de fondo d\'ebiles, ${\cal R} \ll m^2$, es decir en el caso
en que la longitud de onda Compton del campo cu\'antico 
sea mucho menor que la longitud
t\'{\i}pica de variaci\'on del campo de fondo (campos de
fondo lentamente variables).
En ese caso los coeficientes de SDW verifican $a_n \ll m^{2n}$. 
Haciendo el cambio de variables $u = s m^2$ e intercambiando el 
orden de la integral en el tiempo $u$ con el de la sumatoria, 
podemos obtener una expansi\'on anal\'{\i}tica de la acci\'on efectiva en 
potencias inversas de la masa
\be
S_{\rm ef} = S - \frac{1}{2 (4 \pi)^{-w}} \int d^{2w}x \sqrt{g}
\sum_{n=0}^{\infty} \frac{1}{(m^2)^{n-w}} {\rm Tr} a_n(x)
\int_0^{\infty} du e^{-u} u^{n-w-1} .
\ee  
La integral en el tiempo $u$ presenta una divergencia ultravioleta en el
l\'{\i}mite inferior, que debe ser regularizada ya sea mediante 
regularizaci\'on dimensional o mediante un cut-off
$m^2/ L^2$, ($L \rightarrow \infty$).
Para $d=4$ dimensiones, los tres primeros t\'erminos del desarrollo
son divergentes, obteni\'endose
\bea
S_{\rm ef} &=& S - \frac{1}{32 \pi^2} \lim_{\L \rightarrow \infty}
\int d^4x \sqrt{g} {\rm Tr} \left[
\rho_0 a_0(x) + \rho_1 a_1(x) + \rho_2 a_2(x) \right. \nonumber \\
&& ~~~~~~~~~~~~~~~~~~~~~~~~~~~~~~~~~~~~~~ \left. + \sum_{n=3}^{\infty} (n-3)! 
\frac{1}{(m^2)^{n-2}} a_n(x) \right] ,
\eea
donde
\bea
\rho_0 &=& -m^2 L^2 - \frac{1}{2} m^4 \ln (\frac{m^2}{L^2}) + 
        \frac{1}{2} m^4 (\frac{3}{2} - C) \nonumber \\
\rho_1 &=& L^2 + m^2 \ln(\frac{m^2}{L^2}) + m^2 (C-1) \nonumber \\
\rho_2 &=& - \ln(\frac{m^2}{L^2}) - C ,
\label{rhosdiv}
\eea
con $C=- d/dq \ln \Gamma(q)|_{q=1}$.

El desarrollo de SDW es b\'asicamente una expansi\'on en derivadas del campo 
de fondo. Es \'util para estudiar la renormalizabilidad de la teor\'{\i}a
y para analizar los efectos de polarizaci\'on de vac\'{\i}o 
de campos cu\'anticos
masivos. Sin embargo, existen distintos procesos f\'{\i}sicos que esta 
aproximaci\'on no cubre. Es completamente inadecuada para tiempos largos
($s {\cal R} \gg 1$), para campos de fondo intensos (${\cal R} \gg m^2$),
y carece de sentido para teor\'{\i}as no masivas. Tambi\'en es inadecuada
para campos d\'ebiles y r\'apidamente variables.

\section{Resumaci\'on de la expansi\'on de Schwinger-DeWitt}

Cuando $m^2=0$ las integrales en el tiempo $s$ de la expansi\'on de 
Schwinger-DeWitt del heat kernel divergen en el l\'{\i}mite 
superior, y cada t\'ermino del desarrollo en la serie es m\'as divergente que
el anterior. Sin embargo, la integral de partida ec.(\ref{hk}) converge 
en dicho l\'{\i}mite superior. El problema es que la serie de la ec.
(\ref{hksdw}) no puede ser integrada t\'ermino a t\'ermino, como lo hicimos 
en el caso masivo: se requiere alg\'un tipo de resumaci\'on.

El m\'etodo de resumaci\'on que utilizaremos consiste en realizar la suma
de todos los t\'erminos con una dada potencia de la curvatura $\cal R$ y
cualquier n\'umero de derivadas de todos los coeficientes de 
Schwinger-DeWitt \cite{vilk1,vilk2,avra}.
Claramente este m\'etodo de resumaci\'on tendr\'a sentido para situaciones en
las cuales $\nabla \nabla {\cal R} \gg {\cal R} {\cal R}$, es decir para 
campos d\'ebiles y r\'apidamente variables. La t\'ecnica puede ser aplicada
tanto para casos masivos como no masivos. Seguidamente presentaremos la
idea del m\'etodo en forma esquem\'atica y en cap\'{\i}tulos posteriores
haremos c\'alculos concretos bas\'andonos en este proceso de resumaci\'on.

Comencemos entonces a analizar la
resumaci\'on orden a orden en potencias de curvaturas. El orden m\'as 
bajo no posee curvaturas y est\'a dado por el coeficiente $a_0(x)= {\hat 1}$. 
Por lo tanto, a este orden, el heat kernel es
\be
{\rm Tr} K(s)_0 = (4 \pi s)^{-2} e^{-s m^2} \int d^{4}x \sqrt{g} {\rm Tr} 
\left\{ {\hat 1} \right\} .
\ee
El orden uno en curvaturas provendr\'a de la resumaci\'on de los t\'erminos
lineales en curvaturas de los coeficientes de SDW, es decir
\bea
a_1(x) &=& \alpha_1 {\cal R} \nonumber \\
a_2(x) &=& \alpha_2 \Box {\cal R} + {\cal O}({\cal R}^2) \nonumber \\
& \vdots & \nonumber \\
a_n(x) &=& \alpha_n \Box^{n-1} {\cal R} +  {\cal O}({\cal R}^2) .
\eea
Todos los t\'erminos excepto el primero son derivadas totales, de modo que 
pueden ser descartados en la integrales espacio-temporales, 
bajo la suposici\'on de que el espacio-tiempo es asint\'oticamente plano. 
A primer orden el heat kernel queda
\be
{\rm Tr} K(s)_1 = (4 \pi s)^{-2} e^{-s m^2} \int d^{4}x \sqrt{g} {\rm Tr} 
\left\{ s {\cal R} \right\} .
\ee
Teniendo en cuenta la estructura general del en\'esimo coeficiente de SDW 
(ec.(\ref{coefsdw})), usando que el conmutador de derivadas
covariantes da t\'erminos de orden ${\cal O}({\cal R}^3)$ e integrando 
sucesivamente por partes, las contribuciones cuadr\'aticas se pueden reducir a
la forma
\be
\int d^{4}x \sqrt{g} a_n(x) = \int d^{4}x \sqrt{g} \sigma_n {\cal R} 
(-\Box)^{n-2}
{\cal R} + {\cal O}({\cal R}^3) ~~~ (n \ge 2) ,
\label{417}
\ee
con ciertos coeficientes $\sigma_n$, que se calculan al transformar al 
coeficiente de SDW de modo que tenga esta forma particular. Una vez hallados,
definiendo la funci\'on 
\be
\sigma(\eta)=\sum_{n=2}^{\infty} \sigma_n \eta^{n-2} ,
\label{ffgeneral}
\ee
llamada factor de forma, la resumaci\'on del heat kernel 
a segundo orden en curvaturas conduce a una expresi\'on no local de la forma
\be
{\rm Tr} K(s)_2 = (4 \pi s)^{-2} e^{-s m^2} \int d^{4}x \sqrt{g} {\rm Tr} 
\left\{ s^2 {\cal R} \sigma(-s \Box) {\cal R} \right\} .
\ee
Lo dif\'{\i}cil en este proceso es hallar los factores de forma que
acompa\~nan a las distintas contribuciones que provienen de las distintas
posibles curvaturas ${\cal R}$. Dicho c\'alculo ha sido llevado a cabo en
\cite{vilk1,vilk2,avra}, y el resultado final, a orden
cuadr\'atico en curvaturas
\SSfootnote{Recientemente la resumaci\'on ha sido calculada hasta orden
c\'ubico en curvaturas \cite{cubic}, resultando una estructura no local muy
complicada que involucra veintinueve factores de forma. Afortunadamente,
en las aproximaciones en que trabajaremos en los cap\'{\i}tulos siguientes,
estos t\'erminos c\'ubicos son despreciables frente a los cuadr\'aticos, de 
modo que podremos evitar trabajar con semejante cantidad de factores de forma.}
, es
\bea
{\rm Tr} K(s) &=& (4 \pi s)^{-2} e^{- s m^2} \int d^{4}x \sqrt{g} 
{\rm Tr} \left\{ {\hat 1} + s {\hat Q} +
s^2 \left[
R_{\mu\nu} f_1(-s \Box) R^{\mu\nu} {\hat 1} +
R f_2(-s \Box) R {\hat 1}  \right. \right. \nonumber \\
&& + \left. \left. {\hat Q} f_3(-s \Box) R {\hat 1}+
{\hat Q} f_4(-s \Box) {\hat Q} + 
{\hat{\cal R}}_{\mu\nu} f_5(-s \Box) {\hat{\cal R}}^{\mu\nu}
\right]
\right\} + {\cal O}({\cal R}^3) ,
\label{hkruso}
\eea
donde $f_i (i=1,\ldots,5)$ son factores de forma, funci\'on de la
variable $\eta=-s \Box$, dados por
\bea
f_1(\eta) &=& \frac{f(\eta) - 1 + \eta/6}{\eta^2} \\
f_2(\eta) &=& \frac{1}{8} 
\left[
\frac{1}{36} f(\eta) + \frac{f(\eta)-1}{3 \eta} - 
\frac{f(\eta)-1+ \eta/6}{\eta^2}
\right] \\
f_3(\eta) &=& \frac{1}{12} f(\eta) + \frac{f(\eta)-1}{2 \eta} \\
f_4(\eta) &=& \frac{1}{2} f(\eta) \\
f_5(\eta) &=& - \frac{f(\eta)-1}{2 \eta} \\
f(\eta) & \equiv & \int_0^1 dt \, e^{-\frac{1-t^2}{4} \eta} .
\eea
Los t\'erminos de orden cero y uno en curvaturas en la ec.(\ref{hkruso})
son locales, mientras que los de segundo orden son no locales, y tienen la
estructura ${\cal R} \beta(-s \Box) {\cal R}$ de la que habl\'abamos 
anteriormente. La sexta posible estructura,
 $R_{\mu\nu\alpha\beta} f(-s \Box) R^{\mu\nu\alpha\beta}$ no est\'a presente
en la ec.(\ref{hkruso}), pues, en este desarrollo perturbativo 
covariante, es siempre posible eliminar al tensor de Riemann utilizando las
identidades de Bianchi para expresarlo en funci\'on del tensor y del escalar
de Ricci, m\'as t\'erminos no locales de orden superior en curvaturas.
 
Para tiempos cortos $\eta \rightarrow 0$, el comportamiento de
los factores de forma es
\bea
f_1(\eta) = \frac{1}{60} + {\cal O}(\eta) 
 & ~~~ f_2(\eta) = - \frac{1}{180} +{\cal O}(\eta)~~~ &
~~~f_3(\eta) = 0 +{\cal O}(\eta) \nonumber \\
f_4(\eta = \frac{1}{2} +{\cal O}(\eta) 
 & ~~~ f_5(\eta) = \frac{1}{12} +{\cal O}(\eta) . ~~~ &  
\label{smalltimes}
\eea 
En este l\'{\i}mite se recupera el desarrollo de Schwinger-DeWitt, a orden
cuadr\'atico en curvaturas. En el l\'{\i}mite opuesto, de tiempos largos
 $\eta \rightarrow \infty$, los factores de forma se comportan como
\bea
f_1(\eta) = \frac{1}{6 \eta } + {\cal O}(\frac{1}{\eta^2}) 
& ~~~ f_2(\eta) = - \frac{1}{18 \eta } + {\cal O}(\frac{1}{\eta^2})~~~ &
~~~f_3(\eta) = - \frac{1}{3 \eta} + {\cal O}(\frac{1}{\eta^2}) \nonumber \\
f_4(\eta) = \frac{1}{\eta} + {\cal O}(\frac{1}{\eta^2})   
 & ~~~ f_5(\eta) = \frac{1}{2} +  {\cal O}(\frac{1}{\eta^2}) .  &
\label{largetimes} 
\eea 
En este caso, la traza del heat kernel $K(s)$ es proporcional a $s^{-w+1}$
para todos los ordenes en curvatura, expecto el orden cero, que es 
proporcional a $s^{-w}$. 

La acci\'on efectiva resulta de integrar en el tiempo $s$ el heat kernel 
de acuerdo a la ec.(\ref{hk}). Obtenemos as\'{\i}
\bea
S_{\rm ef}&=& S - \frac{1}{32 \pi^2} \int d^{4}x \sqrt{g}
\lim_{L \rightarrow \infty} 
{\rm Tr} \left\{ 
h_0 {\hat 1} + h_1 {\hat Q} +
R_{\mu\nu} F_1(\Box) R^{\mu\nu} {\hat 1} + R F_2(\Box) R {\hat 1}  
\right. \nonumber \\
&& + \left. {\hat Q} F_3(\Box) R {\hat 1} +
{\hat Q} F_4(\Box) {\hat Q} + 
{\hat{\cal R}}_{\mu\nu} F_5(\Box) {\hat{\cal R}}^{\mu\nu}
\right\} + {\cal O}({\cal R}^3) ,
\eea
donde
\bea
h_0 &=& \int_{1/L^2}^{\infty} ds s^{-3} e^{-s m^2} = 
- \frac{ m^4}{2} \ln \frac{m^2}{L^2} \\
h_1 &=& \int_{1/L^2}^{\infty} ds s^{-2} e^{-s m^2} = 
m^2 \ln \frac{m^2}{L^2} ,
\eea
y los factores de forma no locales est\'an dados por
$F_i(\Box) = \int_{1/L^2}^{\infty} ds \frac{e^{-s m^2}}{s^{w-1}} 
f_i(-s \Box)$ y resultan ser, para $d=4$ dimensiones
\bea
F_1(\Box)&=& \int_0^1 dt
\left[
-\frac{1}{32} (1-t^2)^2 \ln \frac{G(\Box)}{L^2} +
\left(
-\frac{m^4}{2 \Box^2} + \frac{1-t^4}{4} \frac{m^2}{\Box}
\right) \ln \frac{G(\Box)}{m^2}
\right] \label{F1} \\
F_2(\Box) &=& \frac{1}{8} \int _0^1 dt
\left[
\left(
-\frac{1}{36}+ \frac{1}{12} (1-t^2) + \frac{1}{32} (1-t^2)^2
\right)  \ln \frac{G(\Box)}{L^2}  \right. \nonumber \\
&& ~~~~~~~~~~ + \left. \left(
-\frac{m^2}{3 \Box} + \frac{m^4}{2 \Box^2} - \frac{m^2}{\Box} \frac{1-t^2}{4}
\right)  \ln \frac{G(\Box)}{m^2}
\right] \\
F_3(\Box) &=& \int _0^1 dt
\left[
\frac{1}{8} (\frac{1}{3} - t^2) \ln \frac{G(\Box)}{L^2} - \frac{m^2}{2 \Box}
\ln \frac{G(\Box)}{m^2}
\right] \\
F_4(\Box) &=& -\frac{1}{2}\int _0^1 dt
\ln \frac{G(\Box)}{L^2} \\
F_5(\Box) &=& \frac{1}{2} \int _0^1 dt
\left[
- \frac{1-t^2}{4} \ln \frac{G(\Box)}{L^2} + \frac{m^2}{\Box} \ln 
\frac{G(\Box)}{m^2}
\right] \label{F5} ,
\eea
donde $G(\Box)=m^2 - \frac{1}{4} (1-t^2) \Box$. El factor de forma
 $F(\Box) \equiv \ln G(\Box)/L^2$ 
admite una representaci\'on espectral en t\'erminos del propagador 
eucl\'{\i}deo masivo $(M^2-\Box)^{-1}$
\be
F(\Box) = \ln 
\left[ \frac{m^2 - \frac{1}{4} (1-t^2) \Box }{L^2} \right] =
\ln \left( \frac{1-t^2}{4} \right)  + \int_0^{\infty} dz 
\left[
\frac{1}{z+L^2} - \frac{1}{z + \frac{4 m^2}{1-t^2} - \Box}
\right] .
\ee
Entonces, $F(\Box)$ es una funci\'on de dos puntos que sobre una funci\'on
de prueba $f(x)$ act\'ua de la forma $F(\Box) f(x) = \int d^4x' F(\Box)(x,x') 
f(x')$.

Tanto las integrales $h_i$ como los factores de forma $F_i(\Box)$ son
cantidades que divergen en el l\'{\i}mite ultravioleta $L \rightarrow \infty$,
siendo proporcionales a $\ln L^2$. La teor\'{\i}a es renormalizable si la 
acci\'on cl\'asica de partida $S$ tiene la misma estructura que las 
divergencias, es decir, si posee t\'erminos independientes de curvaturas
(proporcionales a ${\hat 1}$), 
t\'erminos lineales (proporcionales a ${\hat Q}$),
y t\'erminos cuadr\'aticos (proporcionales a ${\cal R} {\cal R}$).
En ese caso las divergencias pueden ser absorbidas en las constantes
desnudas $\{ \alpha_d^i \}$ de la acci\'on cl\'asica $S$, redefini\'endolas
en la forma  $\alpha_d^i = \alpha_v^i - {\rm cte}^i \ln(L^2/ \mu^2)$, 
donde $\{ \alpha_d^i \}$ son las correspondientes constantes vestidas, $\mu$
es un par\'ametro de escala con unidades de masa, y $\{ {\rm cte}^i \}$
son las constantes que acompa\~nan a las divergencias en $\ln L^2$. 
Los factores de forma renormalizados son los mismos que los dados en las
ecs. (\ref{F1})-(\ref{F5}) con $L^2$ reemplazado por $\mu^2$.
En esta
forma se renormaliza la teor\'{\i}a, resultando un conjunto de constantes 
vestidas que dependen de la escala de energ\'{\i}a, 
 $\alpha_v^i = \alpha_v^i(\mu)$. Usando el hecho que la acci\'on efectiva no
depende de este par\'ametro arbitrario $\mu$, se obtiene un conjunto de 
ecuaciones del grupo de renormalizaci\'on para las constantes vestidas,
 $\mu d \alpha_v^i / d \mu = \beta^i( \{ \alpha_v^j \})$ que rigen la
forma en que dichas constantes cambian a distintas escalas de energ\'{\i}a.

\newpage

\thispagestyle{empty}
~
\newpage

\chapter{Correcciones cu\'anticas al potencial newtoniano}

\thispagestyle{empty}

En este cap\'{\i}tulo calculamos correcciones cu\'anticas al potencial
newtoniano bas\'andonos en la acci\'on efectiva. Con el objeto de dar
un ejemplo sencillo del m\'etodo que utilizaremos, comenzamos por calcular
el apantallamiento de la carga el\'ectrica en electrodin\'amica cu\'antica.
Pasamos luego a tratar los efectos de campos de materia (escalares y
fermi\'onicos) sobre un fondo gravitatorio cl\'asico, en el contexto de
teor\'{\i}a semicl\'asica de la gravedad. A partir de los t\'erminos
no locales de la acci\'on efectiva, obtenidos en el cap\'{\i}tulo anterior,
calculamos c\'omo la constante gravitatoria pasa a ser funci\'on de la
distancia entre part\'{\i}culas masivas \cite{DMnewt}. 
Hacemos un an\'alisis comparativo 
entre los resultados obtenidos con los que resultan de la aplicaci\'on del 
grupo de renormalizaci\'on \cite{DMscaling}. 

\section{Teor\'{\i}as de campos efectivas}

La teor\'{\i}a de la relatividad general es una teor\'{\i}a cl\'asica que
describe en forma satisfactoria di\-fe\-ren\-tes procesos 
gravitacionales a escalas 
de energ\'{\i}as presentes, es decir mucho menores que la escala de Planck. 
Si se desea estudiar procesos a escalas de energ\'{\i}a mayores se deben 
incluir efectos cu\'anticos. Como hemos discutido en la introducci\'on,
es entonces necesario 
tener en cuenta no s\'olo las fluctuaciones de los campos de materia,
sino tambi\'en las fluctuaciones del
propio campo gravitatorio \cite{dewitt}. 
Es aqu\'{\i} donde aparecen serias dificultades
t\'ecnicas que pueden describirse como una incompatibilidad entre la mec\'anica
cu\'antica y la relatividad general. En efecto, debido a que la constante 
gravitatoria tiene dimensiones y a la naturaleza no lineal de la 
teor\'{\i}a, en el c\'alculo de lazos de gravitones aparecen divergencias
ultravioletas que no pueden ser absorbidas en la acci\'on cl\'asica de partida:
la teor\'{\i}a no es renormalizable. Hist\'oricamente esta dificultad fue
considerada como un impedimento para realizar predicciones mediante la 
relatividad general cuantizada, debiendo entonces encontrarse la verdadera
teor\'{\i}a cu\'antica de la gravedad (hasta hoy desconocida), v\'alida
para energ\'{\i}as cercanas y superiores a la de Planck. A pesar de la no
renormalizabilidad de la teor\'{\i}a, los procesos cu\'anticos a bajas
energ\'{\i}as y largas distancias {\it pueden} ser calculados en forma
satisfactoria cuantizando la relatividad general, independientemente de
la eventual forma de la gravedad cu\'antica. La clave es que, a bajas 
energ\'{\i}as, la relatividad general se comporta en la forma en que usualmente
se tratan las teor\'{\i}as de campos efectivas. Seguidamente describiremos
las ideas que subyacen en el tratamiento de teor\'{\i}as efectivas. 
Una descripci\'on m\'as completa de estas ideas puede hallarse 
en \cite{wein,don0} y su aplicaci\'on a la relatividad general 
en \cite{donoghue1,don2}.

Una teor\'{\i}a de campos efectiva est\'a construida en base a ciertos grados
de libertad e interacciones entre los mismos que describen procesos 
f\'{\i}sicos en un cierto rango de energ\'{\i}as y distancias. Estos grados
de libertad pueden resultar insuficientes o completamente inadecuados para
explicar la f\'{\i}sica a escalas de energ\'{\i}as mayores, en cuyo caso
es necesario reemplazar la teor\'{\i}a efectiva por otra mejor. La teor\'{\i}a
de bajas energ\'{\i}as involucra un lagrangiano efectivo (que es una cantidad
local construida con los grados de libertad efectivos) y un tratamiento 
completo de su cuantizaci\'on (lazos, renormalizaci\'on, etc.). El objetivo
final es poder describir en forma satisfactoria todos los procesos cu\'anticos
de bajas energ\'{\i}as. Hay dos elementos que juegan un papel fundamental en
la metodolog\'{\i}a de teor\'{\i}as efectivas. El primero es la separaci\'on
entre el sector de baja y de alta energ\'{\i}a. En un tratamiento 
diagram\'atico, los efectos de las part\'{\i}culas virtuales pesadas 
involucran propagaciones de corto alcance y pueden ser representados por
una serie de lagrangianos locales. Un ejemplo de \'esto es la teor\'{\i}a de
Fermi de las interacciones d\'ebiles. Por el contrario, los efectos no locales
corresponden a la propagaci\'on de part\'{\i}culas virtuales livianas por
grandes distancias, y s\'olo pueden provenir del sector de bajas energ\'{\i}as
de la teor\'{\i}a. 

El segundo aspecto de importancia es la expansi\'on en 
energ\'{\i}as. La idea es escribir el lagrangiano m\'as general compatible
con las simetr\'{\i}as de la teor\'{\i}a y ordenar los infinitos t\'erminos
en potencias del cociente entre la escala de bajas energ\'{\i}as y la de 
altas energ\'{\i}as. El orden m\'as bajo de dicho lagrangiano sirve para
definir los propagadores y v\'ertices, mientras que los restantes ordenes
se tratan perturbativamente. Al cuantizar y calcular lazos, aparecen 
divergencias
ultravioletas (cantidades locales y anal\'{\i}ticas en el espacio de 
momentos) que tienen la forma de alg\'un t\'ermino en dicho lagrangiano y 
renormalizan las constantes de acoplamiento del lagrangiano. En el proceso de
c\'alculo de los lazos aparecen otras contribuciones de un car\'acter
diferente, que surgen del sector de bajas energ\'{\i}as, y que son
no anal\'{\i}ticas, t\'{\i}picamente de la forma $\ln(-q^2)$
\SSfootnote{Esta forma aparece para fluctuaciones no masivas, como por ejemplo
gravitones. Si las fluctuaciones tuviesen masa $\Omega$, entonces 
habr\'{\i}a otra contribuci\'on no anal\'{\i}tica de la forma
 $\Omega/\sqrt{-q^2}$, que tambi\'en es relevante a bajas energ\'{\i}as.}.
Estos efectos
provienen del orden m\'as bajo del lagrangiano y por lo tanto son 
independientes del valor que tomen las constantes de acoplamiento 
de los ordenes 
superiores. Son estos t\'erminos no anal\'{\i}ticos los que conducen
a predicciones cu\'anticas bien definidas de bajas energ\'{\i}as.

Una teor\'{\i}a de campos efectiva cuya estructura es similar a la de 
la relatividad general
es el modelo $\sigma$ no lineal, que describe una teor\'{\i}a de piones
que es el l\'{\i}mite de bajas energ\'{\i}as de QCD
\cite{don0,weinP,gasser}.
Esta teor\'{\i}a
ha sido extensamente estudiada, tanto te\'orica como experimentalmente.
El lagrangiano
de orden m\'as bajo es
\bea
{\cal L}_2 = -\frac{f_{\pi}^2}{4} {\rm Tr} (\partial_{\mu} U \partial^{\mu} 
U^{\dagger}) & ~~~~~~& U=\exp(i \pi^a \tau^a/f_{\pi}) , 
\eea
donde $\tau^a$ son los generadores de ${\rm SU}(n)$. 
Al igual que en la relatividad
general, la constante de acoplamiento $f_{\pi}$ posee dimensiones, la 
teor\'{\i}a es intr\'{\i}nsecamente no lineal, y no es 
renormalizable. En efecto,
a un lazo existe una divergencia ultravioleta de la forma
\be
{\cal L}_{\rm div} \propto 
\left[ {\rm Tr} \partial_{\mu} U \partial^{\mu} U^{\dagger} \right]^2 + 
2 {\rm Tr}  \left[ \partial_{\mu} U \partial^{\mu} U^{\dagger} \right]^2 .
\ee
que no posee la estructura de ${\cal L}_2$.
Adoptando la metodolog\'{\i}a de teor\'{\i}as efectivas, escribimos el
lagrangiano m\'as general compatible con las simetr\'{\i}as de la teor\'{\i}a
\be
{\cal L}_{\rm ef} = {\cal L}_2 + 
\alpha \left[ {\rm Tr} \partial_{\mu} U \partial^{\mu} U^{\dagger} \right]^2 +
\beta {\rm Tr}  \left[ \partial_{\mu} U \partial^{\mu} U^{\dagger} \right]^2
+ \ldots ,
\ee
y a bajas energ\'{\i}as ($| \partial| =p^2 < f_{\pi}^2$), ${\cal L}_2$
ser\'a el t\'ermino dominante. El procedimiento para implementar el m\'etodo
de teor\'{\i}as efectivas a un lazo es el siguiente:

\noindent i) Calcular v\'ertices y propagadores con ${\cal L}_2$,  
$[{\cal O}(p^2)]$.

\noindent ii) Calcular las correcciones a un lazo.

\noindent iii) Combinar los efectos de orden ${\cal O}(p^2)$ y 
 ${\cal O}(p^4)$ del lagrangiano
efectivo con las correcciones a un lazo. Las divergencias (y las 
correspondientes partes finitas) pueden absorberse en las constantes 
 $\alpha$ y $\beta$.

\noindent iv) Medir los coeficientes desconocidos por comparaci\'on con el
experimento. Habiendo determinado los par\'ametros de la teor\'{\i}a,
se pueden hacer predicciones v\'alidas a orden ${\cal O}(p^4)$ en una
expansi\'on en energ\'{\i}as.

\noindent v) Para obtener las correcciones cu\'anticas m\'as relevantes
en la expansi\'on en energ\'{\i}as, es necesario calcular las contribuciones
no anal\'{\i}ticas a un lazo. Estas son independientes de los par\'ametros
 $\alpha$ y $\beta$.

Como veremos, este procedimiento tambi\'en puede aplicarse a la relatividad
general. Sin embargo, dado que 
el c\'alculo de correcciones cu\'anticas por efecto de gravitones en 
relatividad general es complicado
desde un punto de vista t\'ecnico, dejaremos para el pr\'oximo
cap\'{\i}tulo el efecto de las fluctuaciones cu\'anticas del propio campo 
gravitatorio, y en este cap\'{\i}tulo comenzaremos 
por ejemplos m\'as sencillos. Calcularemos las 
correcciones cu\'anticas debidas a campos de materia 
(campos escalares y fermi\'onicos) a la din\'amica de un fondo gravitatorio 
cl\'asico. Pero antes trataremos un caso a\'un m\'as simple en el contexto de
electrodin\'amica, que nos servir\'a para ejemplificar las t\'ecnicas
que utilizaremos.

\section{El apantallamiento de la carga el\'ectrica}

En electrodin\'amica cu\'antica es bien sabido que debido a efectos de
polarizaci\'on de vac\'{\i}o el potencial de interacci\'on electrost\'atico
entre cargas puntuales en reposo difiere del potencial coulombiano cl\'asico
 $r^{-1}$. La forma usual de obtener este efecto es calculando la interacci\'on
entre dos part\'{\i}culas de carga $e$ (con corrientes asociadas $J^{\mu}(p)$,
siendo $p^{\mu}$ el cuadri-impulso) a trav\'es del intercambio de un fot\'on,
e incluir las correcciones al propagador del fot\'on $D_{\mu\nu}(p)$ por 
emisi\'on de pares virtuales electr\'on-positr\'on. Los diagramas de Feynman
necesarios para realizar dicho c\'alculo a primer orden en la constante de
estructura fina $\alpha =e^2/4 \pi \hbar c$ se muestran en la figura 5.1. 

%%%%%%%%%%%%%%%%%%%%%%%%%%%%%%%%%%%%%%%%%%%%%%%%%%%%%%%%%%%%%%%%%%%%%%%%

\begin{figure}[h]
\centering \leavevmode
\epsfxsize=10cm
\epsfbox{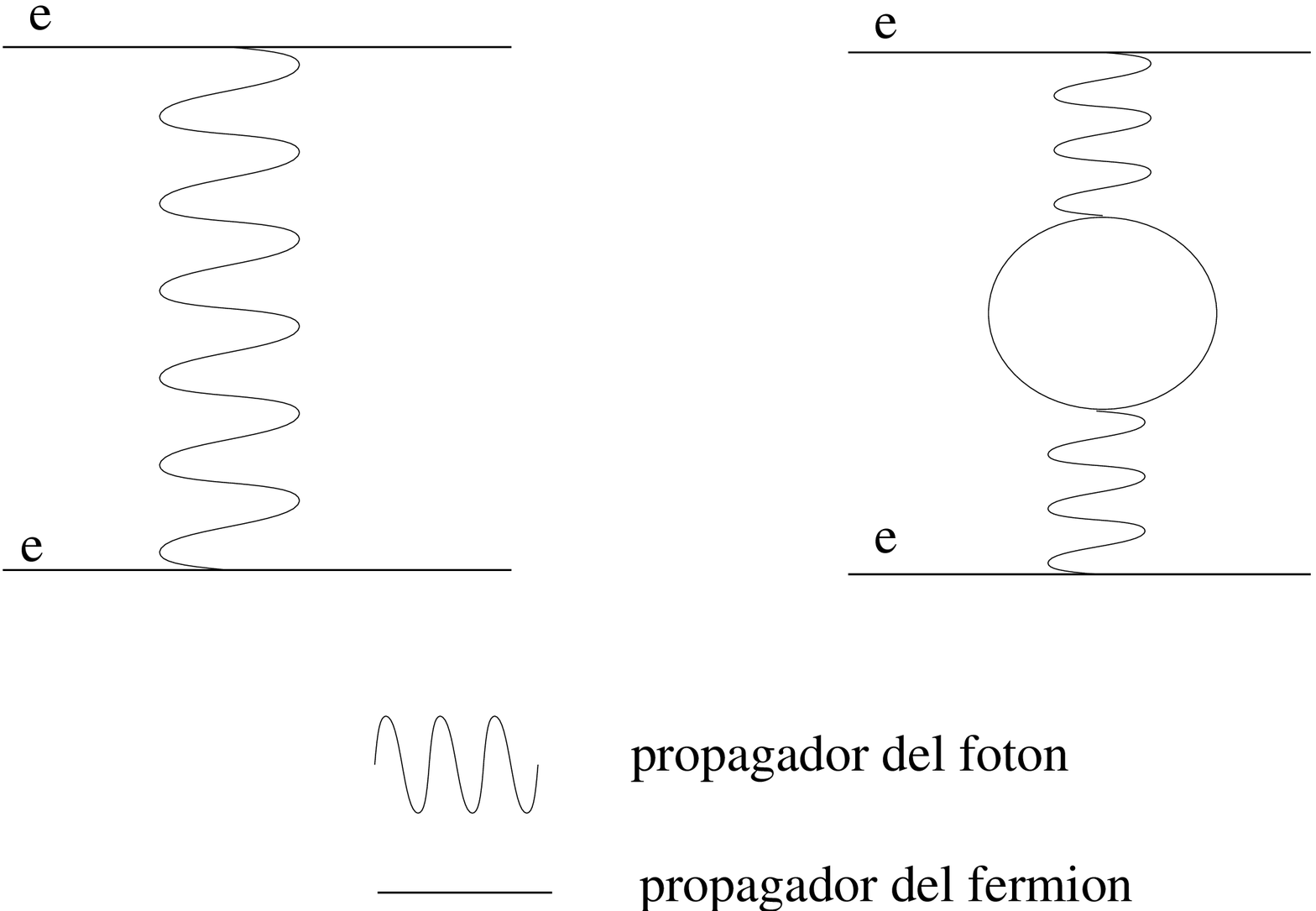}
\setlength{\captwidth}{12cm}
\capt{Diagramas de Feynman para calcular, al orden m\'as bajo en la 
constante de estructura fina, el potencial de interacci\'on 
coulombiano entre cargas puntuales. El segundo diagrama es la contribuci\'on
debida a la polarizaci\'on de vac\'{\i}o.}
\end{figure}

%%%%%%%%%%%%%%%%%%%%%%%%%%%%%%%%%%%%%%%%%%%%%%%%%%%%%%%%%%%%%%%%%%%%%%%%

El potencial de 
interacci\'on  electrost\'atica se encuentra tomando la transformada de 
Fourier del l\'{\i}mite est\'atico de 
 $M_{12}(p) \equiv J^{\mu}(p) D_{\mu\nu}(p) J^{\nu}(-p)$. El resultado final es
\cite{itzykson} 
\be
V_{{\rm int}}(r)=   \frac{e^{2}(r)}{4 \pi} = \frac{e^2}{4 \pi r} 
\left[ 1 + \frac{e^2}{6 \pi^2} \int_{1}^{\infty} du e^{-2 m r u}
(1+\frac{1}{2 u^2}) \frac{\sqrt{u^2-1}}{u^2} + {\cal O}(e^4)\right] .
\label{eq:pot}
\ee
Por los efectos de los pares virtuales, la carga el\'ectrica que ``ve'' una
de las part\'{\i}culas no es igual a $e$, sino que mide una carga $e(r)$ 
que depende de la distancia a la otra part\'{\i}cula cargada. Para grandes
distancias ($m r \gg 1$), dicha carga es aproximadamente igual a $e$,
mientras que para separaciones peque\~nas ($m r \ll 1$), 
\be 
e(r) = e \left[ 1 - \frac{e^2}{12 \pi^2} 
        \ln{\frac{r}{r_{\sst 0}}} + {\cal O}(e^4) \right] ,
\label{eq:itzi}
\ee
donde $r_{\sst 0}$ est\'a definido por $-\ln{m r_{\sst 0}} = 2 \gamma + 
\frac{5}{3}$. La presencia de la carga desnuda $e$ polariza el vac\'{\i}o, 
siendo rodeada por cargas virtuales de signo opuesto, que finalmente hacen
que la carga vestida $e(r)$ sea de magnitud menor. A medida que nos acercamos
al origen, la carga vestida crece logar\'{\i}tmicamente.

Otra manera de obtener el potencial electrost\'atico en el 
l\'{\i}mite de cortas distancias se basa en un argumento ``wilsoniano''.
La idea es partir del potencial coulombiano cl\'asico y reemplazar a la
carga $e$ por la soluci\'on de la ecuaci\'on del grupo de renormalizaci\'on
 $e(\mu)$, y finalmente identificar a la escala de energ\'{\i}a $\mu$ con la
inversa de la distancia, es decir
\be
V_{\rm int}(r) = \frac{e^2(\mu \rightarrow \frac{1}{r})}{4 \pi} .
\ee 
La ecuaci\'on del grupo de renormalizaci\'on en electrodin\'amica es 
 $\mu d e/ d\mu = - e^2/12 + {\cal O}(e^4)$, cuya soluci\'on resulta
\be
e(\mu) = e(\mu_{\sst 0}) \left[ 1-\frac{e^2(\mu_{\sst 0})}{12 \pi^2}
\ln{\frac{\mu_{\sst 0}}{\mu}} + {\cal O}(e^4) \right] .
\label{eq:eren}
\ee 
Es inmediato entonces comprobar que se obtiene la dependencia correcta
con la distancia tanto para la carga vestida como para el potencial
electrost\'atico, en el l\'{\i}mite de cortas distancias. En electrodin\'amica
el argumento wilsoniano conduce al resultado correcto.

A continuaci\'on calcularemos, a partir de la acci\'on efectiva,
las ecuaciones de movimiento para el campo electromagn\'etico en presencia 
de los fermiones y veremos c\'omo resulta el apantallamiento de la carga 
el\'ectrica. La acci\'on cl\'asica de la electrodin\'amica es 
\SSfootnote{Usamos la signatura $(-+++)$ y las definiciones de \cite{wein}.}
\be
S_{\rm clas} = S[A_{\mu}] + S[\psi, \bar{\psi}, A_{\mu}]  = 
-\frac{1}{4} \int d^4x F_{\mu\nu} F^{\mu\nu} - 
\int d^4x \bar{\psi} ( \gamma^{\mu} \partial_{\mu} + m + i e 
\gamma^{\mu} A_{\mu}) \psi ,
\ee
donde las matrices de Dirac satisfacen 
 $\{ \gamma^{\mu},\gamma^{\nu} \} =  2 g^{\mu\nu}$, 
 $\gamma^{\mu \dagger} = \gamma^0 \gamma^{\mu} \gamma^0$, siendo
 $\gamma^0$ antiherm\'{\i}tica y $\gamma^i$ herm\'{\i}tica.

Comencemos por calcular la acci\'on efectiva in-out, para lo cual integramos 
sobre los campos fermi\'onicos. De este modo resulta
\be
S_{\rm ef}[A_{\mu}] = S[A_{\mu}] - i \ln {\rm det}
(\gamma^{\mu} \partial_{\mu} + m + i e \gamma^{\mu} A_{\mu}) .
\ee
En general este determinante fermi\'onico no puede calcularse en forma exacta,
y por ello haremos un desarrollo perturbativo en potencias de la constante
de estructura fina, o equivalentemente en potencias del campo $A_{\mu}$. 
Obtenemos
\be
S_{\rm ef}[A_{\mu}] = S[A_{\mu}] + \frac{i}{2} \int d^4x d^4x'
A_{\mu}(x) \Pi^{\mu\nu}(x,x') A_{\nu}(x') + {\cal O}(A^4) ,
\ee
donde $\Pi^{\mu\nu}(x,x') \equiv - e^2 {\rm Tr}[\gamma^{\mu} F_{++}(x,x')
\gamma^{\nu} F_{++}(x',x)]$ es el tensor de polarizaci\'on de vac\'{\i}o en 
electrodin\'amica cu\'antica, y 
\be
F_{++}(x,x')=i \langle {\rm T} \psi(x) {\bar \psi}(x') \rangle =
\int \frac{d^4p}{(2 \pi)^4} e^{ip(x-x')} \frac{i \gamma^{\mu} p_{\mu} - m}
{p^2 + m^2 - i \epsilon} ,
\ee
es el propagador fermi\'onico. La ecuaci\'on de
movimiento in-out para el campo es
\be
0=-\partial_{\nu} F^{\mu\nu} + i e^2 \int d^4x' {\rm Tr}
[\gamma^{\mu} F_{++}(x,x') \gamma^{\nu} F_{++}(x',x)] A_{\nu}(x') .
\label{ecinout}
\ee
El problema de esta ecuaci\'on es que no es ni real ni causal, debido a
la estructura del propagador $F_{++}$. Para obtener las
ecuaciones de movimiento reales y causales para el campo electromagn\'etico
debemos hallar la acci\'on efectiva in-in. 

La funcional generatriz fermi\'onica in-in para el caso sin interacci\'on es
\bea
Z[\bar{\eta}^+,\eta^+,\bar{\eta}^-,\eta^-] &=& \int_{\rm CTC}
D {\bar \psi}^- D \psi^- D{\bar \psi}^+ D{\psi}^+  \nonumber \\
&& ~~~~ \times
e^{-i (S[\psi^-,\bar{\psi}^-] + \bar{\eta}^- \psi^- + \bar{\psi}^- \eta^-)}
e^{ i (S[\psi^+,\bar{\psi}^+] + \bar{\eta}^+ \psi^+ + \bar{\psi}^+ \eta^+)} ,
\eea
donde la integral se hace con las condiciones de contorno de camino 
temporal cerrado. De aqu\'{\i}
resulta el siguiente propagador fermi\'onico $F_{ab}$,
\bea
F_{++}(x,x') &=& i \left. \frac{\delta}{i {\bar \eta}^+(x)}
                   \frac{\delta}{i {\eta}^+(x')} Z 
                   \right|_{\eta={\bar \eta}=0} 
                   \nonumber \\
             &=& i \langle {\rm T} \psi(x) {\bar \psi}(x') \rangle = 
i [ \theta(x^0-x'^0) \langle \psi(x) {\bar \psi(x')} \rangle -
    \theta(x'^0-x^0) \langle {\bar \psi(x')} \psi(x) \rangle ] \\
F_{+-}(x,x') &=&  i \left. \frac{\delta}{i {\bar \eta}^+(x)}
                   \frac{\delta}{-i {\eta}^-(x')} Z 
                   \right|_{\eta={\bar \eta}=0} 
                  = -i \langle {\bar \psi}(x') \psi(x) \rangle    \\
F_{-+}(x,x') &=&  i \left. \frac{\delta}{- i {\bar \eta}^-(x)}
                   \frac{\delta}{i {\eta}^+(x')} Z 
                   \right|_{\eta={\bar \eta}=0} 
                  = i  \langle \psi(x) {\bar \psi}(x') \rangle \\   
F_{--}(x,x') &=&  i \left. \frac{\delta}{-i {\bar \eta}^-(x)}
                   \frac{\delta}{-i {\eta}^-(x')} Z 
                   \right|_{\eta={\bar \eta}=0} 
                   \nonumber \\
             &=& i \langle {\hat{\rm T}}  \psi(x) {\bar \psi}(x') \rangle =
i [- \theta(x^0-x'^0) \langle {\bar \psi(x')} \psi(x) \rangle +
    \theta(x'^0-x^0) \langle \psi(x') {\bar \psi(x)} \rangle ] .
\eea
Estos propagadores verifican las propiedades siguientes
\bea
F_{++}(x,x') &=& \theta(x^0-x'^0) F_{-+}(x,x') +
                 \theta(x'^0-x^0) F_{+-}(x,x') \label{prop1} \\
\gamma^0 F_{++}^{\dagger}(x,x') \gamma^0 &=&
- \theta(x^0-x'^0) F_{-+}(x',x) -
                 \theta(x'^0-x^0) F_{+-}(x',x) \label{prop2} .
\eea
La acci\'on efectiva in-in para el campo electromagn\'etico la obtenemos
integrando los fermiones con la prescripci\'on del camino temporal cerrado
\be
e^{i S_{\rm ef}[A_{\mu}^+,A_{\mu}^-]} = 
e^{i (S[A_{\mu}^+]-S[A_{\mu}^-])} \int_{\rm CTC} D{\bar \psi}^- D\psi^- 
D{\bar \psi}^+ D\psi^+ 
e^{-i S[\psi^-,{\bar \psi^-},A_{\mu}^-] +
i S[\psi^+,{\bar \psi^+},A_{\mu}^+]} .
\ee
Haciendo perturbaciones en la constante de estructura fina (o equivalentemente
en un desarrollo en potencias del campo electromagn\'etico), la acci\'on 
efectiva CTC adopta la siguiente forma
\be
\hspace{-0.5cm}
S_{\rm ef}[A_{\mu}^+,A_{\mu}^-] =
S[A_{\mu}^+] - S[A_{\mu}^-] + \frac{i}{2} e^2 \int d^4x d^4x' \times
\ee
\[ 
\hspace{-0.5cm}{\rm Tr} \left[ 
\gamma^{\mu} A_{\mu}^+(x) F_{++}(x,x') \gamma^{\nu} A_{\nu}^+(x') 
F_{++}(x',x) 
- \gamma^{\mu} A_{\mu}^+(x) F_{+-}(x,x') \gamma^{\nu} A_{\nu}^-(x') 
F_{-+}(x',x) \right.
\]
\[
\hspace{-0.5cm}
\left. - \gamma^{\mu} A_{\mu}^-(x) F_{-+}(x,x') \gamma^{\nu} A_{\nu}^+(x') 
F_{+-}(x',x) 
+  \gamma^{\mu} A_{\mu}^-(x) F_{--}(x,x') \gamma^{\nu} A_{\nu}^-(x') 
F_{--}(x',x) \right] + {\cal O}(A^4)
\]
Las ecuaciones de movimiento in-in quedan entonces
\bea
\hspace{-0.5cm}
0 &=& \left. \frac{\delta S_{\rm ef}}
{\delta A_{\mu}^+}\right|_{A_{\mu}^+=A_{\mu}^-=A_{\mu}}  
\label{ecinin} \\
&= & - \partial_{\nu} F^{\mu\nu} + i e^2 \int d^4x' {\rm Tr} [
\gamma^{\mu} F_{++}(x,x') \gamma^{\nu} F_{++}(x',x) -
\gamma^{\mu} F_{+-}(x,x') \gamma^{\nu} F_{-+}(x',x) ] A_{\nu}(x') . \nonumber 
\eea
Para mostrar que efectivamente estas ecuaciones son reales y causales,
usamos las propiedades de los propagadores fermi\'onicos 
(ecs.(\ref{prop1}) y (\ref{prop2})) 
y de las matrices de Dirac para demostrar la identidad
\bea
&& i {\rm Tr}[ \gamma^{\mu} F_{++}(x,x') \gamma^{\nu} F_{++}(x',x) - 
\gamma^{\mu} F_{+-}(x,x') \gamma^{\nu} F_{-+}(x',x) ]  \nonumber \\
&& = 2 \theta(x^0-x'^0) {\rm Re} {\rm Tr}
[ i \gamma^{\mu} F_{++}(x,x') \gamma^{\nu} F_{++}(x',x)] .
\eea
Introduciendo esta identidad en la ec.(\ref{ecinin}) y compar\'andola con la
ec.(\ref{ecinout}), vemos que la ecuaci\'on de movimiento in-in 
puede obtenerse a partir
de la in-out si a los t\'erminos no locales de esta \'ultima le tomamos dos
veces su parte real y causal. Esta propiedad que acabamos de demostrar es
tambi\'en v\'alida en el caso bos\'onico. 

Volvamos entonces a concentrarnos en la acci\'on efectiva in-out. 
Como es sabido, el tensor de polarizaci\'on de vac\'{\i}o en QED presenta
divergencias (ultravioletas) en $d=4$, que deben ser tratadas mediante alg\'un
procedimiento adecuado de renormalizaci\'on. Utilizaremos regularizaci\'on
dimensional, para lo cual trabajaremos en dimensi\'on arbitraria $d$.
El tensor de polarizaci\'on de vac\'{\i}o resulta entonces \cite{pt}
\bea
\Pi^{\mu\nu}(x,x') &=& - e^2 \mu^{4-d} 
{\rm Tr}[\gamma^{\mu} F_{++}(x,x') \gamma^{\nu} F_{++}(x',x)] \nonumber \\
&=&
 i e^2 \mu^{4-d} (4 \pi)^{-2} \Gamma(2-\frac{d}{2}) \, 2^{1+\frac{d}{2}} \,
(\partial^{\mu} \partial^{\nu} - \eta^{\mu\nu} \Box) \nonumber \\
&& \times
\int_0^1 dt (1-t^2) 
\left[
\frac{m^2 - \frac{1}{4} (1-t^2) \Box - i \epsilon}{4 \pi}
\right]^{\frac{d}{2}-2} \delta(x,x') ,
\eea
donde $\mu$ es un par\'ametro con unidades de masa. Expandiendo alrededor del
polo en $d=4$ ob\-te\-ne\-mos 
la siguiente forma para la acci\'on efectiva in-out
\bea
S_{\rm ef} &=& -\frac{1}{4} \int d^4x F_{\mu\nu} F^{\mu\nu} -
\frac{e^2}{24 \pi^2} \left[ \frac{2}{d-4} + \gamma + 
\ln \frac{m^2}{4 \pi \mu^2} \right] 
\int d^4x A_{\mu}(x) (\eta^{\mu\nu} \Box - \partial^{\mu} \partial^{\nu})
A_{\nu}(x) \nonumber \\
&& - \frac{e^2}{2 \pi^2} \int d^4x A_{\mu}(x) 
 (\eta^{\mu\nu} \Box - \partial^{\mu} \partial^{\nu}) F(\Box) A_{\nu}(x) ,
\eea
donde el factor de forma no local est\'a dado por
\be
F(\Box) = \frac{1}{8} \int_{0}^{1} (1-t^2)
  \ln \left[\frac{m^2 - \frac{1}{4} (1-t^2) \Box - i \epsilon}{\mu^2} \right] .
\label{fqed}
\ee
La divergencia adopta la misma forma que la acci\'on cl\'asica, de modo que
la absorbemos en el campo desnudo cl\'asico. As\'{\i} obtenemos finalmente
una expresi\'on renormalizada (finita) para la acci\'on efectiva in-out
\be
S_{{\rm ef}}= -\frac{1}{4} \int d^{\sst 4}x F_{\mu\nu} 
 \left[ 1 + \frac{e^2}{\pi^2} F(\Box) \right] F^{\mu\nu} +
 {\cal O}(A^{\sst 4}) .
\label{sqed}
\ee
Las correspondientes ecuaciones de Maxwell son 
\be
\left[1 + \frac{e^2}{\pi^2} F(\Box)\right] \partial_{\mu}F^{\mu\nu} = 
J^{\nu}_{\rm clas} ,
\label{euqed}
\ee
donde hemos incluido una fuente cl\'asica $J_{\rm clas}^{\nu}$. En el
l\'{\i}mite de distancias cortas ($m^2 \ll - \Box$), estas ecuaciones in-out
se reducen a
\be
\left[1 - \frac{e^2}{12 \pi^2} \ln(\frac{-\Box-i \epsilon}{\mu^2}) \right]
\partial_{\mu} F^{\mu\nu} = J^{\nu}_{\rm clas} .
\ee
Frecuentemente encontraremos la distribuci\'on $G(-\frac{\Box}{\mu^2}) \equiv
\ln(\frac{-\Box - i \epsilon}{\mu^2})$ a nivel de ecuaciones in-out.
Para hallar su contrapartida a nivel in-in, $G_{\rm in}$, le debemos tomar 
dos veces su parte real y causal. Sobre una funci\'on de prueba $f(x)$,
este factor de forma act\'ua de la manera siguiente 
\be
G_{\rm in}(- \frac{\Box}{\mu^2}) f(x) = 2 {\rm Re} \int d^4x'
\theta(x^0-x'^0) \int \frac{d^4k}{(2\pi)^4} e^{ik(x-x')}
\ln(\frac{k^2-i \epsilon}{\mu^2}) f(x') .
\ee
Podemos escribir el t\'ermino derecho en forma m\'as compacta si introducimos
el propagador de Feynman no masivo plano 
$G(x,x')= \int \frac{d^4k}{(2\pi)^4} \frac{e^{i k(x-x')}}{k^2-i \epsilon}$
\cite{jordan86,horowitz,jjg}.
Usando que
\bea
\frac{\partial}{\partial x^{\alpha}} G(x,x')&=& 
i \int \frac{d^4k}{(2\pi)^4} 
\frac{ k_{\alpha} e^{i k(x-x')}}{k^2-i \epsilon} =
\frac{i}{2} \int \frac{d^4k}{(2\pi)^4} e^{i k(x-x')}
\frac{\partial}{\partial k^{\alpha}} \ln(\frac{k^2-i \epsilon}{\mu^2}) 
\nonumber \\
&=&  \frac{1}{2} (x_{\alpha}-x'_{\alpha}) 
\int \frac{d^4k}{(2\pi)^4} e^{i k(x-x')} \ln(\frac{k^2-i \epsilon}{\mu^2}) ,
\eea
y el hecho que ${\rm Re} G(x,x') = \delta((x-x')^2)/4 \pi$, obtenemos
\be
{\rm Re} 
\int \frac{d^4k}{(2\pi)^4} e^{i k(x-x')} \ln(\frac{k^2-i \epsilon}{\mu^2}) 
= \frac{1}{\pi} \delta'((x-x')^2) .
\ee
De este modo, la distribuci\'on logar\'{\i}tmica in-in adopta la forma
\be
G_{\rm in}(- \frac{\Box}{\mu^2}) f(x)
= \frac{2}{\pi} \int d^4x' \theta(x^0-x'^0) \delta'((x-x')^2) f(x') .
\ee
Un caso especialmente sencillo es cuando la funci\'on de prueba es
independiente del tiempo
\be
G_{\rm in}(- \frac{\Box}{\mu^2}) f({\bf x}) = 
G(-\frac{\nabla^2}{\mu^2}) f({\bf x}) = 
\int d^3x' \int \frac{d^3k}{(2 \pi)^3} e^{i {\bf k} ({\bf x} - {\bf x}')}
\ln(\frac{{\bf k}^2}{\mu^2}) f({\bf x}') .
\label{factstatic}
\ee

Existe otra manera alternativa de hallar las ecuaciones de movimiento reales y
causales para estados de vac\'{\i}o in, 
que se basa en el c\'alculo de la acci\'on
efectiva 
eucl\'{\i}dea. La acci\'on cl\'asica de la electrodin\'amica en espacio
eucl\'{\i}deo es \cite{ramond}
\be
S^{\rm E}[A_{\mu}] = \int d^4x \left[ \frac{1}{4} F_{\mu\nu} F^{\mu\nu} +
\psi^{\dagger} (\gamma^{\mu} \nabla_{\mu} + i m) 
\psi \right] ,
\ee
donde $\gamma^{\mu}$ son las matrices de Dirac para signatura eucl\'{\i}dea
y $\nabla_{\mu} = \partial_{\mu} + i e A_{\mu}$.
Siguiendo los mismos m\'etodos utilizados para calcular la acci\'on efectiva
in-out renormalizada, 
concluimos que la acci\'on efectiva eucl\'{\i}dea resulta 
(comparar con la ec.(\ref{sqed}))
\be
S^{\rm E}_{{\rm ef}}= S^{\rm E}[A_{\mu}] - {\rm Tr} 
\ln(\gamma^{\mu} \nabla_{\mu} + i m)
= \frac{1}{4} \int d^{\sst 4}x F_{\mu\nu} 
 \left[ 1 + \frac{e^2}{\pi^2} F(\Box_{\rm E}) \right] F^{\mu\nu} +
 {\cal O}(A^{\sst 4}) ,
\label{EAeuc}
\ee
donde el factor de forma $F(\Box_{\rm E})$ es el mismo de la ec.(\ref{fqed})
sin la prescripci\'on $-i \epsilon$ y con el D'Alambertiano eucl\'{\i}deo
\SSfootnote{Esta expresi\'on para 
la acci\'on efectiva eucl\'{\i}dea tambi\'en puede deducirse
en base a las t\'ecnicas del cap\'{\i}tulo anterior. Para ello usamos la
identidad 
${\rm Tr} \ln(\gamma^{\mu} \nabla_{\mu} + i m) = \frac{1}{2} {\rm Tr} \ln(K)$
donde 
$K \equiv (\gamma^{\mu} \nabla_{\mu} + i m) (\gamma^{\mu} \nabla_{\mu} - i m)=
-\Box + m^2 + \frac{e}{2} \sigma_{\mu\nu} F^{\mu\nu}$, siendo
 $\sigma_{\mu\nu} = \frac{i}{2} [\gamma_{\mu},\gamma_{\nu}]$. Vemos entonces
que el determinante fermi\'onico puede escribirse en t\'erminos del 
determinante de un operador diferencial de segundo orden $K$, del tipo de la
ec. (\ref{Fminimo}). Aplicando la t\'ecnica de resumaci\'on del cap\'{\i}tulo
4 obtenemos la expresi\'on de la acci\'on efectiva 
eucl\'{\i}dea (\ref{EAeuc}).}.
Las ecuaciones de movimiento eucl\'{\i}deas para el campo electromagn\'etico
son las mismas ecs.(\ref{euqed}), tambi\'en 
con el D'Alambertiano eucl\'{\i}deo. 

Para hallar las ecuaciones de movimiento in-in, basta reemplazar en las 
ecuaciones de mo\-vi\-mien\-to 
eucl\'{\i}deas el propagador $\Box_{\rm E}^{-1}$ 
por el 
propagador retardado $\Box_{\rm ret}^{-1}$ \cite{gospel}, obteni\'endose 
de esta forma la versi\'on in-in del factor de forma. Sobre
funciones de prueba independientes del tiempo, dicho factor de forma
act\'ua como
\be
F_{\rm in}(\Box) f({\bf x}) = F(\nabla^2) f({\bf x}) =
\int d^3x' \int \frac{d^3k}{(2 \pi)^3} e^{i {\bf k} ({\bf x} - {\bf x}')}
F(- {\bf k}^2) f({\bf x}') ,
\ee
pues la integral temporal del propagador retardado coincide con la funci\'on
de Green del laplaciano. En particular, en el l\'{\i}mite de distancias
cortas, el factor de forma es logar\'{\i}tmico y reobtenemos la 
ec.(\ref{factstatic}).

Estamos ahora listos para calcular las modificaciones al potencial 
electrost\'atico. Tomando como fuente cl\'asica una carga puntual est\'atica,
la ley de Gauss adopta la forma
\be
{\bf{\nabla} \cdot \bf{E}} -\frac{e^2}{12 \pi^2} 
G(-\frac{\nabla^2}{\mu^2}) {\bf{\nabla} \cdot \bf{E}}  = e \delta^{3}(\bf{x}) .
\ee
La soluci\'on para el campo el\'ectrico es esf\'ericamente sim\'etrica
${\bf{E}} = E(r) {\bf{\hat{r}}}$ y la hallaremos perturbativamente en potencias
de $e^2$, $\bf{E} = \bf{E^{\sst (0)}} + \bf{E^{\sst (1)}}$.
El t\'ermino de orden cero es la contribuci\'on cl\'asica 
\begin{eqnarray}
{\bf{\nabla}} \cdot {\bf{E^{\sst (0)}}} = e \delta^{3}({\bf{x}}) &
 \Longrightarrow & E^{\sst (0)} (r) = \frac{e}{4 \pi r^2} ,
\label{e0}
\end{eqnarray}
y el t\'ermino de primer orden es una correcci\'on cu\'antica dada por
\be
{\bf{\nabla}} \cdot {\bf{E^{\sst (1)}}} =
        \frac{e^2}{12 \pi^2} G(-\frac{\nabla^2}{\mu^2})
{\bf{\nabla}} \cdot {\bf{E^{\sst (0)}}}	. 
\ee
En consecuencia, debemos evaluar la acci\'on de $G(-\frac{\nabla^2} 
{\mu^2})$ sobre la delta de Dirac. Usando la ec.(\ref{factstatic}) obtenemos
\be
\ln (-\frac{\nabla^2}{\mu^2}) \delta^{3}({\bf{x}}) =
- \frac{1}{2 \pi r^3} -\ln(\mu^2) \delta^{3}({\bf{x}}) ,
\label{logdelta}
\ee
donde el \'ultimo t\'ermino da una correcci\'on que depende de $\mu$ 
a la soluci\'on cl\'asica, y que ab\-sor\-be\-re\-mos en la fuente cl\'asica
\SSfootnote{Esta expresi\'on tambi\'en puede obtenerse por medio de la
transformada de Fourier $\int \frac{d^3 p}{(2 \pi)^3} e^{-i {\bf p} \cdot
{\bf r}} \ln p^2 = - \frac{1}{2 \pi r^3}$.}. 
La correcci\'on cu\'antica es
\be
E^{\sst (1)} (r) = E^{\sst (1)} (r_{\sst 0}) \frac{r_{\sst 0}^2}{r^2} -
    \frac{e^3}{24 \pi^3 r^2} \ln \frac{r}{r_{\sst 0}} ,
\label{e1}
\ee
donde $r_{\sst 0}$ es un radio de referencia arbitrario. 
Integrando las ecuaciones (\ref{e0}) y (\ref{e1})
y multiplicando por la carga $e$, obtenemos el potencial de
interacci\'on electrost\'atico
\be
V_{\rm int}(r) = \frac{e^2}{4 \pi r}
\left[
1-\frac{e^2}{6 \pi^2} \ln \frac{r}{r_{\sst 0}} + {\cal O}(e^4) 
\right] .
\ee
A partir de esta ecuaci\'on es inmediato obtener la forma en que la
carga el\'ectrica var\'{\i}a con la distancia, y el resultado coincide con el
hallado en la ec.(\ref{eq:itzi}). 
 
%%%%%%%%%%%%%%%%%%%%%%%%%%%%%%%%%%%%%%%%%%%%%%%%%%%%%%%%%%%%%%%%%%%%%

\section{Campos escalares en fondos curvos}

\subsection{Ecuaciones efectivas no locales para el campo gravitatorio}

Consideremos ahora un campo escalar cu\'antico en un fondo gravitatorio
cl\'asico y en presencia de una masa cl\'asica $M$. La acci\'on
eucl\'{\i}dea de la 
teor\'{\i}a es $S=S_M + S_{\rm grav} + S_{\rm materia}$, donde
\bea
S_M &=& - M \int \sqrt{g_{\mu\nu} dx^{\mu} dx^{\nu}} \\
S_{\rm grav} &=& -\int d^4x\sqrt g\left [{1\over 16\pi G}(R-2\Lambda)
+\alpha R^2 +\beta R_{\mu\nu}R^{\mu\nu}\right ] \\
S_{\rm materia} &=& {1\over 2}\int d^4x\sqrt g 
[\partial_{\mu}\phi\partial^{\mu}\phi
+m^2\phi^2+\xi R\phi^2] .
\eea
Aqu\'{\i} $g_{\mu\nu}$ es la m\'etrica del fondo gravitatorio, $x^{\mu}$ es
la trayectoria de la part\'{\i}cula $M$, $m$ es la masa del campo escalar y
 $\xi$ es una constante adimensional asociada al acoplamiento 
del campo escalar 
con la curvatura. En la acci\'on gravitatoria, $G$ es la constante de Newton,
 $\Lambda$ es la constante cosmol\'ogica y $\alpha$ y $\beta$ son constantes
adimensionales que acompa\~nan a los t\'erminos cuadr\'aticos en curvaturas.
Todos estos par\'ametros en la acci\'on gravitatoria son constantes
desnudas que se renormalizar\'an al integrar las fluctuaciones del campo
escalar. Haciendo la integral gaussiana sobre el campo escalar obtenemos
la acci\'on efectiva para el campo gravitatorio. Formalmente el resultado es
\be
S_{\rm ef} = S_M + S_{\rm grav} + \frac{1}{2} \ln {\rm det}
\left[ \frac{-\Box+m^2+\xi R}{\mu^2} \right] \equiv S_M + S_{\rm grav} + 
\Gamma .
\ee
La acci\'on efectiva tiene la forma estudiada en el cap\'{\i}tulo 4, siendo 
en este caso la matriz ${\hat Q}= (\frac{1}{6} - \xi) R {\hat 1}$. 
Como hemos visto, la tarea de evaluar
este determinante funcional en un fondo gravitatorio arbitrario es 
extremadamente dif\'{\i}cil, y se requieren t\'ecnicas de aproximaci\'on.
Para campos gravitatorios lentamente variables, las t\'ecnicas de 
Schwinger-DeWitt (l\'{\i}mite de tiempos cortos del heat kernel) da una
expansi\'on de $\Gamma$ en potencias inversas de la masa. En la notaci\'on
del cap\'{\i}tulo 4, 
\be
\hspace{-0.6cm}
\Gamma =  - \frac{1}{32 \pi^2} \lim_{\L \rightarrow \infty}
\int d^4x \sqrt{g} {\rm Tr} \left[
\rho_0 a_0(x) + \rho_1 a_1(x) + \rho_2 a_2(x) 
+ \sum_{n=3}^{\infty} (n-3)! 
\frac{1}{(m^2)^{n-2}} a_n(x) \right] ,
\ee
donde
\SSfootnote{Usamos la identidad de Gauss-Bonnet 
 $0=\int d^4x \sqrt{g} [R_{\mu\nu\rho\sigma} R^{\mu\nu\rho\sigma}
 - 4 R_{\mu\nu} R^{\mu\nu} + R^2]$, que es un invariante topol\'ogico en $d=4$
dimensiones, para escribir el coeficiente $a_2(x)$ en t\'erminos s\'olo de
 $R$ y de $R_{\mu\nu}$.}
\bea
a_0(x)&=& 1 \\
a_1(x)&=& (\frac{1}{6}-\xi) R \\
a_2(x)&=& \frac{1}{60} R_{\mu\nu} R^{\mu\nu} +
\frac{1}{2} \left[ (\frac{1}{6} -\xi)^2 - \frac{1}{90} \right] R^2
- \frac{1}{6} (\frac{1}{5} -\xi) ,
\eea
y $\rho_0$, $\rho_1$ y $\rho_2$ son cantidades divergentes 
para $L \rightarrow \infty$
dadas en la ec.(\ref{rhosdiv}). 
Las divergencias son absorbidas en una 
redefinici\'on de las constantes desnudas de la acci\'on cl\'asica
gravitatoria.
Las divergencias en $\rho_0$ se renormalizan en la constante cosmol\'ogica,
las de $\rho_1$ en la constante gravitatoria, y las de $\rho_2$ 
en las constantes $\alpha$ y $\beta$. En esta forma obtenemos la versi\'on 
renormalizada de la acci\'on efectiva
\bea
\Gamma &=& {1\over 32\pi^2}\int d^4x\sqrt g 
\left[ {1\over 2} m^4 \ln ({m^2\over \mu^2}) - m^2 a_1(x) \ln({m^2\over \mu^2})
 + a_2(x) \ln({m^2\over \mu^2}) \right. \nonumber \\
&& ~~~~~~~~~~~~~~~~~~~~
\left.- {1\over 2} \sum_{n=3}^{\infty} \frac{(n-3)!}{(m^2)^{n-2}} a_n(x)
\right] ,
\label{sdwe}
\eea
donde ahora todos los par\'ametros son vestidos.
A partir de esta expansi\'on, es f\'acil derivar el escaleo de las constantes
gravitatorias. Dado que la acci\'on efectiva no depende de la escala 
arbitaria $\mu$, es decir $\mu d S_{\rm ef} / d \mu =0 $, se obtiene el 
siguiente conjunto de ecuaciones del grupo de renormalizaci\'on
\bea
\mu{dG\over d\mu}&=&
 \frac{G^2 m^2}{\pi} \left(\xi -  \frac{1}{6}  \right)\label{rge1} \\
\mu{d\alpha\over d\mu}&=&
 -\frac{1}{32 \pi^2}
 \left[
 \left( \frac{1}{6} - \xi \right)^2 -\frac{1}{90}
 \right] \label{rge2}\\
\mu{d\beta\over d\mu}&=&
 -\frac{1}{960 \pi^2} \label{rge3}\\
\mu{d\over d\mu} \frac{\Lambda}{G}&=&
 \frac{m^4}{4 \pi} .
\label{rge4}
\eea
Partiendo de la ec.(\ref{rge1}) 
podemos obtener el potencial gravitatorio `wilsoniano'
generado por la part\'{\i}cula $M$, asumiendo que \'esta es est\'atica y que
est\'a localizada en el origen. Dado que el escaleo de $G(\mu)$ viene dado por
 $G(\mu) =G_0\left (1+{m^2G_0\over\pi}(\xi-{1\over 6})
 \ln{\mu\over\mu_0}\right )$, el potencial wilsoniano resulta
\be
V(r)= - \frac{M G(\mu=r^{-1})}{r} =  
-{G_0 M\over r}\left[ 1-{m^2G_0\over\pi}(\xi-{1\over 6}) 
\ln{r\over r_0} \right] . 
\label{VRG}                        
\ee
Es decir, los efectos cu\'anticos debidos al campo de materia hacen que se 
modifique la ley de Newton cl\'asica $r^{-1}$, y entonces la constante
gravitatoria pasa a ser funci\'on de la distancia. Siguiendo las mismas
t\'ecnicas que en la secci\'on anterior, a continuaci\'on analizaremos si
es posible derivar este potencial bas\'andonos en los t\'erminos no locales
de la acci\'on efectiva.

Como hemos visto en el cap\'{\i}tulo 4, la acci\'on efectiva posee 
contribuciones no locales, que han sido calculadas mediante la resumaci\'on 
de la expansi\'on de Schwinger-DeWitt, v\'alida en el caso de campos d\'ebiles
y r\'apidamente variables. En el presente caso la acci\'on efectiva, a orden
cuadr\'atico en curvaturas, toma la forma
\bea
\Gamma &= & \frac{1}{32 \pi^2} \int d^4x \sqrt{g} 
\lim_{L \rightarrow \infty} \left[
\frac{1}{2} m^4 \ln(\frac{m^2}{L^2}) + m^2 \ln(\frac{m^2}{L^2})
(\xi - \frac{1}{6}) R \right. \nonumber \\
&& ~~~~~~~~~~~~~~~~~~~~~~~~~~~
\left. + R H_1(\Box) R + R_{\mu\nu} H_2(\Box) R^{\mu\nu} 
\right] .
\eea
Los factores de forma $H_i(\Box)$ se expresan en funci\'on de los factores
de forma $F_i(\Box)$ del cap\'{\i}tulo 4 como sigue
\bea
H_1(\Box) &=& F_2(\Box) + (\frac{1}{6} - \xi) F_3(\Box) + (\frac{1}{6}-\xi)^2
              F_4(\Box) \\
H_2(\Box) &=& F_1(\Box) .
\eea
A partir de esta expresi\'on para la acci\'on efectiva, podemos derivar las
ecuaciones de Einstein semicl\'asicas 
 $\delta S_{\rm ef}/\delta g_{\mu\nu} =0$. Dado que estamos despreciando 
t\'erminos de orden ${\cal O}({\cal R}^3)$ en la acci\'on efectiva, 
debemos descartar toda 
contribuci\'on a las ecuaciones de movimiento que sea cuadr\'atica en 
curvaturas. En consecuencia, al tomar la variaci\'on de la acci\'on respecto a
 $g_{\mu\nu}$, no es necesario tener en cuenta la dependencia de los factores
de forma en la m\'etrica. Adem\'as, a este orden, es posible conmutar derivadas
covariantes actuando sobre una curvatura, es decir
 $\nabla_{\mu}\nabla_{\nu}{\cal R} = \nabla_{\nu}\nabla_{\mu}{\cal R}
 +{\cal O}({\cal R}^2)$. 
Obtenemos as\'{\i} las ecuaciones semicl\'asicas (eucl\'{\i}deas)
para el campo gravitatorio
\bea
\left[
\frac{1}{8 \pi G} - \frac{m^2}{16 \pi^2} (\xi-\frac{1}{6}) 
\ln \frac{m^2}{L^2}
\right]
(R_{\mu\nu}-\frac{1}{2} R g_{\mu\nu}) - \alpha H_{\mu\nu}^{(1)} -
\beta H_{\mu\nu}^{(2)} = T_{\mu\nu} + \langle T_{\mu\nu} \rangle ,
\label{euclidSEE}
\eea
donde $T_{\mu\nu}(y) = M \int d\tau {\dot x}^{\mu} {\dot x}^{\nu} 
\delta^4(y-x(\tau))$ es el tensor de energ\'{\i}a-momento asociado a la
part\'{\i}cula $M$. Denotamos con $\cdot$ a la derivada respecto al tiempo
propio $\tau$, definido como $d \tau^2 = g_{\mu\nu} dx^{\mu} dx^{\nu}$.
La contribuci\'on cu\'antica al tensor de energ\'{\i}a-impulso es
\be
\langle T_{\mu\nu} \rangle = \frac{1}{32 \pi^2} [ H_1(\Box) H_{\mu\nu}^{(1)} +
H_2(\Box) H_{\mu\nu}^{(2)} ] .
\ee
Los tensores $H_{\mu\nu}^{(i)}$ valen
\bea
H_{\mu\nu}^{(1)}&=& 4\nabla_{\mu}\nabla_{\nu}R - 4g_{\mu\nu}\Box R
+{\cal O}(R^2)\nonumber\\
H_{\mu\nu}^{(2)}&=& 2\nabla_{\mu}\nabla_{\nu}R - g_{\mu\nu}\Box R
                    -2\Box R_{\mu\nu} +{\cal O}(R^2) ,
\eea
y provienen de derivar los t\'erminos cuadr\'aticos en curvaturas a nivel de
la acci\'on. En las ecuaciones de movimiento hemos asumido que la constante 
cosmol\'ogica vestida es nula, de modo que no aparecen contribuciones 
proporcionales a la m\'etrica $g_{\mu\nu}$ independientes de curvaturas.

Las ecuaciones de movimiento son v\'alidas para campos gravitatorios d\'ebiles
y r\'apidamente variables y, a priori, para cualquier valor de la masa del
campo cu\'antico. En el l\'{\i}mite en que la escala de variaci\'on del campo
gravitatorio es mucho mayor que la inversa de la masa (es decir
 $m^2 {\cal R} \gg \nabla\nabla {\cal R}$), los factores de forma $H_i(\Box)$
se pueden desarrollar en potencias de $-\Box/m^2$ y se reobtiene la ecuaci\'on
de movimiento (local) que resulta del desarrollo de Schwinger-DeWitt. En el
l\'{\i}mite opuesto ($m^2 {\cal R} \ll \nabla\nabla {\cal R}$) se puede hacer
una expansi\'on en $-m^2/\Box$. Al igual que en el caso de la 
electrodin\'amica, estaremos interesados en este \'ultimo caso.

Para estudiar el comportamiento de los factores de forma para el caso
 $z \equiv -m^2/ \Box \ll 1$, partimos de su definici\'on
 $H_i(\Box) = \int_{1/L^2}^{\infty} ds \frac{e^{-s m^2}}{s} h_i(-s \Box)$,
donde
\bea
\hspace{-0.5cm} 
h_{1}(\eta) & = & \frac{f(\eta)}{8} 
\left[
\frac{1}{36} + \frac{1}{3 \eta} -
\frac{1}{\eta^2}
\right] - \frac{1}{16 \eta} + \frac{1}{8 \eta^2} +
(\xi - \frac{1}{6}) \left[
\frac{f(\eta)}{12} + \frac{ f(\eta)-1}{2 \eta} \right] +
\frac{1}{2} (\xi -\frac{1}{6})^2 f(\eta) \nonumber \\
\hspace{-0.5cm}
h_{2}(\eta) & = & \frac{ f(\eta) - 1 + \eta/6 }{\eta^2} . 
\eea

Denotemos en forma gen\'erica con $\sigma$ a cualquiera de las funciones
 $h_i$ y procedamos a estudiar el comportamiento de la integral
 $I \equiv \int_{1/L^2}^{\infty} ds \frac{e^{-s m^2}}{s} \sigma(-s \Box)$ 
en t\'erminos de $z$. Para ello partimos la integral en la forma $I=A+B$, donde
\bea
A(z) &=& \int_{-\Box/L^2}^{C} \frac{d\eta}{\eta} e^{-\eta z} \sigma(\eta) \\
B(z) &=& \int_C^{\infty}  \frac{d\eta}{\eta} e^{-\eta z} \sigma(\eta) ,
\eea
con $C$ una constante elegida de forma tal que $z^{-1} \gg C \gg 1$. En la
primer integral podemos usar la expansi\'on en Taylor del factor de forma,
 $\sigma(\eta) = \sum_{n=2}^{\infty} \sigma_n \eta^{n-2}$, donde las constantes
 $\sigma_n$ se leen del correspondiente coeficiente de Schwinger-DeWitt
(ver ecs.(\ref{417}) y (\ref{ffgeneral})). 
Los t\'erminos con $n \ge 3$ son finitos en el 
l\'{\i}mite $L^2 \rightarrow \infty$ y dan una contribuci\'on anal\'{\i}tica
en la variable $z$, mientras que el t\'ermino en $n=2$ (que corresponde al
segundo coeficiente de Schwinger-DeWitt) da una contribuci\'on divergente.
Expandiendo la exponencial en la integral $A(z)$ en potencias de $\eta z \ll 1$
obtenemos finalmente
\be
A = - \sigma_2 \ln(-\frac{\Box}{L^2}) + {\cal O}((-\frac{m^2}{\Box})^2) ,
\ee
donde el \'ultimo t\'ermino denota contribuciones finitas y anal\'{\i}ticas
en $-m^2/\Box$. Como ya vimos, la divergencia logar\'{\i}tmica se renormaliza con el
contrat\'ermino cl\'asico ${\cal R}^2$. En cuanto a la integral B, su 
comportamiento en potencias de  $-m^2/\Box$ est\'a gobernado por el 
l\'{\i}mite asint\'otico de $\beta(\eta)$ para tiempos grandes. Usando
que $\beta(\eta)=\frac{k}{\eta}$ para $\eta \rightarrow \infty$ 
(ver ec.(\ref{largetimes})), la integral resulta
\be
B = - k \frac{m^2}{\Box} \ln(- \frac{m^2}{\Box}) + 
{\cal O}((-\frac{m^2}{\Box})^2) .
\ee
En conclusi\'on
\be
\lim_{\L \rightarrow \infty} \int_{1/L^2}^{\infty} ds \frac{e^{-s m^2}}{s}
\sigma(-s \Box) = - \sigma_2 \ln(- \frac{\Box}{L^2}) - k \frac{m^2}{\Box}
\ln(- \frac{m^2}{\Box}) + {\cal O}(- \frac{m^2}{\Box})^2 .
\ee
Los coeficientes $\sigma^{(i)}$ los obtenemos del l\'{\i}mite de tiempos
cortos de las funciones $f_i(\eta)$ (ec.(\ref{smalltimes})), y los coeficientes
$f_i(\eta)$ del l\'{\i}mite de tiempos largos (ec.(\ref{largetimes})).
Resulta
\clearpage
\begin{eqnarray}
\sigma_{2}^{(1)} = h_1(0) = \frac{1}{2} 
\left[ \left( \frac{1}{6} - \xi \right)^2 -\frac{1}{90} \right] 
& ~~~~~ & \sigma_{2}^{(2)} = h_2(0)= \frac{1}{60} \nonumber \\
k^{(1)} = \lim_{\eta \rightarrow \infty} \eta h_1(\eta) = 
\xi^2 - \frac{1}{12} & ~~~~~ &
k^{(2)} =\lim_{\eta \rightarrow \infty} \eta h_2(\eta)    =\frac{1}{6} .
\label{eq:coef4d}
\end{eqnarray}
Finalmente los factores de forma renormalizados quedan
\bea
H_1(\Box) &=& - \sigma_2^{(1)} \ln(-\frac{\Box}{\mu^2}) - 
                k^{(1)} \frac{m^2}{\Box} \ln(-\frac{m^2}{\Box}) 
                + {\cal O}(-\frac{m^2}{\Box})^2 \\
H_2(\Box) &=& - \sigma_2^{(2)} \ln(-\frac{\Box}{\mu^2}) - 
                k^{(2)} \frac{m^2}{\Box} \ln(-\frac{m^2}{\Box}) 
                + {\cal O}(-\frac{m^2}{\Box})^2 .
\eea
Los t\'erminos independientes de la masa del campo escalar son proporcionales
al factor de forma no local $\ln(-\Box/\mu^2)$, siendo los factores de 
proporcionalidad las correspondientes constantes del segundo coeficiente de
Schwinger-DeWitt. Cabe mencionar que estos t\'erminos no locales y no masivos
pueden deducirse f\'acilmente partiendo de las ecuaciones del grupo de 
renormalizaci\'on. En efecto, imponiendo el hecho que la acci\'on efectiva
no depende del par\'ametro arbitrario $\mu$, y usando el hecho que
los factores de forma son cantidades adimensionales,
 $H_i=H_i(-\frac{\Box}{\mu^2};\xi)$, el escaleo de las 
constantes $\alpha$ y $\beta$ determina su expresi\'on en forma un\'{\i}voca. 
Notemos que estas contribuciones no masivas de los factores de forma tienen 
una interpretaci\'on clara: corresponden a la acci\'on gravitatoria cl\'asica
en la cual las constantes $\alpha$ y $\beta$ han sido reemplazadas por 
funciones de dos puntos no locales.
En el caso masivo, la situaci\'on es m\'as complicada pues los factores
de forma pueden tambi\'en depender de otra cantidad adimensional $m^2/\mu^2$,
y el grupo de renormalizaci\'on no es suficiente para determinar su 
expresi\'on. En este caso es necesario recurrir al an\'alisis asint\'otico
descripto anteriormente.

El tensor de energ\'{\i}a-momento queda finalmente
\bea
\langle T_{\mu\nu} \rangle &=& -\frac{1}{32 \pi^2} 
\left\{
\log(-\frac{\Box}{\mu^2}) \left[ \sigma_{2}^{(1)} H_{\mu\nu}^{(1)} +
			    \sigma_{2}^{(2)} H_{\mu\nu}^{(2)} \right] 
\right. \nonumber \\
&& ~~~~~~~~~~~~ \left.+
\frac{m^2 k^{(1)}}{\Box} \log(-\frac{m^2}{\Box}) \left[ H_{\mu\nu}^{(1)} +
 \frac{k^{(2)}}{k^{(1)}} H_{\mu\nu}^{(2)} \right]
\right\} .
\eea
El primer t\'ermino (no masivo) puede ser interpretado como una correcci\'on
a las constantes gra\-vi\-ta\-cionales 
$\alpha$ y $\beta$. Como ya hemos dicho, los
factores num\'ericos de esas correcciones provienen del l\'{\i}mite de
tiempos cortos del heat kernel, y por lo tanto coinciden con los que resultan
del segundo coeficiente de Schwinger-DeWitt. El segundo t\'ermino (masivo) 
podr\'{\i}a ser interpretado como una correcci\'on a $G$ si pudiese ser 
expresado en una combinaci\'on proporcional a $m^2 \ln(-\frac{m^2}{\Box})
(R_{\mu\nu} - \frac{1}{2} R g_{\mu\nu})$. Al orden que estamos trabajando
 ${\cal O}({\cal R}^2)$, vale la relaci\'on
\be
H_{\mu\nu}^{(1)} - 2 H_{\mu\nu}^{(2)} = 4 \Box (R_{\mu\nu} - 
\frac{1}{2} R g_{\mu\nu}) ,
\ee
de modo que esta interpretaci\'on ser\'a posible s\'olo si 
 $k^{(2)}/k^{(1)} =-2$. Esta condici\'on se satisface \'unicamente para
acoplamiento
m\'{\i}nimo, $\xi =0$. Cuando resolvamos las ecuaciones de movimiento
veremos c\'omo este hecho se traduce en la dependencia de la constante
gravitatoria con la distancia.

A partir de las ecuaciones de movimiento tambi\'en es posible derivar
las ecuaciones del grupo de renormalizaci\'on, imponiendo que las
primeras no dependan de la
escala $\mu$. Una forma alternativa es hacer la transformaci\'on
 $g_{\mu\nu} \rightarrow s^{-2} g_{\mu\nu}$ y analizar c\'omo se comportan
las constantes gravitatorias en el l\'{\i}mite $s \rightarrow \infty$ 
\cite{pt,nelpan}. Dado que ante dicha transformaci\'on, 
 $\Box \rightarrow s^2 \Box$, los t\'erminos no locales proporcionales a 
 $\ln(- \Box)$ se hacen 
relevantes para $s$ grande. De los t\'erminos no masivos
de la ecuaci\'on de movimiento obtenemos el mismo escaleo que mediante el 
grupo de renormalizaci\'on
\bea
\alpha (s) &=& \alpha (s=1) -\frac{1}{32 \pi^2} \left(
      (\xi-\frac{1}{6})^2 - \frac{1}{90} \right) \ln s 
\label{as}\\
\beta (s) &=& \beta(s=1) -\frac{1}{960 \pi^2} \ln s .
\label{bs}
\eea
Los t\'erminos masivos, en cambio, conducen a un escaleo para $G$ s\'olo
para acoplamiento 
m\'{\i}nimo $\xi=0$, por las mismas razones indicadas m\'as arriba.
En este caso,
\be
G(s) = G(s=1)\left( 1 - \frac{G(s=1) m^2}{6 \pi} \ln{s} 
\right )\;\; ~~~ (\xi=0) ,
\ee
que coincide con lo prescripto por el grupo de renormalizaci\'on para $G(\mu)$
para este valor particular del acoplamiento. Este hecho ya hab\'{\i}a sido 
remarcado en \cite{pt}.

\vspace{-0.5cm}

\subsection{El potencial newtoniano a partir de la acci\'on efectiva}

Al igual que en electrodin\'amica cu\'antica, los efectos de polarizaci\'on de
vac\'{\i}o 
contenidos en $\langle T_{\mu\nu} \rangle$ inducen modificaciones en el
potencial newtoniano. Seguidamente calcularemos estas correcciones partiendo
de las ecuaciones de Einstein semicl\'asicas. Para ello es necesario 
transformar las ecuaciones de movimiento eucl\'{\i}deas (\ref{euclidSEE})
en ecuaciones
in-in. Ello se logra reemplazando en los factores de forma no locales los
propagadores eucl\'{\i}deos masivos por los correspondientes propagadores 
retardados, o bien haciendo una rotaci\'on de Wick para escribir las 
ecuaciones minkowskianas y tomando luego dos veces la parte real y causal
de las mismas. Sin embargo, al calcular el potencial newtoniano consideramos
campos independientes del tiempo, de modo que las ecuaciones in-in son 
simplemente las ecuaciones eucl\'{\i}deas donde $\Box$ es reemplazado por
 $\nabla^2$. La signatura de la m\'etrica en estas ecuaciones in-in ser\'a
 $(-+++)$.

Las ecuaciones de movimiento son covariantes, es decir invariantes ante
cambios generales de coordenadas.
Para resolverlas haremos perturbaciones alrededor del espacio-tiempo
plano, $g_{\mu\nu} = \eta_{\mu\nu} + h_{\mu\nu}$, con
$\eta_{\mu\nu} = {\rm diag}(-+++)$ y $|h_{\mu\nu}| \ll 1$, y fijaremos
la medida mediante la denominada medida arm\'onica, 
$(h_{\mu\nu} - \frac{1}{2} h \eta_{\mu\nu})^{; \nu}=0$. Para esta elecci\'on,
\bea
R_{\mu\nu} &=& - \frac{1}{2} \Box h_{\mu\nu} \\
R &=& - \frac{1}{2} \Box h \\
H_{\mu\nu}^{(1)} &=& (-2 \partial_{\mu} \partial_{\nu} + 2 \eta_{\mu\nu}
\Box) \Box h \\
H_{\mu\nu}^{(2)} &=& (- \partial_{\mu} \partial_{\nu} + \frac{1}{2} 
\eta_{\mu\nu} \Box) \Box h + \Box \Box h_{\mu\nu} , 
\eea
donde $h=\eta^{\mu\nu} h_{\mu\nu}$. Los \'{\i}ndices suben y bajan con la
m\'etrica plana $\eta_{\mu\nu}$. Las ecuaciones de movimiento linealizadas
quedan entonces
\be
\left[
- \frac{1}{16 \pi G} + \frac{m^2}{32 \pi^2} (\xi-\frac{1}{6}) 
\ln \frac{m^2}{\mu^2} \right] \Box {\bar{h}_{\mu\nu}} 
- \alpha H_{\mu\nu}^{(1)} -
\beta H_{\mu\nu}^{(2)} = T_{\mu\nu} + \langle T_{\mu\nu} \rangle ,
\label{htecho}
\ee
con ${\bar h}_{\mu\nu}=h_{\mu\nu} - \frac{1}{2} h \eta_{\mu\nu}$.

Asumiremos que la part\'{\i}cula $M$ es
est\'atica y que est\'a localizada en el origen, 
${\dot x}^{\mu} =(1,0,0,0)$, 
 $T^{\mu\nu}({\bf x}) = \delta^{\mu}_0 \delta^{\nu}_0 M \delta^3({\bf x})$,
 $T^{\mu}_{\mu}({\bf x}) = -M \delta^3({\bf x})$.
Dado que calcularemos correcciones al potencial newtoniano a largas 
distancias, podemos asumir que la fuente $M$ es una ``masa puntual'',
a pesar que su tama\~no debe ser mucho mayor que su radio de Schwarzschild
y que la longitud de Planck, de modo de justificar la aproximaci\'on de 
campo d\'ebil que haremos m\'as adelante.
Para obtener el potencial newtoniano consideraremos las ecuaciones 
geod\'esicas de una part\'{\i}cula de prueba
de masa $m_{\sst{\rm p}}$ con coordenadas $z^{\mu}(\tau)$, 
[$d\tau^2=- g_{\mu\nu} dz^{\mu} dz^{\nu}$] 
que se mueve en el fondo gravitatorio
$g_{\mu\nu}$. Estas ecuaciones son
\be
0 = \frac{d^2 z^{\rho}}{d \tau^2} + \Gamma^{\rho}_{\mu\sigma}
\frac{d z^{\mu}}{d \tau} \frac{d z^{\sigma}}{d \tau} ,
\ee
donde $\Gamma^{\rho}_{\mu\sigma}$ es el s\'{\i}mbolo de 
Christoffel asociado a la
m\'etrica soluci\'on de las ecuaciones semicl\'asicas. En el l\'{\i}mite de 
campo d\'ebil y velocidades no relativistas, las ecuaciones geod\'esicas se
reducen a
\be
\frac{d^2{\bf z}}{dt^2} = -{\bf \nabla} V = \frac{1}{2} {\bf \nabla}h_{00} ,
\ee
de modo que podemos definir el potencial newtoniano como $V(r=|{\bf z}|) =
-\frac{1}{2} h_{00}$. Debemos entonces hallar la componente $00$ de la 
perturbaci\'on a la m\'etrica plana, $h_{00} = {\bar h}_{00} - \frac{1}{2} h$,
y para ello resolveremos la ec.(\ref{htecho}) para ${\bar h}_{00}$ 
en el caso est\'atico, y la ecuaci\'on para la traza $h$, 
\be
\left[
 \frac{1}{16 \pi G} -  \frac{m^2}{32 \pi^2} (\xi-\frac{1}{6}) 
\ln \frac{m^2}{\mu^2} \right] \nabla^2 h - 2(3\alpha+\beta) \nabla^2 \nabla^2
h = T_{\mu}^{\mu} + \langle T_{\mu}^{\mu} \rangle ,
\ee
donde
\be
\langle T_{\mu}^{\mu} \rangle = -\frac{1}{32 \pi^2}
\left[
3 (\xi -\frac{1}{6})^2 \ln(-\frac{\nabla^2}{\mu^2}) \nabla^2 \nabla^2 
- 6 m^2 (\xi^2 - \frac{1}{36}) \ln(-\frac{\nabla^2}{m^2}) \nabla^2
\right] h .
\label{traza}
\ee
Primero resolvemos la ecuaci\'on para $h$ y luego haremos lo propio con
${\bar h}_{00}$.

En forma consistente con la aproximaci\'on semicl\'asica, proponemos una 
soluci\'on perturbativa $h=h^{(0)} + h^{(1)}$, donde el primer t\'ermino
corresponde a la soluci\'on cl\'asica y el segundo es la contribuci\'on 
cu\'antica. La primera parte satisface la ecuaci\'on
\bea
( \nabla^{2} - \sigma^{-2} \nabla^{2} \nabla^{2} ) h^{(0)} = 
- 16 \pi G M \delta^{3}({\bf{x}}) & ~~~~ & 
\sigma^{-2}=32 \pi G (3\alpha+\beta) ,
\label{eqhzero}
\eea
que posee la siguiente soluci\'on esf\'ericamente sim\'etrica \cite{stelle}
\be
h^{(0)} =  \frac{4 G M}{r} (1 - e^{- \sigma r}) .
\label{hzero}
\ee
Vemos que las contribuciones en $\alpha$ y $\beta$, que provienen de los
t\'erminos cuadr\'aticos en curvaturas en la acci\'on cl\'asica, conducen a
una perturbaci\'on de corto alcance del tipo Yukawa, que para grandes 
distancias ($\sigma r \gg 1$) 
\SSfootnote{Las ecuaciones semicl\'asicas fueron deducidas bajo las condiciones
 $\nabla \nabla {\cal R} \gg {\cal R}^2$ y 
 $m^2 {\cal R} \ll \nabla \nabla {\cal R}$, de modo que para el potencial 
cl\'asico $G M /r$ ello implica que el rango de distancias en el cual la
soluci\'on tendr\'a validez es $GM \ll r \ll m^{-1}$.}
es despreciable frente a la contribuci\'on de
largo alcance en $r^{-1}$, que proviene del t\'ermino lineal en curvaturas.
Este hecho es una propiedad de la relatividad general, que discutiremos
en m\'as detalle en el cap\'{\i}tulo 6, al estudiar a la relatividad como una
teor\'{\i}a efectiva de bajas energ\'{\i}as.
En este l\'{\i}mite la soluci\'on se reduce entonces 
al potencial cl\'asico $r^{-1}$. La parte cu\'antica satisface
\be
( \nabla^{2} - \sigma^{-2} \nabla^{2} \nabla^{2} ) h^{(1)} = 
{\cal D}(\nabla^2) h^{(0)} ,
\label{h1}
\ee
donde
\be
\hspace{-1.0cm}
{\cal D}(\nabla^2) = 
-\frac{3 G}{2 \pi} (\xi-\frac{1}{6})^2 \ln(-\frac{\nabla^2}{\mu^2})
\nabla^2 \nabla^2 + \frac{G m^2}{\pi} 
\left[
\frac{1}{2} (\xi -\frac{1}{6})  \ln(\frac{m^2}{\mu^2}) + 
3 (\xi^2 -\frac{1}{36})  \ln(-\frac{\nabla^2}{m^2}) \nabla^2
\right] .
\ee
Usando que $\nabla^2 r^{-1} = - 4 \pi \delta^3({\bf x})$  y recordando la
acci\'on del n\'ucleo 
logar\'{\i}tmico sobre la delta de Dirac (ec.(\ref{logdelta})),
encontramos 
\be
\frac{1}{64 \pi G^2 M} 
( \nabla^{2} - \sigma^{-2} \nabla^{2} \nabla^{2} ) h^{(1)} = 
 \left[ \frac{3 m^2}{8 \pi^2} (\xi^2 -\frac{1}{36}) \right] \frac{1}{r^3} 
 - \left[\frac{9}{8 \pi^2} (\xi-\frac{1}{6})^2 \right] \frac{1}{r^5} . 
\ee
La soluci\'on a esta ecuaci\'on es
\be
h^{(1)} = - \frac{24 G^2 M m^2}{\pi} (\xi^2-\frac{1}{36}) 
 \frac{ \ln{ \frac{r}{r_{\sst 0}} } }{r}  -
\frac{12 G^2 M}{\pi} (\xi-\frac{1}{6})^2 \frac{1}{r^3} + \ldots .
\ee
El primer t\'ermino proviene de la fuente proporcional a $r^{-3}$ y el segundo
de la proporcional a $r^{-5}$. Los puntos suspensivos denotan contribuciones
que decrecen m\'as r\'apidamente que $r^{-3}$.

A continuaci\'on hallaremos la soluci\'on para ${\bar h}_{00}$ en el 
l\'{\i}mite de grandes distancias ($\sigma r \gg 1$). Como ya hemos visto,
los t\'erminos en $\alpha$ y $\beta$ dan contribuciones de corto alcance, y
por lo tanto podemos nuevamente obviar su contribuci\'on. Proponemos una
soluci\'on perturbativa ${\bar h}_{00} ={\bar h}_{00}^{(0)} +
 {\bar h}_{00}^{(1)}$, donde el primer t\'ermino es la parte cl\'asica, que
resulta ser ${\bar h}_{00}^{(0)}=4 G M /r$. El segundo t\'ermino satisface
 $\nabla^2 {\bar h}_{00}^{(1)}=-16 \pi G \langle T_{00} \rangle$, estando
el miembro derecho evaluado en la soluci\'on cl\'asica. Para ello necesitamos
las expresiones linealizadas de $H_{00}^{(1)}$ y $H_{00}^{(2)}$, 
\clearpage
\bea
H_{00}^{(1)} &=& (-2 \partial_0 \partial_0 - 2 \Box) \Box h \nonumber \\
H_{00}^{(2)} &=& (- \partial_0 \partial_0 - \frac{1}{2} \Box) \Box h + 
                 \Box \Box h_{00} .
\eea
Al evaluarlas en la soluci\'on cl\'asica $h^{(0)} = 4G M /r$ y 
 $h_{00}^{(0)} = {\bar h}_{00}^{(0)} - \frac{1}{2} h^{(0)} = 2 G M /r$,
resulta
 $H_{00}^{(1)} = 32 \pi G M \nabla^2 \delta^3({\bf x})$ y 
 $H_{00}^{(2)} = 0$. Finalmente obtenemos 
\be
{\bar h}_{00}^{(1)} = - \frac{4 G^2 M}{\pi} 
\left[ (\xi -\frac{1}{6})^2 - \frac{1}{90}
\right] \frac{1}{r^3} -
\frac{8 G^2 M m^2}{\pi} (\xi^2 - \frac{1}{12}) \frac{\ln(r/r_0)}{r} .
\ee
El potencial newtoniano lo obtenemos a partir de 
$h_{00} = {\bar h}_{00} - \frac{1}{2} h $, y resulta
\be
V(r) = -\frac{1}{2} h_{00} =
- \frac{G M}{r} 
\left\{
1 + \frac{2G}{\pi} \left[\frac{1}{2} (\xi-\frac{1}{6})^2 -\frac{1}{90} \right]
\frac{1}{r^2} + \frac{2 m^2 G}{\pi} (\xi^2 + \frac{1}{12})
\ln \frac{r}{r_0}
\right\} .
\label{VCONF}
\ee
Hay dos tipos de correcciones cu\'anticas al potencial newtoniano:

\noindent a)
Una correcci\'on en $r^{-3}$ que es independiente de la masa del
campo escalar y cuyo coeficiente est\'a asociado al l\'{\i}mite de tiempos
cortos del heat kernel (segundo coeficiente de Schwinger-DeWitt), como hemos
discutido con anterioridad. La forma
en que $\alpha$ y $\beta$ escalean ante el grupo de renormalizaci\'on determina
la estructura no local y no masiva de la acci\'on efectiva, a segundo orden 
en curvaturas. Por lo tanto, es dicho escaleo el responsable de las
correcciones en $r^{-3}$. 

\noindent b)
Una correcci\'on en $m^2 r^{-1} \ln r$, 
que depende de la masa del campo escalar 
y cuyo coeficiente est\'a asociado
al l\'{\i}mite de tiempos largos del heat kernel. Esta contribuci\'on 
logar\'{\i}tmica es cualitativamente similar a la correcci\'on que resulta
del escaleo de $G$ mediante el argumento wilsoniano, 
pero el coeficiente num\'erico no coincide
(comparar la ec.(\ref{VRG}) con la ec.(\ref{VCONF})). La raz\'on
de esta discrepancia es que el potencial wilsoniano se basa en el grupo de
renormalizaci\'on, es decir en el escaleo de las constantes gravitatorias
como aparecen en el segundo coeficiente de Schwinger-DeWitt. En cambio, la
correcci\'on logar\'{\i}tmica 
en el {\it espacio de configuraci\'on} surge de los tiempos
largos del heat kernel, que no est\'a relacionado con dicho coeficiente.
Para el caso particular de acoplamiento m\'{\i}nimo $\xi=0$, nuestro resultado
coincide con el wilsoniano. 
Ello se debe, como hemos visto anteriormente, al hecho
de que s\'olo para este valor particular del acoplamiento se cumple 
 $k^{(2)}/ k^{(1)}=-2$, de modo que los t\'erminos masivos pueden interpretarse
efectivamente como una correcci\'on al escaleo de $G$. 
Notemos que el argumento wilsoniano nunca conduce a la correcci\'on 
proporcional a $r^{-3}$.

%%%%%%%%%%%%%%%%%%%%%%%%%%%%%%%%%%%%%%%%%%%%%%%%%%%%%%%%%%%%%%%%%%%%%%%%%%%%

\section{Campos fermi\'onicos en fondos curvos}

En esta secci\'on estudiamos las correcciones cu\'anticas al potencial 
newtoniano debidas a campos espinoriales. El tratamiento de espinores de
Dirac en espacio-tiempo curvo cuadridimensional se basa en el uso del
formalismo de vierbeins. Se introducen matrices de Dirac $\gamma^{\mu}(x)=
V^{\mu}_a(x) \gamma^a$, donde las $\gamma^a$ son las matrices de Dirac usuales
en el espacio plano, y $V^{\mu}_a(x)$ es el vierbein. Una descripci\'on 
detallada del formalismo puede hallarse en \cite{birrell,odibook}. 

La acci\'on eucl\'{\i}dea para fermiones masivos en espacio-tiempo plano
es \cite{ramond}
\be
S[\psi,\psi^{\dagger}]=\int d^4x \psi^{\dagger} [\gamma^{\mu} 
\partial_{\mu} + i m] \psi ,
\ee
donde $\{ \gamma^{\mu},\gamma^{\nu} \} = -2 \eta^{\mu\nu}$. Su generalizaci\'on
a espacio-tiempo curvo, con una m\'etrica con signatura eucl\'{\i}dea, es
\be
S[\psi,\psi^{\dagger}]=\int d^4x \sqrt{g} \psi^{\dagger} [\gamma^{\mu}(x) 
\nabla_{\mu} + i m] \psi ,
\ee
donde  $\{ \gamma^{\mu}(x),\gamma^{\nu}(x) \} = -2 g^{\mu\nu}(x)$ y
 $\nabla_{\mu}$ es la derivada covariante. El conmutador de derivadas
covariantes es 
 ${\cal R}_{\mu\nu}  = [\nabla_{\mu},\nabla_{\nu}] =
\frac{1}{8} [\gamma^{\alpha}(x),\gamma^{\beta}(x)] 
R_{\alpha\beta\mu\nu}(x)$ \cite{christ}. La contribuci\'on del campo 
fermi\'onico a la acci\'on efectiva a un lazo es
\bea
\Gamma &=& - {\rm Tr} \ln (\gamma^{\mu}(x) \nabla_{\mu} + i m) = 
         - \frac{1}{2}  {\rm Tr} \ln {\hat K} \nonumber \\
{\hat K} \psi &=& (\gamma^{\mu}(x) \nabla_{\mu} + i m)
                (\gamma^{\mu}(x) \nabla_{\mu} - i m) \psi =
(- \Box + m^2 + \frac{1}{4} R) \psi .
\eea
Vemos que evaluar la acci\'on efectiva para los fermiones es similar a  evaluar
la acci\'on efectiva para un campo escalar con $\xi=1/4$, salvo un signo 
global y la traza sobre los indices de las matrices de Dirac. Para calcular
esta acci\'on aplicamos el mismo m\'etodo de resumaci\'on de la secci\'on
anterior. A segundo orden en curvaturas, la acci\'on efectiva
resulta
\clearpage
\bea
\Gamma &=& - \frac{1}{32 \pi^2} \int d^4x \sqrt{g} 
\lim_{L \rightarrow \infty} 
\left\{ {\rm Tr} ( {\cal R}_{\mu\nu} H_3(\Box) {\cal R}^{\mu\nu}) + 4
\times  \right. 
\nonumber \\
&& \left. \left[ \left( 
\frac{1}{2} m^4 \ln(\frac{m^2}{L^2}) + m^2 \ln(\frac{m^2}{L^2}) 
(\frac{1}{4} - \frac{1}{6}) R  + R H_1(\Box) R + R_{\mu\nu} H_2(\Box) 
R^{\mu\nu} \right) \right] \right\} 
\label{EAfermions} , 
\eea
donde el t\'ermino entre corchetes es igual al caso escalar evaluado en
$\xi = 1/4$, y el prefactor $4$ proviene de trazar sobre los indices de Dirac.
Tenemos adem\'as una contribuci\'on extra proporcional a
\bea
H_{3}(\Box) = \int_{1/L^2}^{\infty} ds \frac{e^{-s m^2}}{s} h_3(-s \Box) 
&~~~~~~~~~~~& h_3(\eta) = - \frac{f(\eta)-1}{2 \eta} .
\eea
Usando la expresi\'on para ${\cal R}_{\mu\nu}$ 
y calculando la traza del producto de cuatro matrices de Dirac, esta nueva
contribuci\'on puede escribirse en la forma
${\rm Tr } {\cal R}_{\mu\nu} H_{3}(\Box) {\cal R}^{\mu\nu} = 
- \frac{1}{2} R_{\alpha\beta\mu\nu} H_{3}(\Box) R^{\alpha\beta\mu\nu}
$. Finalmente, usando integraciones por partes, las identidades de Bianchi
y la expansi\'on no local del tensor de Riemann en t\'erminos del tensor de
Ricci \cite{vilk2,manitoba}
\begin{equation}
R_{\alpha\beta\mu\nu} = \frac{1}{\Box} \{
\nabla_{\mu} \nabla_{\alpha} R_{\nu\beta} +  
\nabla_{\nu} \nabla_{\beta} R_{\mu\alpha} -
\nabla_{\nu} \nabla_{\alpha} R_{\mu\beta} -
\nabla_{\mu} \nabla_{\beta} R_{\nu\alpha}  \} + {\cal O}( {\cal R}^2 )\,\,\,  ,
\end{equation}
resulta la siguiente identidad
\begin{equation}
\int d^{4}x {\rm Tr } {\cal R}_{\mu\nu} H_{3}(\Box) {\cal R}^{\mu\nu} = 
\int d^{4}x
\left[ 
\frac{1}{2} R H_{3}(\Box) R - 
2 R_{\mu\nu} H_{3}(\Box) R^{\mu\nu} + {\cal O}({\cal R}^3) 
\right] .
\end{equation}
Entonces el tensor de energ\'{\i}a-momento es b\'asicamente el de un campo
escalar, pero modificado en la forma
\begin{eqnarray}
\langle T_{\mu\nu} \rangle &=&  \frac{1}{32 \pi^2}  \left\{
\log(-\frac{\Box}{\mu^2}) 
\left[
(4 \sigma_{2}^{(1)} + \frac{1}{2} \sigma_{2}^{(3)}) H_{\mu\nu}^{(1)} + 
(4 \sigma_{2}^{(2)} - 2 \sigma_{2}^{(3)}) H_{\mu\nu}^{(2)} 
\right] 
\right. \nonumber \\
& & ~~~~~~~~~ + \left.
\frac{m^2}{\Box} \log(-\frac{m^2}{\Box}) 
\left[
(4 k^{(1)} + \frac{1}{2} k^{(3)}) H_{\mu\nu}^{(1)} +
(4 k^{(2)} - 2 k^{(3)}) H_{\mu\nu}^{(2)}
\right]
\right\} .
\end{eqnarray}
Los nuevos coeficientes, asociados al factor de forma $H_3(\Box)$, est\'an
dados  por  $\sigma_{2}^{(3)} = 1/12$ (comportamiento a tiempos cortos) y
 $k^{(3)} =1/2$ (comportamiento a tiempos largos), mientras que los otros 
coeficientes (ecs.(\ref{eq:coef4d})), asociados al campo escalar, deben ser
evaluados en $\xi=1/4$. En el caso fermi\'onico, los t\'erminos dependientes
de la masa en  $ \langle T_{\mu\nu} \rangle $ pueden interpretarse como una
correcci\'on a la constante de Newton, pues se verifica que
 $(4 k^{(2)} - 2 k^{(3)}) / (4 k^{(1)} + \frac {k^{(3)}}{2} ) = -2$. El campo
espinorial se comporta, en este sentido, como un campo escalar m\'{\i}nimamente
acoplado. Entonces es de esperar que, al igual que en el caso escalar, el
potencial newtoniano wilsoniano coincida con el que surge de las ecuaciones
geod\'esicas para una part\'{\i}cula de prueba. 

Podemos obtener la ecuaci\'on del grupo de renormalizaci\'on para $G(\mu)$
imponiendo que la acci\'on efectiva (la acci\'on cl\'asica mas la
contribuci\'on cu\'antica, ec.(\ref{EAfermions})) sea independiente de $\mu$.
En esta forma obtenemos que, debido a los fermiones, la constante de
gravitaci\'on corre con la escala en la forma
\be
\mu \frac{d G}{d \mu} = - \frac{G^2 m^2}{3 \pi} ,
\ee
de modo que $G(\mu)=G_0 [1- \frac{G_0 m^2}{3 \pi} \ln \frac{\mu}{\mu_0}]$, 
y el potencial wilsoniano asociado es $V(r)=- M G(\mu=r^{-1})/r$. Por
otro lado, siguiendo los mismos pasos que en la secci\'on anterior, podemos
resolver las ecuaciones de Einstein semicl\'asicas en el l\'{\i}mite newtoniano
y obtener la contribuci\'on de los fermiones a las ecuaciones geod\'esicas
de la part\'{\i}cula de prueba. De all\'{\i} leemos la modificaci\'on
al potencial newtoniano, que resulta
\be
V(r)  = - \frac{G M}{r} \left[ 1 + \frac{2 G}{15 \pi r^2} + 
\frac{G m^2}{3 \pi} \ln \frac{r}{r_0} 
\right] .
\ee
Como hab\'{\i}amos 
anticipado, la correcci\'on logar\'{\i}tmica coincide con la 
predicci\'on del argumento wilsoniano. 	

%%%%%%%%%%%%%%%%%%%%%%%%%%%%%%%%%%%%%%%%%%%%%%%%%%%%%%%%%%%%%%%%%%%%%%%%%%%%%

\section{Discusi\'on}

En este cap\'{\i}tulo hemos calculado correcciones cu\'anticas al potencial
newtoniano v\'alidas para bajas energ\'{\i}as y largas distancias, debidas
a campos de materia cu\'anticos. Como hemos visto, hay correcciones del tipo
$r^{-3}$, independientes de la masa de campo de materia, y correcciones del
tipo $m^2 r^{-1} \ln r$. Ambas correcciones son extremadamente peque\~nas, 
del orden de $l_{\rm planck}/r$. 

Para QED y campos fermi\'onicos en espacio-tiempo curvo, el argumento 
wilsoniano coincide con las correcciones logar\'{\i}tmicas que se obtienen de
resolver las correspondientes ecuaciones de movimiento, mientras que para
campos escalares dicha coincidencia se produce s\'olo para
acoplamiento m\'{\i}nimo. Hemos analizado este hecho en funci\'on del 
comportamiento del heat kernel para tiempos largos y cortos. En cuanto a las
correcciones en $r^{-3}$, \'estas no se obtienen mediante un argumento 
wilsoniano. Dado que est\'an presentes a\'un para campos escalares no
masivos, y recordando que los grados de libertad f\'{\i}sicos de los gravitones
pueden ser tratados como campos escalares no masivos, esperamos que tambi\'en
haya correcciones en $r^{-3}$ cuando se incluyen las fluctuaciones 
cu\'anticas del campo gravitatorio. Confirmaremos este hecho en el 
pr\'oximo cap\'{\i}tulo. 

Argumentos del tipo wilsoniano
han sido utilizados para intentar explicar, al menos parcialmente, el 
problema de la materia oscura \cite{goldman}: debido a efectos cu\'anticos, 
el potencial newtoniano se modifica en la forma 
$V=-G(\mu=r^{-1}) M r^{-1}$, donde
 $G(\mu)$ es la soluci\'on a las ecuaciones RG para una particular teor\'{\i}a
renormalizable de la gravedad con t\'erminos $R^2$ en el lagrangiano
\cite{fradkin}. Dado que la teor\'{\i}a es asint\'oticamente libre, $G(r)$ es 
una funci\'on creciente de la distancia, lo cual podr\'{\i}a explicar parte 
de la materia faltante \SSfootnote{Sin embargo, el comportamiento de $G(r)$ no 
es suficiente para explicar la curva de rotaci\'on en la Via Lactea 
\cite{bottino}.}.
Adem\'as, la dependencia de la constante gravitatoria
con la distancia puede inducir interesantes efectos cosmol\'ogicos y 
astrof\'{\i}sicos \cite{bertolami}. Sin embargo, en base a los resultados de 
este cap\'{\i}tulo, podemos decir que los t\'erminos de mayor importancia
en el l\'{\i}mite de grandes distancias/bajas energ\'{\i}as, es decir
los que decaen como $r^{-3}$, {\it nunca} aparecen en un argumento wilsoniano.
Por lo tanto, este tipo de argumentos no s\'olo da resultados cuantitativamente
dis\'{\i}milies (el distinto prefactor 
del escaleo de la constante gravitatoria),
sino tambi\'en resultados cualitativamente diferentes, perdi\'endose las
correcciones de mayor relevancia. Para obtener los resultados correctos,
podemos definir una acci\'on efectiva wilsoniana como la acci\'on cl\'asica
en la cual los par\'ametros $\alpha$ y $\beta$ son reemplazados por 
 $\alpha(\Box)$ y $\beta(\Box)$. Estos son los n\'ucleos no locales,
independientes de la masa del campo cu\'antico, y que, como vimos, 
est\'an un\'{\i}vocamente determinados por el escaleo de estas constantes 
que resulta de las ecuaciones del grupo de renormalizaci\'on.

\newpage

\thispagestyle{empty}
~
\newpage

\chapter{Correcciones al potencial newtoniano por efecto de gravitones}
  
\thispagestyle{empty}

En este cap\'{\i}tulo describimos el tratamiento de la relatividad general
como una teor\'{\i}a cu\'antica de campos efectiva, v\'alida para energ\'{\i}as
mucho menores 
que la de Planck. Aplicamos esta t\'ecnica y los m\'etodos de la
acci\'on efectiva para hallar las correcciones cu\'anticas al potencial 
newtoniano en el l\'{\i}mite de grandes distancias debidas a la inclusi\'on de
gravitones \cite{DMgrav}. Mostramos que las ecuaciones de
Einstein a un lazo en gravitones dependen (param\'etricamente) del fijado de
medida de los gravitones. La soluci\'on a dichas ecuaciones tambi\'en 
depende de tales par\'ametros, y como tal no posee relevancia f\'{\i}sica.
Consideramos en cambio un observable f\'{\i}sico, que corresponde a la 
trayectoria de una part\'{\i}cula de prueba en presencia de gravitones.
Derivamos las correcciones cu\'anticas a las ecuaciones geod\'esicas para 
dicha part\'{\i}cula y mostramos que son expl\'{\i}citamente independientes
del fijado de medida. A partir de estas ecuaciones calculamos finalmente el
potencial newtoniano modificado.

\section{La relatividad general como una teor\'{\i}a efectiva}

El lagrangiano m\'as general compatible con la
invariancia general de coordenadas conduce a una acci\'on para la gravedad
de la forma
\be
S=\int d^4x \sqrt{- {\bar g}} 
\left[
\lambda + \frac{2}{\kappa^2} {\bar R} + \alpha {\bar R}^2 +
\beta {\bar R}_{\mu\nu} {\bar R}^{\mu\nu} + \ldots 
\right] ,
\ee
donde $\kappa^2 = 32 \pi G$ y los puntos suspensivos indican t\'erminos 
c\'ubicos y de orden superior en las curvaturas. El lagrangiano queda
ordenado en una expansi\'on en derivadas: $\lambda$ es de orden $\partial^0$
(orden $p^0$ en el espacio de momentos),
 $R$ es de orden $\partial^2$ (dos derivadas respecto a la m\'etrica, orden
$p^2$), los
t\'erminos cuadr\'aticos son de orden $\partial^4$ ($\propto p^4$), etc. 
El primer t\'ermino
est\'a relacionado con la constante cosmol\'ogica, $\Lambda= -8 \pi G \lambda$
y, en principio, deber ser incluido en el lagrangiano efectivo.
Sin embargo, las cotas cosmol\'ogicas dan valores muy peque\~nos para esta
constante, de modo que pondremos $\lambda=0$ de ahora en m\'as.
Por otro lado, el t\'ermino de Einstein lineal en $R$ domina sobre los
t\'erminos cuadr\'aticos, c\'ubicos, etc. en el l\'{\i}mite de bajas
energ\'{\i}as $p^2 \rightarrow 0$ (energ\'{\i}as mucho menores que la 
energ\'{\i}a de Planck, $E_{\rm planck}=G^{-1}$).
Como consecuencia de ello,
los experimentos que involucran la interacci\'on gravitatoria a escalas 
presentes son insensibles a los t\'erminos de orden superior. S\'olo es 
posible establecer cotas m\'as bien pobres sobre los coeficientes,
 $\alpha, \beta < 10^{74}$ \cite{stelle}. Otra forma equivalente de expresar
el dominio del t\'ermino de Einstein es notar que la curvatura de Planck es
$R_{\rm planck} \propto 1/\kappa^2$, de modo que para curvaturas peque\~nas
 $R \ll R_{\rm planck}$ (o distancias grandes $l \gg l_{\rm planck}$), dicho
t\'ermino es el m\'as importante.

Una vez construido el lagrangiano efectivo procedemos a cuantizarlo. Para ello
usaremos la cuantizaci\'on covariante basada en el m\'etodo de campos de
fondo, que tiene la ventaja de preservar la invariancia de medida ante 
transformaciones del campo de fondo. Escribimos las fluctuaciones del campo
gravitatorio alrededor de una m\'etrica de fondo, 
 ${\bar g}_{\mu\nu}=g_{\mu\nu} + \kappa s_{\mu\nu}$ y expandimos el t\'ermino
de Einstein del lagrangiano efectivo en t\'erminos de las
fluctuaciones $s_{\mu\nu}$ (gravitones)
\begin{eqnarray}
S_G & \equiv & \frac{2}{\kappa^2} \int d^4x \sqrt{-{\bar g}} {\bar R} =  
\int d^4x \sqrt{-g} \left\{ 
\frac{2}{\kappa^2} R + \frac{1}{\kappa} s_{\mu\nu} (g^{\mu\nu}R-2 R^{\mu\nu})
\right. \nonumber \\
&& +  \left[ 
-\frac{1}{2} \nabla_{\alpha} s_{\mu\nu} \nabla^{\alpha} s^{\mu\nu} +
\frac{1}{2} \nabla_{\alpha}s \nabla^{\alpha} s - 
\nabla_{\alpha} s \nabla_{\beta} s^{\alpha\beta} +  
\nabla_{\alpha} s_{\mu\beta} \nabla^{\beta} s^{\mu\alpha} 
\right. \nonumber \\
&& \left. \left. + R (\frac{1}{4} s^2 - \frac{1}{2} s_{\mu\nu} s^{\mu\nu}) +
R^{\mu\nu} (2 s^{\lambda}_{\mu} s_{\nu\lambda} - s s_{\mu\nu})
 \right] + {\cal O}(s_{\mu\nu}^3)
\right\} ,
\label{cuadGR}
\end{eqnarray}
donde $s=g^{\mu\nu} s_{\mu\nu}$. Luego fijamos la medida mediante una 
funci\'on $\chi^{\mu}[g,s]$ e introducimos una acci\'on de fijado de medida
\be
S_{{\rm gf}}[g,s] = - \int d^4x \sqrt{-g}  \chi^{\mu} g_{\mu\nu} 
\chi^{\nu} , 
\ee
y agregamos la correspondiente acci\'on de los fantasmas. La acci\'on efectiva
para el campo de fondo $g_{\mu\nu}$ se obtiene de integrar los gravitones y
los fantasmas. A un lazo, su expresi\'on es
\begin{equation}
S_{{\rm ef}} = S_G +
\frac{i}{2} {\rm Tr} \ln 
\left[
\frac{\delta^2 S_G[g]}{\delta g^{\alpha\beta} g^{\gamma\delta}} - 2
\frac{\delta \chi^{\mu}}{\delta g^{\alpha\beta}} g_{\mu\nu}
\frac{\delta \chi^{\nu}}{\delta g^{\gamma\delta}}
\right]
- i {\rm Tr} \ln 
\left[
-2 g_{\nu\alpha} \nabla_{\beta} 
\frac{\delta \chi^{\mu}}{\delta g^{\alpha\beta}} 
\right] ,
\label{EAlorent}
\end{equation}
donde el primer t\'ermino es la acci\'on cl\'asica de Einstein-Hilbert, 
el segundo proviene de las fluctuaciones de gravitones y el tercero de los
fantasmas. Estos dos \'ultimos t\'erminos son lineales en $\hbar$.

Las divergencias ultravioletas pueden ser calculadas una vez elegida la 
funci\'on de fijado de medida, y seg\'un su forma el c\'alculo puede resultar
m\'as o menos engorroso. En la secci\'on siguiente estudiaremos 
\'esto con detalle. Para describir la metodolog\'{\i}a de teor\'{\i}as 
efectivas,
nos basta con considerar el caso m\'as simple (medida de DeWitt), estudiado 
originalmente en \cite{THV}.
La divergencia de la acci\'on efectiva resulta
\SSfootnote{Si se utiliza regularizaci\'on dimensional, el factor $\ln L^2$
en la ec.(\ref{THVdiv}) debe ser reemplazado por $2/(4-d)$.}
\be
\hspace{-1.0cm}
\Delta S_G^{\rm div}(\lambda=0) =
\frac{\ln L^2}{96 \pi^2} \int d^4x \sqrt{-g}
\left[ 
\frac{53}{15} (R_{\mu\nu\rho\sigma} R^{\mu\nu\rho\sigma} - 
4 R_{\mu\nu} R^{\mu\nu} + R^2) 
+ \frac{21}{10} R_{\mu\nu} R^{\mu\nu} +
\frac{1}{20} R^2 
\right] .
\label{THVdiv}
\ee
Si adem\'as de la gravedad hay otros campos de materia, \'estos tambi\'en
proveer\'an contribuciones adicionales en $R^2$ y $R_{\mu\nu} R^{\mu\nu}$,
como veremos luego. El hecho que las divergencias no sean proporcionales a
la acci\'on de Einstein original indica que la teor\'{\i}a 
no es renormalizable. Sin embargo, {\it s\'{\i} lo es} en el esp\'{\i}ritu
de teor\'{\i}as efectivas. 
Las divergencias a un lazo pueden ser absorbidas en las
constantes $\alpha$ y $\beta$ de los t\'erminos cuadr\'aticos del lagrangiano
efectivo. An\'alogamente, las divergencias a dos lazos \cite{goroff} pueden
ser absorbidas en los t\'erminos c\'ubicos, etc. Sin embargo, estas 
divergencias {\it no son} predicciones de la teor\'{\i}a efectiva, pues 
provienen de momentos altos en los diagramas de Feynman, y justamente la 
teor\'{\i}a no pretende describir tales rangos de energ\'{\i}a. Los 
valores de los 
par\'ametros del lagrangiano efectivo {\it tampoco} son predicciones, pues
engloban nuestra ignorancia sobre la verdadera teor\'{\i}a de altas
energ\'{\i}as. 

?` Cu\'ales son 
entonces las predicciones cu\'anticas de la relatividad general
como teor\'{\i}a efectiva? Corresponden a los efectos cu\'anticos debidos
a la porci\'on de bajas energ\'{\i}as de la teor\'{\i}a. Ellos se deben 
a la propagaci\'on de part\'{\i}culas no masivas, que dan contribuciones no 
locales a la acci\'on efectiva. 
En el espacio de momentos estas contribuciones son no anal\'{\i}ticas,
proporcionales a $\ln(-q^2)$, independientes de los par\'ametros contenidos
en los ordenes superiores del lagrangiano efectivo, y claramente distinguibles
de las contribuciones locales de altas energ\'{\i}as. 

En este punto conviene hacer una aclaraci\'on sobre la forma de extraer los
t\'erminos no locales logar\'{\i}tmicos de la acci\'on efectiva 
y obtener ecuaciones
reales y causales para la m\'etrica de fondo. Siguiendo la
l\'{\i}nea de razonamiento de los cap\'{\i}tulos 4 y 5, deber\'{\i}amos 
partir de
la acci\'on eucl\'{\i}dea, resumar la serie de Schwinger-DeWitt, identificar
los t\'erminos logar\'{\i}tmicos en los factores de forma, y finalmente
calcular las ecuaciones de movimiento in-in reemplazando los propagadores
eucl\'{\i}deos por los retardados a nivel de las ecuaciones de movimiento. 
En vez de seguir este camino, en este cap\'{\i}tulo utilizaremos otro 
m\'as corto que nos
conducir\'a a las mismas ecuaciones in-in. Partimos de la acci\'on efectiva
escrita en signatura lorentziana $(-+++)$ (ec.(\ref{EAlorent})), 
y calculamos las divergencias.
Como ya vimos en el cap\'{\i}tulo 5, los coeficientes que acompa\~nan los
factores de forma $\ln(-\Box)$ est\'an dados por las divergencias (segundo
coeficiente de Schwinger-DeWitt), de modo que calculando \'estas podemos
obtener 
inmediatamente las correcciones logar\'{\i}tmicas a la acci\'on efectiva.
Luego escribimos las ecuaciones de movimiento in-out y finalmente, tomando
dos veces su parte real y causal, llegamos a las correctas ecuaciones in-in.

\section{La acci\'on efectiva para gravedad + materia: divergencias}

Distintos autores han estudiado correcciones cu\'anticas
al potencial newtoniano debida a efectos de gravitones,
cuantizando la m\'etrica alrededor
del espacio plano y definiendo el potencial a partir
de diferentes conjuntos de diagramas de Feynman \cite{donoghue1,don2,VM,HL}.
En vez de evaluar diagramas y elementos de matriz de dispersi\'on, nuestra
intenci\'on es usar la acci\'on efectiva para hacer un c\'alculo covariante.
Este m\'etodo co\-va\-rian\-te es m\'as adecuado para estudiar problemas que 
involucren fluctuaciones alrededor de campos de fondo no planos.

Al igual que en el cap\'{\i}tulo anterior, consideremos una part\'{\i}cula
masiva $M$, que trataremos como una fuente cl\'asica. La acci\'on 
(minkowskiana) de partida es $S=S_G + S_M$, con
\bea
S_G &=& \frac{2}{\kappa^2} \int d^4x \sqrt{- \bar{g}} {\bar R}  \\
S_M &=& - M \int \sqrt{-{\bar g}_{\mu\nu}dx^{\mu} dx^{\nu} } .
\eea
Para calcular la acci\'on efectiva a un lazo, debemos expandir estas acciones
en las fluctuaciones $s_{\mu\nu}$ y extraer los t\'erminos cuadr\'aticos.
Para fijar la medida utilizaremos principalmente una familia de funciones
(que llamaremos ``familia $\lambda$'') $\chi^{\mu}(\lambda)$, parametrizadas 
por un n\'umero real $\lambda$, dadas por
\begin{equation}
\chi^{\mu}(\lambda) = \frac{1}{\sqrt{1+\lambda}} \left[
g^{\mu\gamma} \nabla^{\sigma} s_{\gamma\sigma} - 
\frac{1}{2} g^{\gamma\sigma} \nabla^{\mu} s_{\gamma\sigma} \right] .
\end{equation} 
El caso $\lambda=0$ corresponde a la medida de DeWitt de la que habl\'abamos 
anteriormente. Para este tipo de medidas, que son lineales en las 
fluctuaciones, los fantasmas se desacoplan de los gravitones y se acoplan
s\'olo al campo de fondo. El desarrollo a orden cuadr\'atico de la gravedad
pura est\'a dado en la ec.(\ref{cuadGR}) y, para esta elecci\'on de la
medida, la acci\'on efectiva a un lazo queda
\begin{equation}
S_{{\rm ef}} = S_G +
\frac{i}{2} {\rm Tr} \ln F^{\alpha\beta,\mu\nu}(\nabla) 
- i {\rm Tr} \ln (\Box \delta^{\mu}_{\nu} + R^{\mu}_{\nu}) ,
\end{equation} 
donde el operador diferencial de segundo orden es
\begin{equation}
F^{\alpha\beta,\mu\nu}(\nabla) = \sqrt{-g} C^{\alpha\beta,\lambda\sigma} 
\left\{ 
\Box \delta^{\mu}_{(\lambda} \delta^{\nu}_{\sigma)} -
\frac{2 \lambda}{1+\lambda} \delta^{(\mu}_{(\lambda} \nabla_{\sigma)}
\nabla^{\nu)} + 
\frac{\lambda}{1+\lambda} g^{\mu\nu} \nabla_{(\lambda} \nabla_{\sigma)}
+ P^{\mu\nu}_{\lambda\sigma}
\right\} ,
\end{equation}
con
\begin{eqnarray}
C^{\alpha\beta,\lambda\sigma} &=& \frac{1}{4} 
(g^{\lambda \alpha} g^{\sigma\beta} + g^{\lambda\beta} g^{\sigma\alpha} -
g^{\lambda \sigma} g^{\alpha \beta} )  \nonumber \\
P^{\mu\nu}_{\lambda\sigma} &=&
2 R_{\lambda \,\, \cdot \,\, \sigma \cdot}^{~(\mu ~\nu)}
+ 2 \delta^{(\mu}_{(\lambda} R^{\nu)}_{\sigma)}
- g^{\mu\nu} R_{\lambda\sigma} - g_{\lambda\sigma} R^{\mu\nu}
- R \delta^{\mu}_{(\lambda} \delta^{\nu}_{\sigma)}
+ \frac{1}{2} g^{\mu\nu} g_{\lambda\sigma} R . 
\end{eqnarray}
Los par\'entesis denotan simetrizaci\'on con un factor $1/2$. Notemos que para
$\lambda=0$ el operador diferencial tiene la forma de un operador m\'{\i}nimo
 $F_{AB}(\nabla) = {\hat C}_{AB} \, g^{\mu\nu} \nabla_{\mu} \nabla_{\nu} +
{\hat Q}_{AB}$.

Para deducir la contribuci\'on de la masa $M$ a la acci\'on efectiva, 
desarrollamos la acci\'on $S_M$ a orden cuadr\'atico en las fluctuaciones. 
Obtenemos
\begin{equation}
S_M = - M \int d\tau \left[
1 - \frac{\kappa}{2} s_{\mu\nu} {\dot x}^{\mu} {\dot x}^{\nu} 
-  \frac{\kappa^2}{8} s_{\mu\nu} s_{\rho\sigma} 
{\dot x}^{\mu}  {\dot x}^{\nu}  
 {\dot x}^{\rho} {\dot x}^{\sigma} + {\cal O}(s_{\mu\nu}^3) \right] ,
\end{equation}
donde $\cdot$ representa derivada respecto al tiempo propio $\tau$, definido
como  $d\tau^2=-g_{\mu\nu} dx^{\mu} dx^{\nu}$. Introduciendo una identidad,
$1=\int d^4y \sqrt{-g}  \delta^4(y-x(\tau))$, la acci\'on puede ser reescrita
en la siguiente forma
\be
S_M = -M \int d\tau + 
\frac{\kappa}{2} \int d^4y \sqrt{-g} s_{\mu\nu}(y) T^{\mu\nu}(y) +
\int d^4y \sqrt{-g} s_{\mu\nu}(y) s_{\rho\sigma}(y) 
{\tilde M}^{\mu\nu\rho\sigma}(y) + \ldots 
\label{eq:smcuad}
\ee
donde
\bea
T^{\mu\nu}(y) &=& M \int d\tau {\dot x}^{\mu} {\dot x}^{\nu} 
\delta^4(y-x(\tau)) 
\label{eq:tmunu} \\
{\tilde M}^{\mu\nu\rho\sigma}(y) &=& \frac{M \kappa^2}{8} 
\int d\tau \delta^4(y-x(\tau))
{\dot x}^{\mu} {\dot x}^{\nu} {\dot x}^{\rho} {\dot x}^{\sigma} .
\eea
Los t\'erminos cuadr\'aticos de la ec.(\ref{eq:smcuad}) introducen 
una nueva contribuci\'on al operador diferencial
$F(\nabla)$, que finalmente toma la forma
\be
\hspace{-0.8cm}
F^{\alpha\beta,\mu\nu}(\nabla) = 
\sqrt{-g} C^{\alpha\beta,\lambda\sigma} 
\left\{ 
\Box \delta^{\mu}_{(\lambda} \delta^{\nu}_{\sigma)} -
\frac{2 \lambda}{1+\lambda} \delta^{(\mu}_{(\lambda} \nabla_{\sigma)}
\nabla^{\nu)} + 
\frac{\lambda}{1+\lambda} g^{\mu\nu} \nabla_{(\lambda} \nabla_{\sigma)}
+ P^{\mu\nu}_{\lambda\sigma} + M^{\mu\nu}_{\lambda\sigma}
\right\} ,
\label{opdif}
\ee
siendo
\begin{equation}
M^{\mu\nu}_{\lambda\sigma}(y) = (C^{-1})^{\mu\nu\alpha\beta} 
{\tilde M}_{\alpha\beta\lambda\sigma}(y)
= \frac{M \kappa^2}{8} \int d\tau \delta^4(y-x(\tau)) 
\left[ g^{\mu\nu} {\dot x}_{\lambda} {\dot x}_{\sigma} +
2 {\dot x}^{\mu} {\dot x}^{\nu} {\dot x}_{\lambda} {\dot x}_{\sigma} 
\right] .
\label{eq:meff}
\end{equation}

Una vez que tenemos la expresi\'on de los operadores diferenciales para
los gravitones y fantasmas, podemos calcular las divergencias de la acci\'on
efectiva. Para ello conviene distinguir el caso de medidas m\'{\i}nimas y no 
m\'{\i}nimas. El resultado final de la divergencia para cualquier valor de
$\lambda$ puede verse en la ec.(\ref{eq:divergenciesL}).

\subsection{Divergencias para medidas m\'{\i}nimas}

Para $\lambda=0$ (medida de DeWitt) los operadores diferenciales, tanto para
los gravitones como para los fantasmas, tienen la forma m\'{\i}nima, que en 
notaci\'on matricial es
\begin{equation}
\hat{\cal F}(\nabla) = \Box + \hat{\cal Q} - \frac{1}{6} R \hat{1} .
\end{equation}
En efecto, para los gravitones la matriz $\hat{\cal Q}$ est\'a dada por 
 $\hat{\cal Q}= \hat{P} + \hat{M} + \frac{1}{6} R \hat{1}$, mientras que
para los fantasmas, $\hat{\cal Q}= \hat{R} + \frac{1}{6} R \hat{1}$. 
Para calcular las trazas funcionales usamos la t\'ecnica de SDW,
\begin{equation}
{\rm Tr} \ln \hat{\cal F} = \lim_{L \rightarrow \infty} 
\frac{i}{ (4 \pi)^{\frac{d}{2}} } 
\int_{1/L^2}^{\infty}
\frac{ds}{s^{\frac{d}{2}+1}} \int d^dx 
{\rm Tr} \sum_{n=0}^{\infty} (i s)^n \hat{a}_n(x) ,
\label{swdmink}
\end{equation}
donde los $\hat{a}_n(x)$'s son los l\'{\i}mites de coincidencia de los 
coeficientes de SDW. Como ya sabemos, en $d=4$ dimensiones las divergencias
est\'an en los coeficientes ${\hat a}_0$, ${\hat a}_1$ y ${\hat a}_2$.
Los dos primeros renormalizan la constante cosmol\'ogica y la de Newton, 
mientras que el \'ultimo es
\begin{equation}
\hat{a}_2(x) = \frac{1}{180} (R_{\mu\nu\alpha\beta} R^{\mu\nu\alpha\beta} -
R_{\mu\nu} R^{\mu\nu} + \Box R) \hat{1} + \frac{1}{2} \hat{\cal Q}^2 
+ \frac{1}{12} \hat{\cal R}_{\mu\nu} \hat{\cal R}^{\mu\nu} + \frac{1}{6} 
\Box \hat{\cal Q} ,
\end{equation}
con $\hat{\cal R}_{\mu\nu}$ el conmutador de derivadas covariantes. Insertando
las expresiones de $\hat Q$ para gravitones y fantasmas obtenemos las 
divergencias de la acci\'on efectiva debidas a la gravedad pura y a la masa 
$M$. La primera da el resultado de la ec.(\ref{THVdiv})
\bea
\Delta S_G^{\rm div}(\lambda=0) &=&
\frac{\ln L^2}{96 \pi^2} \int d^4x \sqrt{-g}
\left[ 
\frac{53}{15} (R_{\mu\nu\rho\sigma} R^{\mu\nu\rho\sigma} - 
4 R_{\mu\nu} R^{\mu\nu} + R^2) \right. \nonumber \\
&& ~~~~~~~~~~~~~~~~~~~~~~~ \left. + \frac{21}{10} R_{\mu\nu} R^{\mu\nu} +
\frac{1}{20} R^2 
\right] ,
\label{divG}
\eea
mientras que la segunda se lee de la contribuci\'on de $M$ al ${\hat a}_2$, 
a saber $\frac{1}{6} \Box \hat{M} + \frac{1}{2} \hat{M}^2 + \hat{M} 
(\hat{P} + \frac{1}{6} R \hat{1})$. Obtenemos 
\be
\hspace{-0.5cm}
\Delta S_M^{\rm div}(\lambda=0) = 
\frac{\ln L^2}{64 \pi^2} \int d^4x \sqrt{-g}
\left[
M_{\mu\nu\rho\sigma} M^{\mu\nu\rho\sigma}  + 2  M_{\mu\nu\rho\sigma}  
\left( P^{\rho\sigma\mu\nu} + \frac{1}{6} R \delta^{\rho(\mu} 
\delta^{\sigma\nu)} \right) \right] .
\label{divM}
\ee
La divergencia para $\lambda=0$ es entonces
\be
\Delta S^{\rm div}(\lambda=0) = \Delta S_G^{\rm div}(\lambda=0) + 
\Delta S_M^{\rm div}(\lambda=0) .
\ee

\subsection{Divergencias para medidas no m\'{\i}nimas}

Cuando $\lambda \neq 0$ el operador ${\hat F}(\nabla)$ no tiene forma
m\'{\i}nima y por lo tanto no se puede aplicar la t\'ecnica de SDW. Sin 
embargo, en \cite{BV2} ha sido desarrollado un m\'etodo de reducci\'on que 
generaliza esa t\'ecnica y que permite hallar las divergencias de operadores
no m\'{\i}nimos. Al igual que la expansi\'on de SDW, este m\'etodo de
reducci\'on  consiste en un
desarrollo local en los campos de fondo, y ha sido calculado hasta segundo
orden en los tensores de curvatura. El punto de partida es notar que, dado que
la teor\'{\i}a es independiente del fijado de medida en la capa de masa, 
la diferencia de la acci\'on efectiva calculada en dos medidas distintas es
siempre proporcional a las ecuaciones de movimiento cl\'asicas (extremal).
Con esta idea en mente, esa diferencia puede ser expresada en t\'erminos de
funciones de Green no m\'{\i}nimas para gravitones y fantasmas, que luego se
expanden en los campos de fondo. 

Un caso especialmente simple es cuando la acci\'on de fijado de medida 
difiere en un factor global de la m\'{\i}nima. Este es justamente el caso para
la familia $\lambda$, pues 
 $\chi^{\mu}(\lambda) = \frac{1}{\sqrt{1+\lambda}} \chi^{\mu}(\lambda=0)$.
Siguiendo los m\'etodos de \cite{BV2}, podemos escribir la acci\'on efectiva
para cualquier valor del par\'ametro de medida, 
\begin{equation}
S_{{\rm ef}}(\lambda) = S_{{\rm ef}}(\lambda=0) + \frac{i}{2} \lambda 
\left[ 
{\rm Tr} V_{1\nu}^{~\mu}(\nabla) - {\rm Tr} V_{2\nu}^{~\mu}(\nabla)  
\right]
-\frac{i}{4} \lambda^2 {\rm Tr} [ V_{1\nu}^{~\mu}(\nabla)]^2
+ {\cal O} (({\cal E}^{\mu\nu})^2) ,
\end{equation}
donde el extremal est\'a dado por 
\begin{equation}
{\cal E}^{\mu\nu} = \frac{\delta(S_G+S_M)}{\delta g^{\mu\nu}} = 
-\frac{2}{\kappa^2} (R^{\mu\nu}-\frac{1}{2} R g^{\mu\nu})
+\frac{1}{2} T^{\mu\nu} ,
\end{equation}
y $V_{1\nu}^{~\mu}(\nabla)$ y $V_{2\nu}^{~\mu}(\nabla)$ son tensores
lineales y cuadr\'aticos en el extremal. Su acci\'on sobre una funci\'on de
prueba $\zeta^{\nu}$ es
\begin{eqnarray}
V_{1\nu}^{~\mu}(\nabla) \zeta^{\nu} &=& 
2 \kappa^2 \, Q^{\mu}_{\alpha} \,  \nabla_{\beta} 
\, \Gamma^{(\alpha}_{\rho\sigma}(\nabla) \, {\cal E}^{\rho\beta)} \,
Q^{\sigma}_{\nu} \, \zeta^{\nu}  \nonumber\\
V_{2\nu}^{~\mu}(\nabla) \zeta^{\nu} &=&
- \kappa^2 g^{\mu\omega} \, Q^{\gamma}_{\omega} 
\, {\cal E}^{(\alpha\rho} \, \Gamma^{\beta)}_{\rho\gamma}(\nabla) \,
G_{\alpha\beta,\varphi\theta}(\nabla)  \, 
\Gamma^{(\varphi}_{\delta\sigma}(\nabla) \, {\cal E}^{\theta)\delta} \,
Q^{\sigma}_{\nu} \, \zeta^{\nu} .
 \end{eqnarray}
En estas expresiones, $\Gamma^{\nu}_{\rho\sigma}(\nabla)=
\delta^{\nu}_{\rho} \nabla_{\rho} - 2 \delta^{\nu}_{\sigma} \nabla_{\rho}$, 
y $G_{\alpha\beta,\varphi\theta}(\nabla)$ y 
$Q^{\sigma}_{\mu}$ son respectivamente funciones de Green para gravitones y 
fantasmas, evaluadas en la medida de DeWitt
\clearpage
\begin{eqnarray}
F^{\gamma\sigma,\alpha\beta}(\nabla|\lambda=0) \,
G_{\alpha\beta,\varphi\theta}(\nabla)= -
\delta^{\gamma\sigma}_{\varphi\theta} 
& ~~~~~~ &
(\Box \delta^{\mu}_{\alpha} + R^{\mu}_{\alpha}) Q^{\sigma}_{\mu} =
\delta^{\sigma}_{\alpha} .
\end{eqnarray}

Como discutiremos cuando resolvamos las ecuaciones para el campo de fondo,
nos basta con considerar los t\'erminos lineales en el extremal a nivel de
la acci\'on efectiva. Nos concentramos entonces en
${\rm Tr} V_{1\nu}^{~\mu}(\nabla)$, que viene dada por
\begin{equation}
{\rm Tr} V_{1\nu}^{~\mu}(\nabla) = 2 \kappa^2 \int d^4x 
\left[
R^{\alpha}_{\, \cdot \, \gamma\beta\sigma} {\cal E}^{\gamma\beta} -
{\cal E}^{\beta\gamma} \delta^{\alpha}_{\sigma} \nabla_{\beta} 
\nabla_{\gamma}
\right]  
(\Box \delta^{\sigma}_{\alpha} + R^{\sigma}_{\alpha})^{-2}
\delta(x,y) |_{y=x} .
\end{equation}
Para calcular su divergencia usamos los m\'etodos de \cite{BV2}. Dado que
estamos trabajando a orden cuadr\'atico en curvaturas, en el primer t\'ermino
en el corchete podemos aproximar
$(\Box \delta^{\sigma}_{\alpha} + R^{\sigma}_{\alpha})^{-2}$ por
 $\Box^{-2} \delta^{\sigma}_{\alpha}$. Aparecen dos divergencias, que son
\begin{eqnarray}
\Box^{-2} \delta^{\sigma}_{\alpha} \delta(x,y) |^{{\rm div}}_{y=x} &=&
\frac{i}{16 \pi^2} \ln L^2 \sqrt{-g}  \\
\nabla_{\beta} \nabla_{\gamma}
(\Box \delta^{\sigma}_{\alpha} + R^{\sigma}_{\alpha})^{-2}
\delta(x,y)|^{{\rm div}}_{y=x} &=&
\frac{i}{16 \pi^2} \ln L^2 \sqrt{-g} 
\left[
\frac{1}{6} (R_{\beta\gamma}-\frac{1}{2} g_{\beta\gamma} R) 
\delta^{\sigma}_{\alpha}  \right. \nonumber \\ 
&& ~~~~~~~~~~~~~~~~~~~~~
\left. + \frac{1}{2} R^{\sigma}_{\, \cdot \, \alpha\beta\gamma} - 
\frac{1}{2} g_{\beta\gamma} R^{\sigma}_{\alpha}
\right] .
\end{eqnarray}
de modo que
\begin{equation}
\left. {\rm Tr} V_{1\nu}^{~\mu}(\nabla) \right|^{{\rm div}} =
\frac{i \kappa^2}{24 \pi^2} \ln L^2 \int d^4x \sqrt{-g}
\left[ -5 R_{\mu\nu} {\cal E}^{\mu\nu} +
\frac{5}{2} R g_{\mu\nu}  {\cal E}^{\mu\nu} \right] .
\end{equation}

En definitiva, la divergencia total de la acci\'on efectiva a un lazo resulta
\begin{equation}
\Delta S^{\rm div}(\lambda) = \Delta S^{\rm div}(\lambda=0)
- \frac{\lambda \kappa^2}{48 \pi^2} \ln L^2 \int d^4x \sqrt{-g}
\left[ -5 R_{\mu\nu} {\cal E}^{\mu\nu} +
\frac{5}{2} R g_{\mu\nu}  {\cal E}^{\mu\nu} \right] ,
\label{eq:divergenciesL}
\end{equation}

\section{Ecuaciones de Einstein. Problema del fijado de medida}

Como ya vimos, dadas las divergencias de la acci\'on efectiva, podemos extraer
los t\'erminos no locales logar\'{\i}tmicos 
que dominan la f\'{\i}sica de bajas 
energ\'{\i}as / largas distancias. Las constantes de proporcionalidad que 
acompa\~nan al factor de forma $\ln(-\Box)$ se leen de la divergencia 
como sigue
\begin{equation}
\alpha \ln L^2 \int d^4x \sqrt{-g} (\ldots) \rightarrow
- \alpha \int d^4x \sqrt{-g} (\ldots) \ln(-\Box) .
\end{equation}
La parte no local de la acci\'on efectiva es entonces
 $\Delta S = \Delta S^{{\rm nl}}_G(\lambda = 0) 
 + \Delta S^{{\rm nl}}_M(\lambda = 0)  + 
 \Delta S^{{\rm nl}}(\lambda \neq 0)$, con
\begin{eqnarray}
\Delta S^{{\rm nl}}_G(\lambda=0) &=&
- \frac{1}{96 \pi^2} \int d^4x \sqrt{-g}  \left[  \frac{21}{10} R_{\mu\nu} 
\ln(-\Box) R^{\mu\nu} +  \frac{1}{20} R \ln(-\Box) R \right]  \\
\Delta S_M^{{\rm nl}}(\lambda=0) &=& -\frac{1}{64 \pi^2} \int d^4x \sqrt{-g}
\left[ M_{\mu\nu\rho\sigma} \ln(-\Box)  M^{\rho\sigma\mu\nu}  \right.
\nonumber \\
&& \left. ~~~~~~~~~~~~~~~~~~~~~~~~
+ 2  M_{\mu\nu\rho\sigma} \ln(-\Box) \left( P^{\rho\sigma\mu\nu} +
\frac{1}{6} R \delta^{\rho(\mu} \delta^{\sigma\nu)} \right) \right] , 
\label{eq:nolocM}
\end{eqnarray}
\begin{eqnarray}
\Delta S^{{\rm nl}}(\lambda \neq 0) &=&
\int d^4x \sqrt{-g} 
\left[ a(\lambda) R_{\mu\nu} \ln(-\Box) {\cal E}^{\mu\nu} +
b(\lambda) R g_{\mu\nu} \ln(-\Box) {\cal E}^{\mu\nu} \right] ,
\label{eq:nolocL}
\end{eqnarray} 
donde $a(\lambda)=-\frac{5 \lambda \kappa^2}{48 \pi^2}$ y
$b(\lambda)=\frac{5 \lambda \kappa^2}{96 \pi^2}$.

Consideremos ahora el caso en que $M$ es est\'atica y est\'a localizada en el
origen:  ${\dot x}^{\mu} =(1,0,0,0)$, 
 $T^{\mu\nu}(x)=M \delta^{\mu}_0 \delta^{\nu}_0 \delta^3({\bf x})$, 
 $T^{\mu}_{\mu}=-M \delta^3({\bf x})$. Para esta elecci\'on de la fuente,
los distintos tensores que aparecen en la parte masiva no local son
\begin{eqnarray}
M^{\mu\nu\lambda\sigma}(y) &=&
\frac{M \kappa^2}{8} \delta^3({\bf y}) \,
[g^{\mu\nu}+2 \delta^{\mu}_0 \delta^{\nu}_0] \, \delta^{\lambda}_0 
\delta^{\sigma}_0  \nonumber\\
M_{\mu\nu\rho\sigma}  R \delta^{\rho(\mu} \delta^{\sigma\nu)} &=&
\frac{M \kappa^2}{8} R  \delta^3({\bf y}) \\
M_{\mu\nu\rho\sigma} P^{\rho\sigma\mu\nu} &=&
\frac{M \kappa^2}{8}  \delta^3({\bf y}) 
[g^{\mu\nu} P_{00\mu\nu} + 2 P_{0000}]=
-  \frac{M \kappa^2}{8} R  \delta^3({\bf y}). \nonumber
\end{eqnarray}
de modo que
\begin{equation}
\Delta S_M^{{\rm nl}} (\lambda=0) = \frac{5 M \kappa^2}{1536 \pi^2} \int d^4x 
\sqrt{-g} R \ln(-\nabla^2) \delta^3({\bf x}) ,
\end{equation}
donde hemos usado que $M$ es est\'atica para reemplazar
 $\Box$ por $\nabla^2$. Omitimos el t\'ermino cuadr\'atico en $M$
pues es irrelevante en el l\'{\i}mite de distancias largas.

Las ecuaciones de movimiento in-out se obtienen derivando funcionalmente la
acci\'on efectiva respecto a la m\'etrica de fondo $g_{\mu\nu}$, y no son
ni reales ni causales. Las ecuaciones in-in (reales y causales) las deducimos
tomando dos veces la parte real y causal de las ecuaciones in-out.
En el caso de campos est\'aticos, la situaci\'on se simplifica pues basta
reemplazar el D'Alambertiano por el laplaciano. A orden cuadr\'atico en
curvaturas, las ecuaciones para el valor medio del campo de fondo son
\begin{equation}
\frac{1}{8\pi G} (R_{\mu\nu}-\frac{1}{2} R g_{\mu\nu}) =
T_{\mu\nu} + \langle T_{\mu\nu}\rangle^{G}_{\lambda=0} +
\langle T_{\mu\nu}\rangle^{M}_{\lambda=0} +
\langle T_{\mu\nu}\rangle_{a(\lambda)} +
\langle T_{\mu\nu}\rangle_{b(\lambda)} ,
\end{equation}
donde
\begin{eqnarray}
\langle T_{\mu\nu}\rangle^{G}_{\lambda=0} &=&
-\frac{1}{96 \pi^2} \left[ \frac{21}{10} \ln(-\nabla^2) H_{\mu\nu}^{(2)} +
\frac{1}{20} \ln(-\nabla^2) H_{\mu\nu}^{(1)} \right] \nonumber\\
\langle T_{\mu\nu}\rangle^{M}_{\lambda=0} &=&
\frac{5 M \kappa^2}{768 \pi^2} (\nabla_{\mu} \nabla_{\nu} - g_{\mu\nu} 
\nabla^2) \ln(-\nabla^2) \delta^3({\bf x}) \nonumber\\
\langle T_{\mu\nu}\rangle_{a(\lambda)} &=&
a(\lambda) \left[-\frac{2}{\kappa^2} \ln(-\nabla^2) H_{\mu\nu}^{(2)} +
\frac{1}{\kappa^2}  \ln(-\nabla^2) H_{\mu\nu}^{(1)} -\frac{1}{2}
\nabla^2 \ln(-\nabla^2) T_{\mu\nu} \right] \nonumber\\
\langle T_{\mu\nu}\rangle_{b(\lambda)} &=&
b(\lambda) \left[ \frac{2}{\kappa^2} \ln(-\nabla^2) H_{\mu\nu}^{(1)}
+ \nabla_{\mu} \nabla_{\nu} \ln(-\nabla^2) T^{\alpha}_{\alpha} -
g_{\mu\nu} \nabla^2 \ln(-\nabla^2) T^{\alpha}_{\alpha} \right] .
\label{eq:tensors}
\end{eqnarray}

Estas son las ecuaciones de Einstein modificadas por los efectos cu\'anticos
de los gravitones. Su soluci\'on da la m\'etrica de fondo del espacio-tiempo.
Para encontrarla procedemos en forma an\'aloga al cap\'{\i}tulo anterior.
Perturbamos alrededor del espacio plano 
 $g_{\mu\nu}=\eta_{\mu\nu}+h_{\mu\nu}$ ($\eta_{\mu\nu}={\rm diag}(-+++)$)
y elegimos la medida arm\'onica 
 $(h_{\mu\nu} - \frac{1}{2} h \eta_{\mu\nu})^{;\nu}=0$
para la perturbaci\'on $h_{\mu\nu}$. Vale la pena recalcar que esta elecci\'on
es completamente independiente de la libertad de fijado de medida para las
fluctuaciones cu\'anticas $s_{\mu\nu}$. Las ecuaciones de movimiento 
linealizadas quedan
\begin{equation}
\nabla^2 {\bar h}_{\mu\nu} = -16 \pi G \left[
T_{\mu\nu} + \langle T_{\mu\nu}\rangle^{G}_{\lambda=0} +
\langle T_{\mu\nu}\rangle^{M}_{\lambda=0} +
\langle T_{\mu\nu}\rangle_{a(\lambda)} +
\langle T_{\mu\nu}\rangle_{b(\lambda)} \right] ,
\label{eq:htecho}
\end{equation}
siendo ${\bar h}_{\mu\nu} = h_{\mu\nu} - \frac{1}{2} h \eta_{\mu\nu}$.
Resolvemos estas ecuaciones en forma perturbativa alrededor de la soluci\'on
cl\'asica. Es justamente este m\'etodo perturbativo el que nos permite 
omitir a nivel de la acci\'on efectiva todo t\'ermino cuadr\'atico (o de orden
superior) en el extremal ${\cal E}^{\mu\nu}$. Estos t\'erminos dan 
contribuciones al lado derecho de la ec.(\ref{eq:htecho}) que se anulan
id\'enticamente cuando resolvemos las ecuaciones perturbativamente.

El potencial newtoniano depende de la componente $00$ de la m\'etrica, y
para ello debemos hallar ${\bar h}_{00}$ y $h$.
El resultado es
\bea
{\bar h}_{00} &=& \frac{4 G M}{r} - \frac{2}{15 \pi} \frac{G^2 M}{r^3} +
\frac{5}{3 \pi} \frac{G^2 M}{r^3} + 4 a(\lambda) \frac{G M}{r^3} +
8 b(\lambda) \frac{G M}{r^3} \\
h &=& \frac{4 G M}{r}  -\frac{18}{3 \pi} \frac{G^2 M}{r^3} + 
\frac{5}{\pi} \frac{G^2 M}{r^3} + 4 a(\lambda) \frac{G M}{r^3} 
+ 24 b(\lambda)  \frac{G M}{r^3} ,
\eea
donde, en ambos casos, el primer t\'ermino es el cl\'asico, el segundo y el 
tercero provienen respectivamente de la parte gravitatoria y masiva en la 
medida $\lambda=0$, y los \'ultimos dos corresponden a otras medidas 
 $\lambda \neq 0$. Finalmente
\begin{equation}
h_{00} = {\bar h}_{00} - \frac{1}{2} h  =
 \frac{2 G M}{r} \left[ 1 + \frac{43 G}{30 \pi r^2} - \frac{5 G}{12 \pi r^2} +
\frac{ a(\lambda) - 2 b(\lambda)}{r^2} \right] . 
\label{eq:BRmetric}
\end{equation}
En resumen, el efecto de gravitones (fluctuaciones no
masivas) introduce correcciones a la m\'etrica en $r^{-3}$. Estas correcciones
son cualitativamente las mismas del cap\'{\i}tulo anterior, en el caso que
campos escalares no masivos. Adem\'as son las m\'as importantes
a bajas energ\'{\i}as / largas distancias. En efecto, si 
hubi\'esemos incluido en el 
lagrangiano efectivo los t\'erminos locales cuadr\'aticos en curvatura, 
hubi\'esemos obtenido modificaciones a la m\'etrica que decrecen 
exponencialmente a largas distancias.

Tanto la acci\'on efectiva, las ecuaciones semicl\'asicas, y su soluci\'on 
dependen param\'etricamente de la elecci\'on del fijado de medida para la 
teor\'{\i}a cu\'antica. Usualmente se argumenta que la gravedad pura (es decir,
en ausencia de campos de materia) no presenta esta patolog\'{\i}a, pues
las correcciones cu\'anticas a las ecuaciones de movimiento se anulan
al evaluarse en l!s soluciones cl\'asicas $R_{\mu\nu}=0$. Sin embargo, la
inclusi\'on de campos de materia hace que este argumento pierda validez. Las
partes cu\'anticas de las ecuaciones de Einstein modificadas no son 
simplemente proporcionales al extremal, de modo que la dependencia en la medida
persiste. En conclusi\'on, la soluci\'on a dichas ecuaciones no puede
ser f\'{\i}sica. La raz\'on de ello es simple: cualquier aparato cl\'asico
utilizado para medir la geometr\'{\i}a del espacio-tiempo tambi\'en sentir\'a
las fluctuaciones de gravitones. Como el acoplamiento entre dicho aparato y la 
m\'etrica es no lineal, el aparato {\it no} medir\'a la geometr\'{\i}a de
fondo $g_{\mu\nu}$ (que resulta de resolver las ecuaciones de Einstein). En 
par\-ti\-cular, 
una part\'{\i}cula de prueba {\it no} seguir\'a las geod\'esicas
de dicha m\'etrica, sino que su movimiento estar\'a determinado por otras
ecuaciones geod\'esicas que incluyan los efectos del acoplamiento
mencionado. Estas nuevas ecuaciones geod\'esicas ser\'an el 
tema que describiremos en la pr\'oxima secci\'on.

El an\'alisis precedente nos conduce a una soluci\'on del denominado problema
de fijado de medida, sobre el cual discutimos en el cap\'{\i}tulo 2. La 
dependencia en el par\'ametro $\lambda$ de la acci\'on y de las ecuaciones 
de movimiento es un ejemplo particular de dicho problema ``t\'ecnico'' del
formalismo de la acci\'on efectiva (in-out o in-in). El m\'etodo est\'andar
para remediar la arbitrariedad en el fijado de medida es introducir la 
acci\'on efectiva de Vilkovisky-DeWitt, que est\'a espec\'{\i}ficamente
construida para solucionar el problema. Sin embargo, como ya vimos en el
cap\'{\i}tulo 2, esa acci\'on tiene otro tipo de arbitrariedad, que es la
dependencia en la superm\'etrica de la variedad de campos. 
Por lo tanto, tampoco
es una soluci\'on satisfactoria al problema del fijado de medida. En realidad,
este problema no es ``t\'ecnico'', sino f\'{\i}sico: a causa del
acoplamiento entre
el aparato cl\'asico y los gravitones, la soluci\'on a las ecuaciones de 
movimiento no tiene una interpretaci\'on clara. La verdadera soluci\'on del
problema pasa por la identificaci\'on de los observables de la teor\'{\i}a.
Dadas las ecuaciones de Einstein, es necesario extraer de su soluci\'on las 
cantidades f\'{\i}sicas, que obviamente deben ser independientes del fijado 
de medida.

\section{Correcciones cu\'anticas a las ecuaciones de las geod\'esicas}

Consideremos una part\'{\i}cula de prueba cl\'asica de masa $m_{\sst{\rm p}}$ 
en presencia
del campo gravitatorio cu\'antico ${\bar g}_{\mu\nu}$. Proponemos como un
observable f\'{\i}sico la trayectoria de esta part\'{\i}cula. 
Supondremos que $m_{\sst{\rm p}} \ll M$, 
lo cual nos permite despreciar todas las 
contribuciones de la part\'{\i}cula de prueba a la soluci\'on 
ec.(\ref{eq:BRmetric}) de las ecuaciones de Einstein. 
Ahora introducimos el punto clave: para determinar c\'omo se
mueve la masa $m_{\sst{\rm p}}$, debemos tener 
en cuenta que se acopla a ${\bar g}_{\mu\nu}$
a trav\'es de un t\'ermino en la acci\'on dado por
 $-m_{\sst{\rm p}} \int \sqrt{- {\bar g}_{\mu\nu}(z) dz^{\mu} dz^{\nu} }$, 
denotando
 $z^{\mu}$ la trayectoria de la part\'{\i}cula. Habr\'a entonces una 
contribuci\'on extra a la acci\'on efectiva debida al acoplamiento 
con los gravitones,
la cual a su vez llevar\'a a una correcci\'on a las ecuaciones geod\'esicas
para la part\'{\i}cula. La modificaci\'on a la acci\'on efectiva la obtenemos
de las ecs.(\ref{eq:nolocM}) y (\ref{eq:nolocL}) reemplazando
 $M_{\mu\nu\rho\sigma}$ por $m_{\mu\nu\rho\sigma} + M_{\mu\nu\rho\sigma}$
y $T^{\mu\nu}$ por  $T^{\mu\nu} + T^{\mu\nu}_{m_{\sst {\rm p}}}$ y 
guardando t\'erminos 
lineales en $m_{\sst{\rm p}}$.   
Aqu\'{\i} el tensor $m_{\mu\nu\rho\sigma}$ est\'a dado por la 
ec.(\ref{eq:meff}) con $M$ reemplazada por $m_{\sst{ \rm p}}$ y
 $x_{\mu}$ por $z_{\mu}$.
$T^{\mu\nu}_{m_{\sst {\rm p}}}$ es el tensor energ\'{\i}a-momento de 
la part\'{\i}cula
de prueba, dado por la ec.(\ref{eq:tmunu}), con el mismo reemplazo. Esta 
contribuci\'on es
\begin{eqnarray}
\Delta S_{m_{\sst {\rm p}}} &=& \int d^4x \sqrt{-g}
\left[ 
-\frac{1}{32 \pi^2} m_{\mu\nu\rho\sigma} \ln(-\Box) M^{\rho\sigma\mu\nu} 
\right. \nonumber \\
&& ~~~~~~~~~~~~~~~  - 
\frac{1}{32 \pi^2}  m_{\mu\nu\rho\sigma} \ln(-\Box) 
\left( P^{\rho\sigma\mu\nu} +
\frac{1}{6} R \delta^{\rho(\mu} \delta^{\sigma\nu)} \right) \nonumber \\ 
&& ~~~~~~~~~~~~~~~  \left . + \frac{a(\lambda)}{2} R_{\mu\nu} \ln(-\Box) 
T^{\mu\nu}_{m_{\sst {\rm p}}} + 
\frac{b(\lambda)}{2} R g_{\mu\nu} \ln(-\Box) T^{\mu\nu}_{m_{\sst {\rm p}}} 
\right] ,
\label{eq:eseeme}
\end{eqnarray}
Los primeros dos t\'erminos corresponden a la medida $\lambda=0$, y los 
\'ultimos dos son t\'erminos extras que aparecen para cualquier otra medida.

Las ecuaciones geod\'esicas para la part\'{\i}cula de prueba las obtenemos 
derivando funcionalmente la acci\'on respecto a las coordenadas de la 
part\'{\i}cula
\begin{equation}
0=\frac{1}{m_{\sst{\rm p}}} \frac{\delta S_{{\rm ef}}}{\delta z_{\rho}}
= - \left[
\frac{d^2 z^{\rho}}{d \tau^2} + \Gamma^{\rho}_{\mu\sigma}
\frac{d z^{\mu}}{d \tau} \frac{d z^{\sigma}}{d \tau} \right]
+\frac{1}{m_{\sst{\rm p}}} \frac{\delta 
\Delta S_{m_{\sst{\rm p}}}}{\delta z_{\rho}} ,
\end{equation}
donde $\Gamma^{\rho}_{\mu\sigma}$ es el s\'{\i}mbolo de Christoffel y 
$d\tau^2=-g_{\mu\nu} dz^{\mu} dz^{\nu}$. En el l\'{\i}mite de campo d\'ebil 
y velocidades no relativistas, las ecuaciones geod\'esicas con correcciones 
cu\'anticas adoptan la forma
\begin{equation}
\frac{d^2{\bf z}}{dt^2} - \frac{1}{2} {\bf \nabla}h_{00} =
\frac{1}{m_{\sst {\rm p}}} \frac{\delta 
\Delta S_{m_{\sst{\rm p}}}}{\delta {\bf z}} .
\end{equation}
Notar que $h_{00}$, dado en la ec.(\ref{eq:BRmetric}), depende de $a(\lambda)$ 
y $b(\lambda)$.

Procedemos a calcular el t\'ermino derecho de esta ecuaci\'on. Para este fin
calculamos los distintos t\'erminos en $\Delta S_{m_{\sst{\rm p}}}$. 
Usando la expresi\'on
para $M^{\mu\nu\rho\sigma}$ correspondiente a la fuente est\'atica, el primer
t\'ermino de la ecuaci\'on (\ref{eq:eseeme}) es
\begin{eqnarray}
\Delta S_{m_{\sst {\rm p}},M}(\lambda=0) &\equiv& 
-\frac{1}{32 \pi^2} \int d^4y \sqrt{-g} m_{\mu\nu\rho\sigma} \ln(-\Box)
M^{\rho\sigma\mu\nu}  \nonumber \\
&=& -\frac{1}{32 \pi^2} \frac{m_{\sst {\rm p}} M \kappa^2}{64} \int d^4y 
\sqrt{-g} 
\ln(-\Box) \delta^3({\bf y}) \nonumber \\
&& ~~ \times
\int d\tau \delta^4(y-z(\tau)) 
[2 {\dot z}_0 {\dot z}_0 + 2 g_{00} {\dot z}_0 {\dot z}_0 +
4 {\dot z}_0 {\dot z}_0 {\dot z}_0 {\dot z}_0 ]  \nonumber\\
& \approx & - \frac{m_{\sst {\rm p}} M  \kappa^4}{512 \pi^2} \int d\tau 
\ln(-\Box) 
\delta^3(z(\tau)) .
\label{eq:mM}
\end{eqnarray}
Ac\'a hemos usamos el hecho que, en el l\'{\i}mite no relativista,
 ${\dot z}_0 \approx -1$. Como $\Delta S_{m_{\sst {\rm p}}}$ es proporcional 
a $\hbar$,
en esta ecuaci\'on tambi\'en hemos tomado que la m\'etrica 
$g_{\mu\nu}$ es igual a la cl\'asica $\eta_{\mu\nu}$ - cualquier otra
correcci\'on dar\'{\i}a t\'erminos ${\cal O}(\hbar^2)$. 
En una forma similar
\begin{eqnarray}
\Delta S_{m_{\sst {\rm p}},R}(\lambda=0) & \equiv & 
- \frac{1}{192 \pi^2} \int d^4y \sqrt{-g} m_{\mu\nu\rho\sigma} \ln(-\Box)
R \delta^{\rho(\mu} \delta^{\sigma\nu)}  \nonumber\\
&=& - \frac{m_{\sst {\rm p}} \kappa^2}{1536 \pi^2} \int d\tau 
\ln(-\Box) R(z(\tau)) .
\end{eqnarray}
Los otros t\'erminos de la ec.(\ref{eq:eseeme}) son 
\begin{eqnarray} 
\Delta S_{m_{\sst {\rm p}},P}(\lambda=0) & \equiv & 
- \frac{1}{32 \pi^2} \int d^4y \sqrt{-g} m^{\mu\nu\rho\sigma} \ln(-\Box) 
P_{\rho\sigma\mu\nu}  \nonumber\\
&=& - \frac{m_{\sst {\rm p}} \kappa^2}{256 \pi^2} \int d\tau 
[g^{\mu\nu} {\dot z}^{\rho}  {\dot z}^{\sigma} + 2
{\dot z}^{\mu} {\dot z}^{\nu} {\dot z}^{\rho} {\dot z}^{\sigma}]
\ln(-\Box) P_{\rho\sigma\mu\nu} 
\end{eqnarray}
\begin{equation}
\Delta S_{m_{\sst {\rm p}},a(\lambda)}  \equiv 
\frac{a(\lambda)}{2} \int d^4y \sqrt{-g} 
R_{\mu\nu} \ln(-\Box) T^{\mu\nu}_{m_{\sst {\rm p}}} = 
a(\lambda) \frac{m_{\sst {\rm p}}}{2} \int d\tau {\dot z}^{\mu}{\dot z}^{\nu} 
\ln(-\Box) R_{\mu\nu}(z(\tau))  
\end{equation} 
\begin{equation}
\Delta S_{m_{\sst {\rm p}},b(\lambda)} \equiv 
\frac{b(\lambda)}{2} \int d^4y \sqrt{-g} 
R g_{\mu\nu} \ln(-\Box) T^{\mu\nu}_{m_{\sst {\rm p}}} = 
- b(\lambda) \frac{m_{\sst {\rm p}}}{2} \int d\tau \ln(-\Box) R(z(\tau)) .
\end{equation}
En estas ecuaciones el escalar de Ricci es el de la m\'etrica cl\'asica,
 $R(z(\tau))=-\frac{1}{2} \nabla^2 h^{(0)}(z(\tau))=8\pi G M 
  \delta^3({\bf z}(\tau))$. Lo mismo vale para el tensor de Ricci
 $R_{\mu\nu}(z(\tau))$. 

Ahora tomamos la variaci\'on respecto de ${\bf z}$. Obtenemos
\begin{eqnarray}
\frac{\delta}{\delta {\bf z}} \Delta S_{m_{\sst {\rm p}},M}(\lambda=0)  &=& 
\frac{m_{\sst {\rm p}} M G^2}{\pi} {\bf \nabla}(\frac{1}{r^3})  \nonumber\\
\frac{\delta}{\delta {\bf z}} \Delta S_{m_{\sst {\rm p}},R}(\lambda=0)  &=& 
 \frac{m_{\sst {\rm p}} M G^2}{12 \pi} {\bf \nabla}(\frac{1}{r^3})  \nonumber\\
\frac{\delta}{\delta {\bf z}} \Delta S_{m_{\sst {\rm p}},P}(\lambda=0)  &=& 
- \frac{m_{\sst {\rm p}} M G^2}{2 \pi} {\bf \nabla}(\frac{1}{r^3})  \nonumber\\
\frac{\delta}{\delta {\bf z}} \Delta S_{m_{\sst {\rm p}},a(\lambda)} &=& 
- a(\lambda) m_{\sst {\rm p}} M G  {\bf \nabla}(\frac{1}{r^3})  \nonumber\\
\frac{\delta}{\delta {\bf z}} \Delta S_{m_{\sst {\rm p}},b(\lambda)} &=& 
2 b(\lambda) m_{\sst {\rm p}} M G  {\bf \nabla}(\frac{1}{r^3}) ,
\end{eqnarray}
donde $r=|{\bf z}|$. En consecuencia
\begin{equation}
\frac{d^2 {\bf z}}{dt^2} - \frac{1}{2} {\bf{\nabla}} h_{00} = 
\frac{1}{m_{\sst {\rm p}}} \frac{\delta 
\Delta S_{m_{\sst {\rm p}}}}{\delta {\bf z}} = 
\left[ \frac{7 G}{12 \pi} - a(\lambda) + 2 b(\lambda) \right]  
{\bf \nabla} \left( \frac{G M}{r^3} \right) .
\end{equation} 
Introduciendo la ec.(\ref{eq:BRmetric}) en esta expresi\'on vemos que 
aquellos t\'erminos dependientes del fijado de medida que provienen de
 $h_{00}$ {\it se cancelan} exactamente con aquellos que salen del acoplamiento
de la part\'{\i}cula de prueba con los gravitones. N\'otese adem\'as que
los t\'erminos en  $a(\lambda)$ y $b(\lambda)$ se cancelan en forma separada.

En una manera relativamente sencilla podemos extender este tipo de c\'alculo
a cualquier medida que no pertenezca a la familia $\lambda$. Como ya hemos
mencionado, la diferencia de la acci\'on efectiva para $\lambda=0$ y para
cualquier otra medida es proporcional al extremal ${\cal E}^{\mu\nu}$, que se
anula en la capa de masa. A orden cuadr\'atico en curvaturas, este hecho fija
la forma m\'as general de esta diferencia. Si nos concentramos en los 
t\'erminos logar\'{\i}tmicos no anal\'{\i}ticos, entonces
\begin{equation}
\hspace{-0.5cm}
\Delta S|^{\rm cualquier}_{\rm medida} - \Delta S(\lambda=0) =
\int d^4x \sqrt{-g} [a R_{\mu\nu} \ln(-\Box) {\cal E}^{\mu\nu} +
b R g_{\mu\nu} \ln(-\Box) {\cal E}^{\mu\nu} +
{\cal O} (({\cal E}^{\mu\nu})^2) ] ,
\end{equation}
donde $a$ y $b$ son constantes que dependen de la particular medida que se
est\'e utilizando. As\'{\i}, por ejemplo, para la familia $\lambda$,
 $a=a(\lambda)=- 5 \lambda \kappa^2/48 \pi^2$ y  
$b=b(\lambda)= 5 \lambda \kappa^2/96 \pi^2$. En virtud de los c\'alculos
precedentes, concluimos que sin importar los valores espec\'{\i}ficos de
 $a$ y $b$, en las ecuaciones geod\'esicas se produce la cancelaci\'on de 
todos los t\'erminos que dependen de estos valores. En esta forma obtenemos
finalmente un potencial newtoniano $V(r)$ que es f\'{\i}sico, independiente
del fijado de medida, que lo extraemos de 
 $d^2 {\bf z}/dt^2 = - {\bf \nabla} V$. Nuestro resultado es
\begin{equation}
V(r) = -\frac{G M}{r} \left[1 + \frac{43 G}{30 \pi r^2} - 
\frac{5 G}{12 \pi r^2} +\frac{7 G}{12 \pi r^2} \right] .
\label{eq:potnew2}
\end{equation}
donde los tres \'ultimos t\'erminos son correcciones cu\'anticas.
Comparando las
ecs.(\ref{eq:BRmetric}) y (\ref{eq:potnew2}) concluimos que el acoplamiento
de la
part\'{\i}cula de prueba con los gravitones produce un t\'ermino adicional
(el \'ultimo en la ec.(\ref{eq:potnew2})) y hace que el potencial sea
independiente del fijado de medida. Al igual que en los casos anteriores,
estas correcciones a largas distancias son extremadamente peque\~nas como para
ser medidas. Sin embargo, la importancia del resultado radica en que tanto
el movimiento de la part\'{\i}cula de prueba como el potencial newtoniano
son independientes del fijado de medida.

%%%%%%%%%%%%%%%%%%%%%%%%%%%%%%%%%%%%%%%%%%%%%%%%%%%%%%%%%%%%%%%%%%%%%%%%%

\section{Discusi\'on}

Para resolver el problema del ``backreaction'' incluyendo el efecto de 
gravitones, {\it no basta} con resolver las ecuaciones de Einstein
semicl\'asicas, dado que dependen del fijado de medida y no son
f\'{\i}sicas. Es necesario adem\'as encontrar los observables f\'{\i}sicos.
Como ilustraci\'on de este hecho hemos elegido la trayectoria de una 
part\'{\i}cula de prueba y hemos mostrado expl\'{\i}citamente que, en el
l\'{\i}mite newtoniano, la acci\'on efectiva usual conduce a un resultado
que es independiente del fijado de medida. 

Cabe preguntarse si el potencial newtoniano que obtuvimos depende de la
parametrizaci\'on para los gravitones. En la parametrizaci\'on usada por
nosotros, y para la medida de DeWitt ($\lambda=0$), 
el t\'ermino cin\'etico de la
acci\'on cl\'asica es diagonal (ver ec.(\ref{opdif})). Entonces la m\'etrica
de la variedad de campos en la acci\'on efectiva de Vilkovisky-DeWitt
es la plana y por ende dicha acci\'on coincide con la acci\'on efectiva
que calculamos aqu\'{\i}. Por lo tanto es de esperar que la acci\'on efectiva
de VDW reproduzca el mismo potencial newtoniano en todas
las parametrizaciones. M\'as a\'un, tambi\'en es de esperar que,
debido al acoplamiento entre la part\'{\i}cula de prueba y los gravitones, se 
produzca una cancelaci\'on
no s\'olo de los par\'ametros de fijado de medida (lo cual es inmediato en
el formalismo de VDW), sino tambi\'en de los par\'ametros arbitrarios de la
m\'etrica de la variedad de campos. As\'{\i}, el resultado final para
el potencial newtoniano deber\'{\i}a coincidir con el obtenido en este 
cap\'{\i}tulo.

\newpage

\thispagestyle{empty}
~
\newpage

\chapter{Conclusiones}

\thispagestyle{empty}

En esta Tesis hemos estudiado correcciones cu\'anticas a la din\'amica
cl\'asica en teor\'{\i}a cu\'antica de campos. Para ello nos hemos basado
en el formalismo de la acci\'on efectiva. La representaci\'on usual de la
acci\'on efectiva (in-out) presenta dos problemas a la hora de estudiar
la evoluci\'on temporal de valores medios. Por un lado est\'a el problema
de que las ecuaciones de movimiento que se obtienen no poseen estructura 
causal. Por otro lado, en teor\'{\i}as de medida, dependen de los par\'ametros
que fijan la medida. M\'as en general, no son invariantes ante 
reparametrizaciones de los campos.

Hemos descripto otra representaci\'on de la acci\'on efectiva, de camino
temporal cerrado (CTC), que es adecuada para analizar problemas con datos 
de Cauchy. Mediante esta representaci\'on, hemos definido una acci\'on efectiva
de granulado grueso, utilizando una frecuencia de corte 
en el espacio de momentos.
Esta acci\'on es \'util para analizar transiciones de fase y la transici\'on
cu\'antico-cl\'asica en teor\'{\i}a de campos. Hemos deducido una ecuaci\'on
exacta del grupo de renormalizaci\'on (RG), 
que describe en forma no perturbativa
(de ah\'{\i} el nombre de ``exacta'') c\'omo la
acci\'on CTC  var\'{\i}a con la escala de granulado 
grueso, y que tiene una complicada estructura integro-diferencial. 
Esta ecuaci\'on contiene toda la informaci\'on de la influencia de los
modos de longitud de onda corta sobre los de larga longitud de onda, y es
el an\'alogo CTC de las diversas formulaciones eucl\'{\i}deas presentes en la
literatura. Al igual que su pariente eucl\'{\i}deo, debido a su extrema 
complejidad, la \'unica esperanza de obtener soluciones es mediante m\'etodos 
de aproximaci\'on. Nosotros hemos utilizado una aproximaci\'on en derivadas,
que nos permiti\'o obtener una mejora ante el grupo de renormalizaci\'on
para el potencial efectivo. Esta aproximaci\'on, muy usada a nivel 
eucl\'{\i}deo,
tiene un serio defecto a nivel CTC: pierde completamente los aspectos
estoc\'asticos que surgen de la interacci\'on entre los modos de longitud
de onda larga (el ``sistema'') y los de corta longitud de onda 
(el ``entorno''). 

Para rescatar estos importantes efectos contenidos en la
acci\'on de granulado grueso es necesario usar otras aproximaciones. En una
aproximaci\'on perturbativa, es relativamente sencillo calcular la acci\'on,
ya que los n\'ucleos de ruido y disipaci\'on tienen distribuciones de 
probabilidad  gaussiana. Sin embargo, no bien se intenta hacer un c\'alculo
no perturbativo basado en la ecuaci\'on exacta RG, los c\'alculos se
vuelven extremadamente complicados, pues los n\'ucleos son ahora funciones 
de $n$ puntos, que no poseen distribuciones de probabilidad gaussianas, y
que dependen de la escala de granulado grueso. El Ansatz m\'as general
posible para la acci\'on, que posee la estructura de la acci\'on de
camino temporal cerrado, tiene una infinitud de n\'ucleos de ruido y 
disipaci\'on. Introduciendo este Ansatz en la ecuaci\'on exacta RG, esperamos
encontrar una jerarqu\'{\i}a entre los distintos tipos de n\'ucleos que 
describir\'a c\'omo, al variar la escala de renor\-ma\-li\-za\-ci\'on desde la
escala ultravioleta hasta la escala f\'{\i}sica, se van generando los
efectos estoc\'asticos, y el ruido y la disipaci\'on se interconectan 
a trav\'es de una relaci\'on de fluctuaci\'on-disipaci\'on no perturbativa.
Para resolver dicha jerarqu\'{\i}a habr\'a que aplicar alg\'un tipo de
truncamiento a nivel de la ecuaci\'on exacta RG. Relacionado
con \'esto, tambi\'en resulta interesante estudiar c\'omo afecta la elecci\'on
de la frecuencia de corte para el granulado. 
Nosotros hemos elegido una frecuencia de corte tipo 
delta (o ``sharp cut-off''). Otras posibilidades, frecuentemente usadas
en las formulaciones eucl\'{\i}deas de la ecuaci\'on exacta RG, son frecuencias
de corte suaves. 
Ser\'{\i}a interesante estudiar la dependencia de los resultados 
con la elecci\'on de la frecuencia de corte a nivel CTC. 
Esta ideas son posibles puntos 
de partida para extensiones de esta parte de la Tesis.

El resto de la Tesis est\'a orientado a tratar el segundo problema de la
acci\'on efectiva, esto es, la dependencia de la acci\'on (in-out o in-in)
y de las ecuaciones de movimiento en los par\'ametros de medida. Con el fin
de describir el m\'etodo de c\'alculo, hemos estudiado ejemplos de dificultad
creciente. Primero tratamos un caso muy conocido, que consiste en el 
apantallamiento de la carga el\'ectrica debido a fluctuaciones de vac\'{\i}o en
QED. Hemos mostrado tres m\'etodos equivalentes para obtener ecuaciones
de movimiento reales y causales a partir de la acci\'on efectiva. El primero
se basa en el c\'alculo de las ecuaciones de movimiento in-in, el segundo
en las in-out, a las cuales hay que tomarles dos veces su parte real y causal,
y un tercer m\'etodo basado en las ecuaciones 
eucl\'{\i}deas, en las cuales hay que
reemplazar los propagadores eucl\'{\i}deos por los retardados. Mostramos que la
resoluci\'on de las ecuaciones de Maxwell corregidas por efectos cu\'anticos
conduce a la misma dependencia del potencial electrost\'atico con la 
distancia que se obtiene mediante argumentos wilsonianos, basados en el
grupo de renormalizaci\'on. Pasamos luego al c\'alculo de correcciones 
cu\'anticas al potencial newtoniano en gravedad semicl\'asica, debidas a
campos de materia cu\'anticos, escalares y espinoriales. Mediante una
t\'ecnica de resumaci\'on del desarrollo de Schwinger-DeWitt, calculamos
las partes no locales de las ecuaciones de Einstein semicl\'asicas. En el
l\'{\i}mite de largas distancias y bajas energ\'{\i}as, en el cual un 
tratamiento semicl\'asico tiene sentido, hallamos dos tipos de correcciones
cu\'anticas, una que decrece como una potencia de la distancia a la fuente
y otra que decrece logar\'{\i}tmicamente. Hemos comparado estos resultados
con los que se obtienen via argumentos wilsonianos y concluimos que, en
general, este tipo de argumentos no dan resultados satisfactorios, ni 
cualitativa ni cuantitativamente, ya que pierden el comportamiento de 
potencias, que es el m\'as importante. Finalmente nos embarcamos en c\'alculos
similares para una teor\'{\i}a de medida, la gravedad cu\'antica. Describimos
las ideas de teor\'{\i}as efectivas y mostramos que la relatividad general
posee la estructura de una teor\'{\i}a efectiva. Por ello, si bien la 
relatividad general no es renormalizable, es posible hacer predicciones
cu\'anticas bien definidas para largas distancias y bajas energ\'{\i}as,
que son independientes de la verdadera teor\'{\i}a cu\'antica de la gravedad,
hasta hoy desconocida. Mostramos que, debido a la inclusi\'on de gravitones,
las ecuaciones de Einstein semicl\'asicas dependen del fijado de medida,
y por lo tanto carecen de interpretaci\'on f\'{\i}sica. Una part\'{\i}cula
de prueba no sigue las geod\'esicas de la m\'etrica que resulta de resolver
esas ecuaciones. La soluci\'on a este problema es encontrar observables
f\'{\i}sicos, que obviamente son independientes de los par\'ametros de fijado
de medida. Hemos estudiado las ecuaciones geod\'esicas de una part\'{\i}cula
de prueba y, teniendo en cuenta su acoplamiento con los gravitones, hemos
mostrado expl\'{\i}citamente que tales ecuaciones resultan independientes
del fijado de medida. Con ellas hallamos las correspondientes correcciones
cu\'anticas al potencial newton\-iano por efectos de gravitones, que son del
tipo de potencias.

Entre las posibles extensiones de las ideas expuestas en esta parte de
la Tesis,
ser\'{\i}a interesante estudiar las correcciones cu\'anticas a las ecuaciones
geod\'esicas en un contexto cosmol\'ogico, de ser posible m\'as all\'a
de la aproximaci\'on newtoniana. Otro posible \'ambito de aplicaci\'on es en el
an\'alisis de las ecuaciones para valores medios en cualquier teor\'{\i}a de
medida. Por ejemplo, en teor\'{\i}as de Yang-Mills, para estudiar efectos de
gluones sobre soluciones cl\'asicas.

\clearpage

\renewcommand{\baselinestretch}{1} 

\small
\fancyhead{}
\fancyhead[LE]{\bf \thepage}
\fancyhead[RE]{\sl Bibliograf\'{\i}a}
\fancyhead[RO]{\bf \thepage}

\renewcommand{\baselinestretch}{1.5}

\newpage

\normalsize

\renewcommand{\baselinestretch}{1.2}

\thispagestyle{empty}

~
\newpage
\thispagestyle{empty}
~
\newpage
\thispagestyle{empty}

\vspace{1cm}

{\it Ante todo deseo agradecerle muy especialmente a Diego por 
todo su apoyo, dedicaci\'on y disposici\'on durante todos estos a\~nos. A
su incre\'{\i}ble capacidad did\'actica le debo haber comprendido temas que
cre\'{\i}a imposibles. Tambi\'en a Juan Pablo, de quien aprend\'{\i} mucha
f\'{\i}sica y con quien discut\'{\i} varios de los temas de esta tesis.
A los miembros de la cueva, Fernando, Sting, Matius, Fecho y Gabriela,   
a Dora, Mercedes, Silvina, Lucho, N\'estor y Rolo, con todos ellos 
compart\'{\i} muy buenos momentos. 
A C\'esar y Miguel, por su infinita paciencia para responder preguntas sobre
computaci\'on; muchachos, prometo leer m\'as los manuales!
A Jimmy, por la aventura del libro, y a
Gabriel, por todos sus consejos durante la escritura del mismo. 
A Marcela, Marina, Carlos, Leo, Alfonso y Mart\'{\i}n. A mi 
familia. A todos ellos, vielen Dank!}

\end{document}